\documentclass[preprint,12pt]{elsarticle}
\usepackage{line no,hyperref}
\usepackage{fullpage}
\usepackage{amsmath,amsthm,amssymb,graphics}
\usepackage{epsfig}
\usepackage{mathrsfs}
\usepackage{color}
\modulolinenumbers[5]
\usepackage[margin=0.75in]{geometry}
\usepackage[figuresright]{rotating}
\usepackage{latexsym}
\usepackage{lscape}
\usepackage{xurl}
\usepackage[rightcaption]{sidecap}
\graphicspath{ {images/} }

\usepackage{mathtools}

\mathtoolsset{showonlyrefs}

\usepackage{array}
\usepackage{enumitem}
\usepackage{float}
\usepackage{graphicx}
\usepackage{subfig}
\usepackage[font=small,labelfont=bf]{caption}
\usepackage{txfonts}
\usepackage{caption}  
\usepackage[utf8]{inputenc}
\usepackage{hyperref}  
\usepackage{comment}
\newcommand{\Sh}{{S\!}}                        
\newcommand{\Vh}{{V}}                        
\newcommand{\Ah}{{I}_{a}}                    
\newcommand{\Ih}{{I}_{s}}                    
\newcommand{\Hh}{{H}}                        
\newcommand{\Rh}{{R}}                        
\renewcommand{\P}{\mathbb{P}}   
\newcommand{\N}{\mathcal{N}}    
\newcommand{\Nh}{\mathcal{N}_{h}}                    
\newcommand{\Nht}{\mathcal{N}_{h}(t)}                    
\newcommand{\Pn}{p} 
\newcommand{\Sm}{S_{\!m}}
\newcommand{\sm}{s_{m}}
\newcommand{\im}{i_{m}}
\renewcommand{\Im}{I_{m}}                           
\newcommand{\Nm}{\mathcal{N}_{m}(t)}                    
\newcommand{\per}[1]{\grave{#1}}

\def\d{\mathrm{d}}

\def\kappaa{\kappa_a}
\def\kappah{\kappa_H}
\def\kappas{\kappa_s}
\def\muh{\mu_{h}}
\def\muH{\mu_{H}}
\def\mum{\mu_m}

\def\R{\mathbb R}
\def\Rfo{R_{\text{eff}}^0}
\def\Rf{R_{\text{eff}}}
\def\RL0{R_{Lotka}^{0}}
\def\RN0{R_{NGM}^{0}}
\def\RMC{R_{\text{eff}}^{M,C}}
\def\RD{R_{\text{eff}}^{D}}
\def\RMD{R_{\text{eff}}^{M,D}}
\def\RMDC{R_{\text{eff}}^{M,D,C}}

\def\L{{L}}
\def\D{{\mathbb{D}}}
\def\K{{\mathcal{K}}}

\newtheorem{thm}{Theorem}[section]

\newtheorem{remark}[thm]{Remark}

\newtheorem{lemma}[thm]{Lemma}
\newtheorem{Assumption}[thm]{Assumption}

\large\normalsize

\makeatletter
\def\ps@pprintTitle{%
  \let\@oddhead\@empty
  \let\@evenhead\@empty
  \let\@oddfoot\@empty
  \let\@evenfoot\@oddfoot
}
\makeatother

\allowdisplaybreaks
\begin{document}
	
\begin{frontmatter}

\title{{Age-structured model of dengue transmission dynamics with time-varying parameters, and its application to Brazil}}

		\author[a]{Ihtisham Ul Haq\corref{cora}}
		\ead{ihtisham.ul.haq.d2@math.nagoya-u.ac.jp}
		\author[b]{Serge Richard}
		\ead{richard@math.nagoya-u.ac.jp}
    \cortext[cora]{Corresponding author}
    \address[a]{ Graduate School of Mathematics, Nagoya University, Furo-cho, Chikusa-ku, Nagoya, 464-8602, Japan}
    \address[b]{ Institute for Liberal Arts and Sciences, Nagoya University, Furo-cho, Chikusa-ku, Nagoya, 464-8601, Japan}

\begin{abstract}
An age-structured mathematical model with time-dependent parameters is developed to investigate the dynamics of dengue transmission. 
Its properties are thoroughly analyzed in the first part of this work, as for example its disease-free steady state,
the corresponding effective reproduction numbers, its basic reproduction number (obtained via the Euler–Lotka equation 
and the next-generation matrix approach). We also provide formulas for the time-varying effective reproduction number,
and draw relations with the instantaneous growth rate.
In the second part, we apply this model to Brazil and use weekly time series data from this country.
Various medical parameters are firstly evaluated from these data, 
and an extensive numerical simulations for the period 2021--2024 is then carried out. 
Estimation of the transmission rates are derived both from epidemiological data 
and from environmental data such as temperature and humidity. 
The time-varying effective reproduction numbers are then estimated on these data,
following the theoretical investigations performed in the first part.
The sensitive parameters that significantly affect the model dynamics are presented graphically. 
Model predictions for following year by using different transmission rates are finally presented.
Our findings show the importance of population age‑distribution, vector population dynamics, and climate, 
contributing  to a deeper understanding of dengue transmission dynamics in Brazil. 
\end{abstract}


\end{frontmatter}

\section{Introduction}
\begin{figure}[htbp]
    \centering
    \includegraphics[scale=0.58]{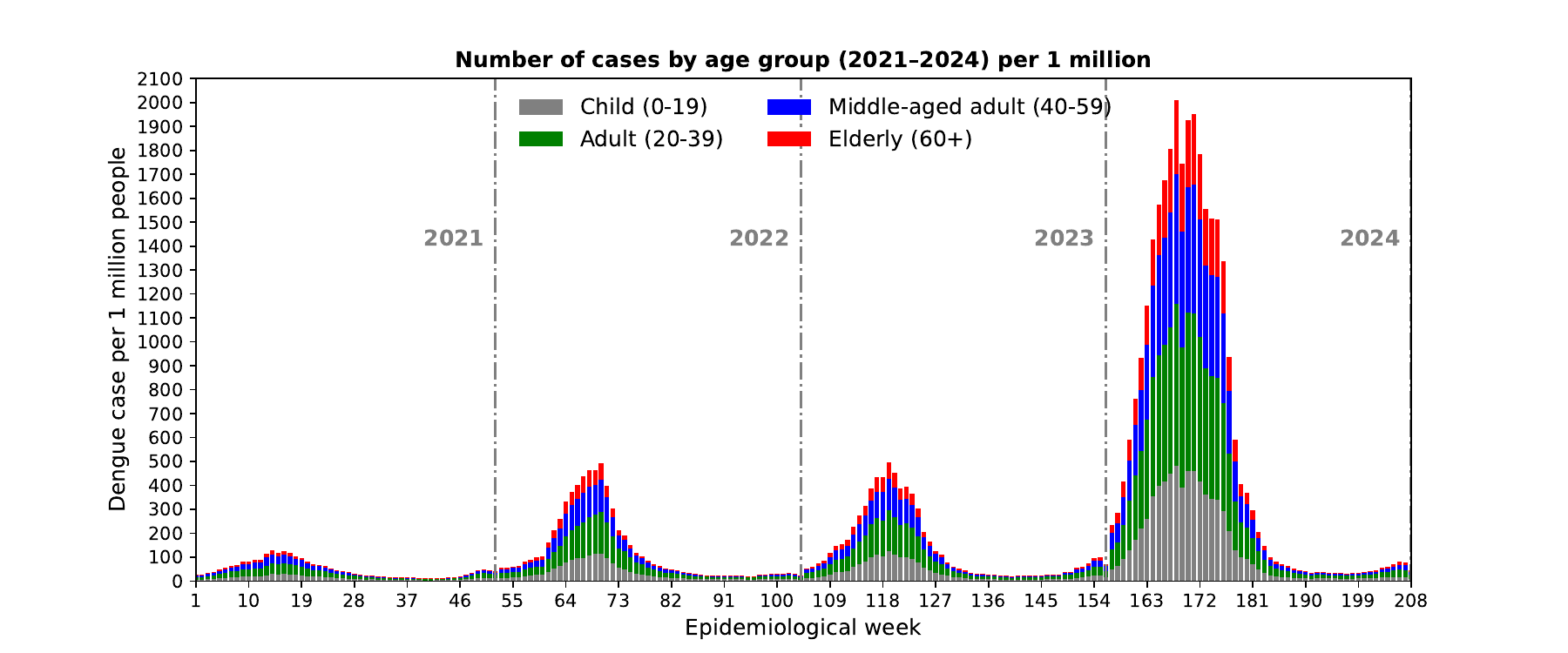}
    \caption[Age-wise distribution of confirmed dengue cases in Brazil]{\small
    \textbf{Age-wise distribution of confirmed dengue cases in Brazil.} Number of confirmed dengue cases per 1 million people 
    across various age groups from 2021 to 2024, reported on a weekly basis. 
    The data were obtained from the Brazilian Ministry of Health’s official database \cite{Brazil_source}.}
    \label{fig1}
\end{figure}

Dengue is a mosquito borne disease that has become a global problem in recent years and an increasing public health concern.
Approximately 390 million dengue infections occur each year, including more than $96$ million symptomatic cases \cite{WHO2024}. 
It affects mainly tropical and subtropical regions. 
In particular, Brazil has seen a significant increase in the past decade with more than 11 million cases reported between 
2000 and 2016  \cite{delpino2023}. 
In 2023, Brazil experienced an alarming situation of 1,508,640 confirmed dengue cases with 1,096 deaths reported. 
The majority of the cases were individuals between the ages of 20 and 39. 
This trend of increasing cases continued in 2024 with 6,442,087 confirmed cases and 187,715 hospitalized, see Figure \ref{fig1}.
In the same year, 6,236 deaths were recorded, corresponding to a death rate of 0.096\% for confirmed cases. 
The mortality rate among hospitalized patients was significantly higher and reached 2.74\% \cite{goncalves2025}. 
The highest mortality rate was observed among individuals aged 70+, followed by those aged between 40 and 59 years. 
These figures highlight the continued severity of dengue as a public health concern in Brazil. 
As a result, the Brazilian Ministry of Health provides weekly age-specific data on dengue, 
which are useful for observation and analysis \cite{sansone2024}.  

The disease is transmitted through the bite of infected mosquitoes, 
especially through Aedes albopictus and Aedes aegypti mosquitoes \cite{harjule2024}. 
Currently, there is no vaccine available that is 100\% effective against this dengue disease. 
There is a vaccine for the dengue called Dengvaxia, but it only works well for people who have had dengue before. 
It is not recommended for people who have not been infected yet \cite{Zhang2023}.

Understanding and addressing the complexity of these transformable diseases requires 
the use of mathematical modeling to predict outbreaks, allocate resources effectively, 
and develop practical solutions to assess the impact of the disease \cite{goncalves2025}. 
By combining computational and mathematical tools with epidemiology, 
researchers have gained important insight into the factors facilitating dengue spread. 
These approaches play an important role for predicting outbreaks and for managing the burden of the disease  \cite{bajpai2020}. 
Numerous studies have utilized mathematical models for describing the dynamics of dengue disease 
and for understanding the interactions between the host, the vector, and the pathogen \cite{Yi2021}. 
These models help predicting infectious disease trends under various conditions,
and are frequently used for assessing the efficacy of public health interventions  \cite{alonso2017}. 
For example, in \cite{buonomo2018}, the authors developed a mathematical model to address dengue 
by including personal use of insecticide-treated bed nets and pesticide spraying. 
Other models capture the interaction of the infectious disease dynamics and of the environmental conditions \cite{stojanovic2020}. 
Note also that evaluating model results can be challenging due to the high correlation among model parameters \cite{berman2020, riley2021}
or due to the lack of enough data \cite{zhou2020}. 

Among various models for the study of dengue, age-structured models provide a more comprehensive understanding 
of disease transmission, since various features of the epidemic depend on age groups and since this knowledge is crucial 
for determining the best prevention strategies, such as for implementing future vaccine plans \cite{nguyen2022}.
In the seminal paper \cite{Inaba1990}, the author presents thresholds and stability results for an age-structured SIR model, 
while this analysis is extended in \cite{Fister2006} for incorporating the immigration of infected individuals 
across all epidemiological compartments.  
In \cite{Zhang2023} another age-structured model is also investigated for predicting the spread of diseases age-wise
and improving control strategies. 
Age is also the main factor in recommending vaccines in countries with high dengue rates, 
as studied by  \cite{aguiar2018dengvaxia}, see also  \cite{thomas2019review}. 

One crucial factor for the study of epidemics is the time-varying effective reproduction number $\Rf(t)$.
This indicator helps to understand how the transmission dynamics of a disease changes over time, and allows to 
predict future trend to timely manage interventions and control strategies. 
In \cite{Nishiura2009} the authors analyze an age-infection SIRS model and focus 
on the time-varying effective reproduction number. They also develop a renewal equation to age infection model. 
Another important and less commonly used concept is the instantaneous growth rate $r_t$.
This indicator has recently emerged as a good alternative to $\Rf(t)$ for evaluating 
the spread of the disease over time \cite{dushoff2021}. 
The instantaneous growth rate, which measures the rate of change of the log-transformed of the incidence curve, 
has been suggested as an informative indicator of transmission dynamics of the disease \cite{pellis2022}.

With the present work, our aim is to provide a deeper understanding of dengue transmission and prevention, 
with a special emphasis on Brazil. 
As mentioned above, the age of the host population is a key factor in the spread and strategies of infectious diseases. 
Different age groups interact in different ways, vaccination programs are often targeted at specific age groups, 
and epidemiological data is typically reported age-wise. 
Studying disease spread models with time-dependent parameters is also important but challenging. 
It helps assessing the effectiveness of health measures and provides valuable information for decision-making during outbreaks. 
Dengue transmission is also influenced by factors such as climate, which affects mosquito life cycles and human mobility, 
which is responsible for virus spread. 
Therefore, we develop a time-dependent host-vector model with an age-structured in human population. 
The model accounts for both mosquito-to-human and human-to-mosquito transmission rates, 
includes both asymptomatic and symptomatic infectious hosts,
and integrates these components within an age-structured framework.
This study also integrates environmental factors, such as temperature and humidity, 
along with epidemiological data to estimate important model parameters. 
Therefore, this study differs from previous models by incorporating these critical factors
that enhance our understanding of dengue dynamics. 

One important contribution of this work is the derivation of renewal equations for dengue transmission 
based on an age-structured model and their application to real-world data. 
We also compute and analyze the evolution of the time-varying effective reproduction number $\Rf(t)$ 
and of the epidemic growth rate $r_t$ during the four dengue outbreaks in Brazil from 2021 to 2024. 
To the best of our knowledge, this is the first derivation of $\Rf(t)$ for an age-structured model.
We also provide a comparison between various approaches for estimating $\Rf(t)$:
either data based, or model based with data based transmission rates, or model-based with temperature-humidity based transmission rates.

Predicting the magnitude and timing of dengue outbreaks is crucial for disease control. 
While previous studies have focused on individual factors such as mosquito population density and humans, 
they have not integrated these factors to capture the complex nonlinear relations driving dengue transmission. 
Our study addresses this gap by accurately predicting dengue cases in Brazil using climate forecasts and historical outbreak data. 
By combining climate predictions with past data our model effectively captures the timing and magnitude of outbreaks.
Thus, this study presents a significant progress in understanding and managing dengue outbreaks in Brazil. 
By forecasting dengue outbreaks based on climate conditions, we can anticipate future disease trends. 
By including environmental factors in the model, we also provide important results on dengue dynamics 
which can help improve future disease strategies.

Our work is divided into two main parts: firstly the theoretical investigations, and secondly
the applications to Brazil. 
For the theoretical part, we introduce in Section \ref{Sec_model} the mathematical model underlying the disease transmission dynamics. 
Section \ref{sec_rescaled} provides a rescaled model which is more suitable for the analysis. 
In Section \ref{sec_df}, we investigate the disease-free steady state of the model. 
The concept and computation of the basic and effective reproduction numbers are addressed in Section \ref{sec_bern}, 
where we develop three distinct approaches.
In Section \ref{sec_comparison} we compare the basic reproduction number obtained via the Euler–Lotka equation 
and the next-generation matrix approach. 
A stability analysis is performed in Section \ref{sec_stability},
while in Section \ref{sec_tvern} we introduce and estimate the time-varying effective reproduction number.
Its relation with the instantaneous growth rate is further investigated in Section \ref{sec_relations}.

For the part related to the application to Brazil,
We start by estimating some medical parameters in Section \ref{sec_med_pre} based on data, as for example death, recovery, 
and hospitalization rates. 
Section \ref{sec_transmission_r} is devoted to the estimation of transmission rates, derived both from epidemiological data 
and from environmental data such as temperature and humidity. 
In Section \ref{sec_21_24}, we carry out extensive numerical simulations for the period 2021--2024, 
and provide sensitivity analysis and evolution of the time-varying effective reproduction numbers. 
In Section \ref{sec_pred} we present the model predictions for next year by using different transmission rates:
(i) transmission rates inferred from temperature and humidity forecasts, 
(ii) four-year average of previously estimated data based transmission rates, 
and (iii) mixed transmission rates combining temperature based and data based estimates. 
Finally, in Section \ref{sec_dis_con} we discuss and provide a conclusion of our study.
In the appendix, Section \ref{sec_app}, some technical outcomes have been gathered. 

\part{Theoretical investigations}

\section{The model}\label{Sec_model}

We develop a ${\Sh}{\Vh}{\Ah}{\Ih}{\Hh}{\Rh}-{\Sm}{\Im}$ model to describe the transmission dynamics of the dengue virus between host and vector populations.
The model focuses on two types of populations: the human population and the mosquito population. 
The human population at age $\zeta$ and time $t$ is ${\P}(t,\zeta)$, which is divided into the compartments: 
susceptible ${S}(t,\zeta)$,
vaccinated ${\Vh}(t,\zeta)$,
asymptomatic infectious $\Ah(t,\zeta)$,
symptomatic infectious ${\Ih}(t,\zeta)$,
hospitalized ${\Hh}(t,\zeta)$,
and recovered ${\Rh}(t,\zeta)$. 
These compartments represent individuals in each state and age.
Therefore, the total host population at age $\zeta$ and time $t$ can be expressed as:
\begin{equation}\label{section1.1}
{\P}(\zeta,t)={\Sh}(t,\zeta)+{\Vh}(t,\zeta)+{\Ah}(t,\zeta)+{\Ih}(t,\zeta)+{\Hh}(t,\zeta)+{\Rh}(t,\zeta).
\end{equation}
The overall human population ${\Nht}$ of all ages is
\begin{equation}\label{eq_total_pop}
{\Nht}:=\intop_{0}^{\infty}{\P}(t,\zeta)\, \d\zeta.
\end{equation}

Susceptible humans become infected through bites by infected mosquitoes. 
The biting rate depends on human mobility (exposure to mosquito aggregation sites) and skin exposure. 
Dengue vaccines are available and have varying levels of efficiency. 
The Dengvaxia dengue vaccine is considered very safe for individuals with laboratory evidence of past dengue virus infection or serotype positivity, 
compared to those who are seronegative. 
The parameter $\alpha(t,\zeta)$ represents the vaccination rate of susceptible individuals 
and reflects the rate at which individuals in the susceptible population are vaccinated at time $t$. 
We assume that dengue vaccination has a limited impact, 
as it is not widely accessible to everyone because the number of cases has continued to rise consistently each year \cite{WHO2024}. 
The parameter $\sigma(t,\zeta)$ denotes the vaccination failure rate and represents the proportion of vaccinated individuals 
who do not develop immunity and may remain susceptible to infection. 

According to the Centers for Disease Control and Prevention (CDC) \cite{CDC2024}, approximately 25\% of dengue virus infections are symptomatic, 
meaning 1 in 4 infected individuals develop symptoms. 
Therefore, we divide the infected individuals into two groups to account the effects of asymptomatic individuals. 
Once susceptible individuals become infected, there is a probability $f(t,\zeta) \in [0,1]$ 
that they develop symptoms and a probability $\Big(1 - f(t,\zeta)\Big)$ that they remain asymptomatic. 
This probability depends on their age $\zeta$ and on the time $t$. 
The likelihood of developing symptoms varies across different age groups because individuals have varying immune system responses. 
Younger individuals may have a stronger immune response, making them less likely to develop severe symptoms, 
whereas older individuals or those with weaker immune systems may have a higher probability of becoming symptomatic. 
The function $f(t,\zeta)$ accounts for these variations and reflects the age dependent nature of immune responses. 

The parameter $\delta(t,\zeta)$ represents the age dependent hospitalization rate of symptomatic cases.
This division helps capture the dynamics of symptomatic and asymptomatic cases. 
Symptomatic individuals with severe symptoms are hospitalized ${\Hh}(t,\zeta)$ \cite{CDC2024}. 
Approximately 500,000 severe cases occur each year, with a mortality rate of 10\%  among hospitalized patients \cite{Tejo2024}. 
Hospitalized individuals avoid public places, reducing their contact with mosquitoes; 
therefore, they do not contribute to the spread of the disease anymore. 
In contrast, the remaining mildly symptomatic individuals often fail to take adequate precautions, 
leading to increased mosquito contact and consequently greater disease transmission. 

The parameter $\mu_h(t,\zeta)$ represents the age-specific natural mortality or death rate, so 
$$
\int_{0}^{\infty} \mu_h(t,\zeta) \P(t,\zeta) \, \d\zeta
$$ 
is the total number of deaths at time $t$. 
In addition, hospitalized individuals die from the disease at a death rate of $\mu_H(t,\zeta)$, 
which represents the mortality rate due to the disease. 
Asymptomatic hosts recover from the disease at a rate of $\kappa_a(t,\zeta)$, symptomatic individuals recover at a rate of $\kappa_s(t,\zeta)$, 
and hospitalized hosts recover at a rate of $\kappa_h(t,\zeta)$. 

For mosquitoes, we divide the total vector population $\Nm$ at time $t$ into two subclasses: 
susceptible mosquitoes $\Sm(t)$ and infectious mosquitoes $\Im(t)$, such that 
$$
\Nm=\Sm(t)+\Im(t).
$$
The mosquito population change over time is given by
\begin{equation}\label{section1.00}
\frac{\d\Nm}{\d t}=\varLambda_{m}(t)-\mum\Nm,
\end{equation} 
where $\varLambda_{m}(t)$ is the time-dependent mosquito recruitment rate.
This rate is time dependent because they vary seasonally with higher values in warm and rainy months 
and lower values in dry or cold months. 
The parameter $\mum$ represents the mortality rate of mosquitoes. 
We assume that once mosquitoes become infected, they remain infected for their entire lifespan and continue to carry the dengue virus.

As a consequence, our model for the dynamics for the human population reads:
\begin{align}
 & \frac{\partial\Sh(t,\zeta)}{\partial t} + \frac{\partial\Sh(t,\zeta)}{\partial \zeta} = -\Big(\beta_{h}(t,\zeta)\frac{\Im(t)}{\Nm} + \alpha(t,\zeta)\Big)\Sh(t,\zeta) - \muh(t,\zeta)\Sh(t,\zeta) \label{section1.2} \\
 & \frac{\partial\Vh(t,\zeta)}{\partial t} + \frac{\partial\Vh(t,\zeta)}{\partial \zeta} = \alpha(t,\zeta)\Sh(t,\zeta) -  \sigma(t,\zeta)\beta_{h}(t,\zeta)\frac{\Im(t)}{\Nm}\Vh(t,\zeta)-\muh(t,\zeta) \Vh (t,\zeta) \label{section1.3} \\
 & \frac{\partial\Ah(t,\zeta)}{\partial t} + \frac{\partial\Ah(t,\zeta)}{\partial \zeta} = \Big(1-f(t,\zeta)\Big)\beta_{h}(t,\zeta)\frac{\Im(t)}{\Nm}\Big(\Sh(t,\zeta) + \sigma(t,\zeta)\Vh(t,\zeta)\Big) - \Big(\kappaa(t,\zeta)+\muh(t,\zeta) \Big)\Ah(t,\zeta) \label{section1.4} \\
 & \frac{\partial\Ih(t,\zeta)}{\partial t} + \frac{\partial\Ih(t,\zeta)}{\partial \zeta}= f(t,\zeta)\beta_{h}(t,\zeta)\frac{\Im(t)}{\Nm}\Big(\Sh(t,\zeta) + \sigma(t,\zeta)\Vh(t,\zeta)\Big) - \Big(\kappas(t,\zeta) + \delta(t,\zeta)+\muh(t,\zeta) \Big)\Ih(t,\zeta) \label{section1.5} \\
 & \frac{\partial\Hh(t,\zeta)}{\partial t} + \frac{\partial\Hh(t,\zeta)}{\partial \zeta} = \delta(t,\zeta)\Ih(t,\zeta) - \Big( \kappah(t,\zeta) + \muh(t,\zeta)+\muH(t,\zeta)\Big)\Hh(t,\zeta) \label{section1.6} \\
 & \frac{\partial\Rh(t,\zeta) }{\partial t} + \frac{\partial\Rh(t,\zeta) }{\partial \zeta}= \kappaa(t,\zeta)\Ah(t,\zeta) + \kappas(t,\zeta)\Ih(t,\zeta) + \kappah(t,\zeta)\Hh(t,\zeta) - \muh(t,\zeta)\Rh(t,\zeta) \label{section1.7}
\end{align}
and the dynamics for the mosquito population is given by:
\begin{align}
 & \frac{\d{\Sm}(t)}{{\d}t}=\varLambda_{m}(t)-{\Sm}(t)\int_{0}^{\infty}\beta_m(t,\zeta)\Big(\frac{{\Ih}(t,\zeta)+{\Ah}(t,\zeta)}{{\Nht}}\Big)\,\d\zeta-\mum{\Sm}(t),\label{section1.8}\\
 & \frac{\d{\Im}(t)}{{\d}t}={\Sm}(t)\int_{0}^{\infty}\beta_m(t,\zeta)\Big(\frac{{\Ih}(t,\zeta)+{\Ah}(t,\zeta)}{{\Nht}}\Big)\,\d\zeta-\mum{\Im}(t).\label{section1.9}
\end{align}

The total infectivity in human is represented by $\frac{\beta_h(t,\zeta){\Im}(t){\Sh}(t,\zeta)}{\Nm},$ where $\beta_h(t,\zeta)=b(t)p_{h}(\zeta)$. 
Here, $b(t)$ represents the mosquitoes time dependent biting rate (the number of bites per mosquito at time $t$) 
and $p_{h}(\zeta)$ is the probability of transmission from an infected mosquito 
to a susceptible human during a single bite.
We do not include mosquitoes age in the model because their short lifespan means age has little effect on disease transmission. 
The transmission rate from an infectious human of age $\zeta$ to a susceptible mosquito is given by 
$\beta_m(t,\zeta) =b(t)p_{m}(\zeta)$, where $p_{m}(\zeta)$ is the probability of transmission from an infected human 
to a susceptible mosquito during a single bite. Thus, the force of infection acting on susceptible mosquitoes from infected hosts is given by
\begin{equation*}
{\Sm}(t)\int_{0}^{\infty}\beta_m(t,\zeta)\Big(\frac{{\Ih}(t,\zeta)+{\Ah}(t,\zeta)}{{\Nht}}\Big)\, \d\zeta,
\end{equation*}
with the assumption that individuals when treated in a hospital do not spread the disease.

\begin{figure}[htbp]
\centering \includegraphics[scale=0.5]{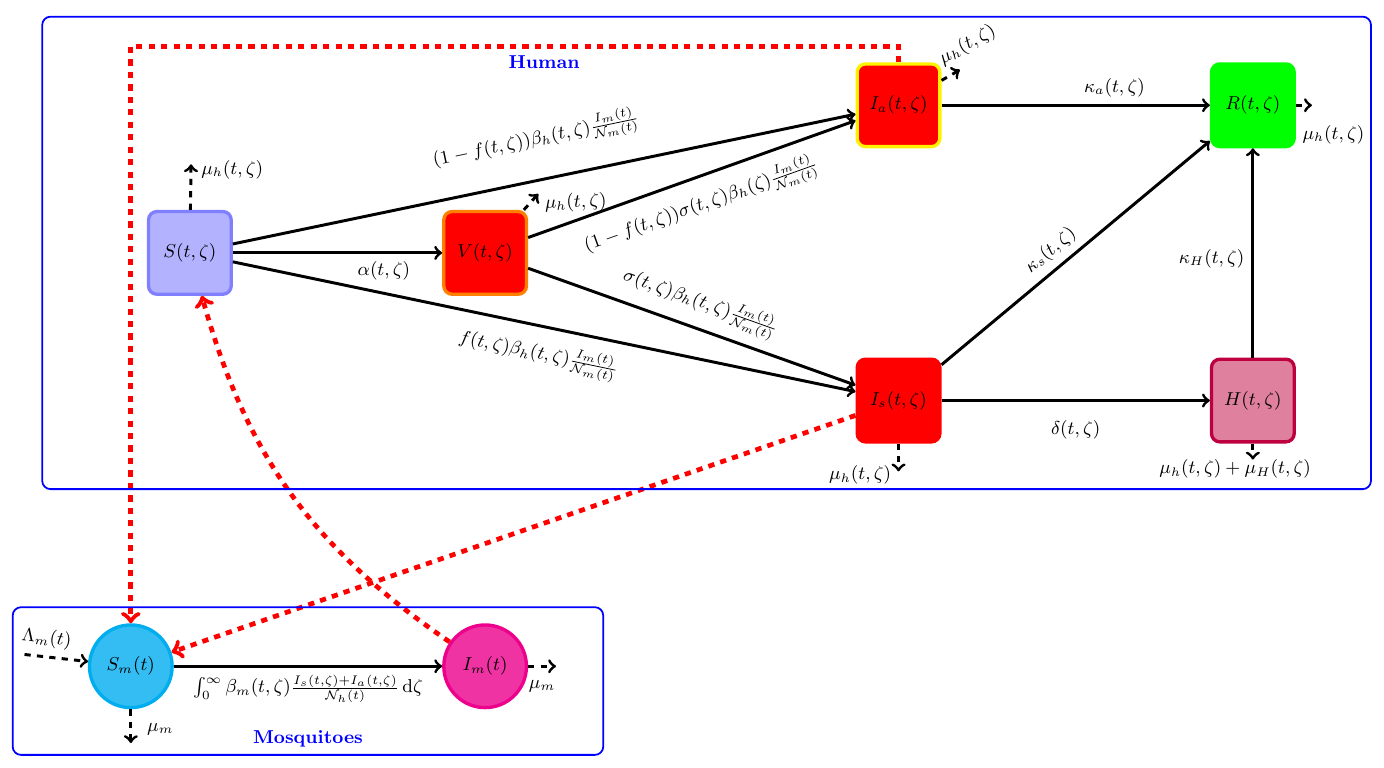} \caption{Flow chart of the proposed model.}
\label{fig2} 
\end{figure}

Hence, \eqref{section1.2} to \eqref{section1.9} represent our mathematical model for dengue fever, 
as shown in the flow chart of Figure \ref{fig2}. 
The initial conditions are given by
\begin{align*}
 & {\Sh}(0,\zeta)={\Sh}_{0}(\zeta),\quad{\Vh}(0,\zeta)={\Vh}_{0}(\zeta),\quad{\Ah}(0,\zeta)={\Ah}_{0}(\zeta),\quad{\Ih}(0,\zeta)={\Ih}_{0}(\zeta),\\
 & {\Hh}(0,\zeta)={\Hh}_{0}(\zeta), \quad{\Rh}(0,\zeta)={\Rh}_{0}(\zeta), \quad {\Sm}(0)={\Sm}_{0},\quad \quad{\Im}(0)={\Im}_{0}.
\end{align*}
Note that initial conditions will always refer to conditions with respect to time.
On the other hand, boundary conditions will always refer to conditions with respect to age, and more precisely to the age $\zeta=0$.
In other terms boundary conditions at $\zeta = 0$ represent individuals at birth or age zero, the time at which newborns enter the model.
For these conditions, let us set $\pi(\zeta)$ for the age-specific fertility rate or birth rate. Then 
$$
B(t):= \int_{0}^{\infty} \pi(\zeta) \P(t,\zeta) \, \d\zeta
$$  
represents the total number of offspring produced by individuals at time $t$.
The boundary conditions then read
\begin{align*}
&{\Sh}(t,0)=B(t),\quad{\Vh}(t,0)=0,\quad{\Ah}(t,0)=0,\quad{\Ih}(t,0)=0,
{\Hh}(t,0)=0, \quad{\Rh}(t,0)=0.
\end{align*}

About the parameters, a summary of them is provided in Table~\eqref{tabel1} below.
\begin{table}[htbp]
\label{table1} \centering {\small{}\caption{Description of the parameters used in the proposed model}
\scalebox{0.74}{ }%
\begin{tabular}{cl}
\hline 
Notation  & Parameters description \tabularnewline
\hline 
\hline 
$\beta_{h}(t,\zeta)$ & Transmission rate from infected mosquitoes to susceptible
hosts \tabularnewline
$\beta_{m}(t,\zeta)$  & Transmission rate from infected host to susceptible mosquitoes\tabularnewline
$\muh(t,\zeta)$  & Natural mortality rate of humans \tabularnewline
$\muH(t,\zeta)$  & Mortality rate of humans due to disease \tabularnewline
$\kappaa(t,\zeta)$  & Recovery rate of asymptomatic individuals \tabularnewline
$\kappas(t,\zeta)$  & Recovery rate of symptomatic individuals \tabularnewline 
$\kappah(t,\zeta)$  & Recovery rate of hospitalized individuals \tabularnewline
$\alpha(t,\zeta)$  & Vaccination rate of susceptible individuals  \tabularnewline
$\sigma(t,\zeta)$  & Failure rate of vaccination \tabularnewline
$\delta(t,\zeta)$ & Hospitalization rate of symptomatic individuals\tabularnewline
$f(t,\zeta)$  & Probability that infected individuals become symptomatic \tabularnewline
$\varLambda_{m}(t) $ & Time dependent mosquito recruitment rate \tabularnewline
$\mum$  & Death rate of mosquitoes \tabularnewline
\hline  
\end{tabular}{\small{}} \label{tabel1} }
\end{table}

In order to be precise, let us make the following assumptions:
\begin{Assumption} \label{as1}
Throughout the paper, we assume that all variable parameters or functions satisfy the conditions:
\begin{itemize} 
\item The parameters $\beta_h,\,\beta_m,\, \muh, \,\muH,\, \kappaa,\,\kappas, \,\kappah,\, \alpha,\, \sigma,\, \delta$,
and $f$ are bounded, non-negative functions of $(t,\zeta)$,
\item The initial functions ${\Sh}_{0},{\Vh}_{0},{\Ah}_{0},{\Ih}_{0}, {\Hh}_{0}$, and ${\Rh}_{0}$
are bounded, non-negative functions of $\zeta$, 
\item The initial conditions for mosquitoes ${\Sm}_{0}$ and ${\im}_{0}$ are non-negative.
\end{itemize} 
\end{Assumption}

Recall that the total human population at age $\zeta$ and time $t$ is given by ${\P}(t,\zeta)$.
By using \eqref{section1.1} to \eqref{section1.9}  we can derive the following dynamics of the total population:
\begin{equation}
\begin{cases}
\frac{\partial{\P}(t,\zeta)}{\partial t}+\frac{\partial{\P}(t,\zeta)}{\partial\zeta}=-\muh(t,\zeta){\P}(t,\zeta)-\muH(t,\zeta)\Hh(t,\zeta),\\
{\P}(t,0)=\int_{0}^{\infty}\pi(\zeta)\P(t,\zeta)\,\d\zeta=B(t),\\
{\P}(0,\zeta)\equiv {\P}_{0}(\zeta)={\Sh}_{0}(\zeta)+{\Vh}_{0}(\zeta)+{\Ah}_{0}(\zeta)+{\Ih}_{0}(\zeta)+{\Hh}_{0}(\zeta)+{\Rh}_{0}(\zeta).
\end{cases}
\label{section1.10}
\end{equation}

\begin{remark}[Idealistic population]\label{rem_ideal_pop}
If there are no demographic changes and if $\muH=0$, then the total human population at age $\zeta$ and time $t$ 
will remain constant, namely ${\P}(t,\zeta)\equiv \P(\zeta)$ is independent of $t$, and 
$\Nh=\intop_{0}^{\infty}{\P}(\zeta)\, \d\zeta$ is a constant independent of $t$, see \eqref{eq_total_pop}.
In this situation, birth and death rates are balanced. 
Ignoring the time variation in \eqref{section1.10} and setting $\muh(t,\zeta)=\muh(\zeta)$, 
we have the age distribution
\begin{equation} \label{section3.2}
\P(\zeta)=Be^{-\int_{0}^{\zeta}\muh(\eta)\,\d\eta}, 
\end{equation}
where $B=\int_{0}^{\infty}\pi(\zeta)\P(\zeta)\,\d\zeta$ denotes the total number of newborns.
In addition, if the death rate is constant and does not depend on $\zeta$, similar to the ODE model, then we have
\begin{equation}\label{section3.3}
\P(\zeta) = B  e^{-\muh \zeta}.
\end{equation}
Finally, if we set $\Pn(\zeta):=\frac{\P(\zeta)}{\Nh}$, then $\Pn(\zeta)$ represents the distribution (according to age) 
of this idealistic population.  However, as we can easily see in figure \ref{fig3}, a realistic population
never corresponds to an idealistic one.
\begin{figure}[htbp]
    \centering
    \includegraphics[scale=0.45]{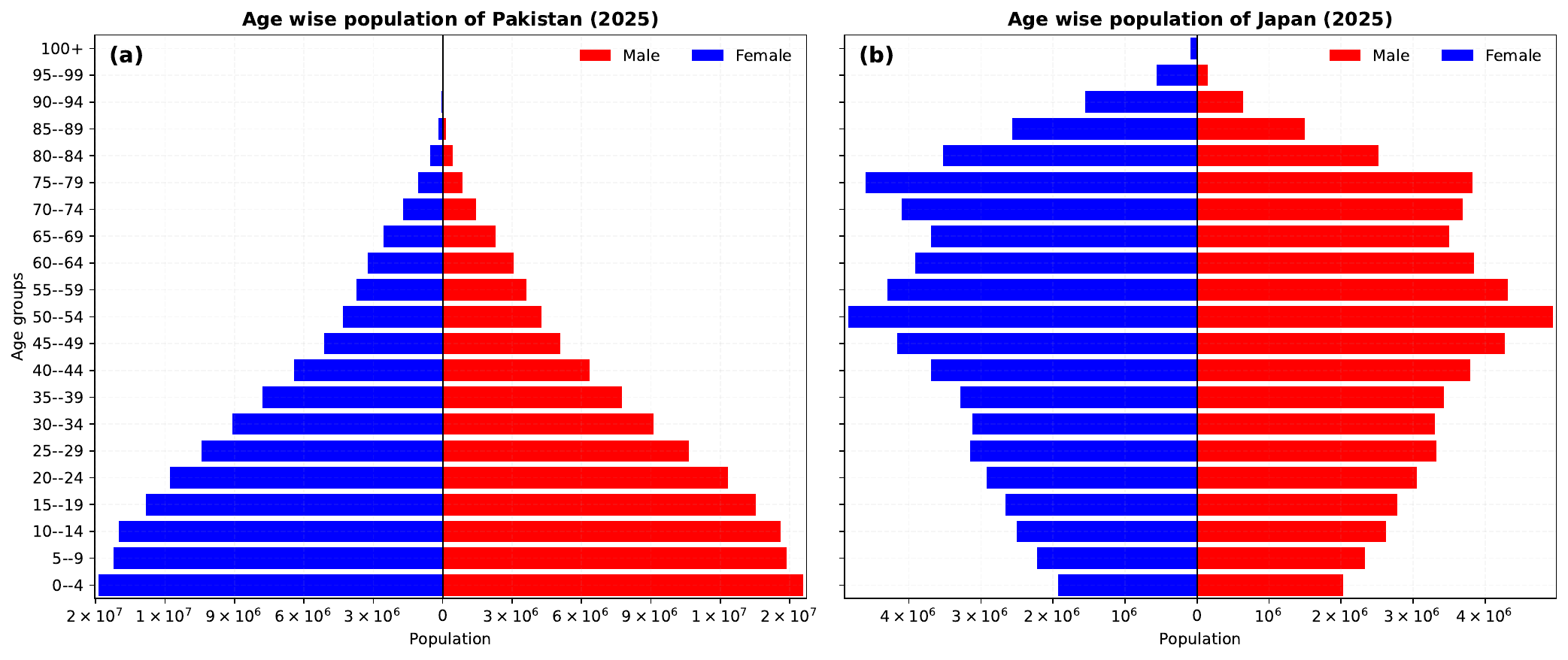}
    \caption[Age-wise population distribution of Pakistan and Japan]{\small
    \textbf{Age-wise population distribution of Pakistan and Japan.} Distribution of the population across various age groups in Pakistan and Japan for the year 2025.}
    \label{fig3}
\end{figure}
\end{remark}

\section{The rescaled model}\label{sec_rescaled}

To simplify the forthcoming analysis, we rewrite the model using population proportions. Therefore, we perform the following normalization of the system:
\begin{align*}
s(t,\zeta)     & := \frac{\Sh(t,\zeta)}{\P(t,\zeta)},       & 
v(t,\zeta)     & := \frac{\Vh(t,\zeta)}{\P(t,\zeta)},       & 
i_{a}(t,\zeta) & := \frac{\Ah(t,\zeta)}{\P(t,\zeta)}, \\
i_{s}(t,\zeta) & := \frac{\Ih(t,\zeta)}{\P(t,\zeta)},       & 
h(t,\zeta)     & := \frac{\Hh(t,\zeta)}{\P(t,\zeta)},       & 
r(t,\zeta)     & := \frac{\Rh(t,\zeta)}{\P(t,\zeta)}, \\
\Pn(t,\zeta)   & := \frac{\P(t,\zeta)}{\Nht},                & 
\sm(t)         & := \frac{\Sm(t)}{\Nm},                     & 
\im(t)         & := \frac{\Im(t)}{\Nm}.
\end{align*}

By using these new definitions, we can express the system as
\begin{align}\label{section1.11}
\begin{split}
\Big(\frac{\partial}{\partial t}+\frac{\partial}{\partial\zeta}\Big)s(t,\zeta){\P}(t,\zeta)
= & -\Big(\beta_{h}(t,\zeta)\im(t) + \alpha(t,\zeta)\Big)s(t,\zeta){\P}(t,\zeta) - \muh(t,\zeta)s(t,\zeta){\P}(t,\zeta)\\
\Big(\frac{\partial}{\partial t}+\frac{\partial}{\partial\zeta}\Big)v(t,\zeta){\P}(t,\zeta)
= & \alpha(t,\zeta)s(t,\zeta){\P}(t,\zeta) -\sigma(t,\zeta)\beta_{h}(t,\zeta)\im(t)v(t,\zeta){\P}(t,\zeta)-\muh(t,\zeta) v (t,\zeta){\P}(t,\zeta)\\
\Big(\frac{\partial}{\partial t}+\frac{\partial}{\partial\zeta}\Big)i_{a}(t,\zeta){\P}(t,\zeta)
= & \Big(1-f(t,\zeta)\Big)\beta_{h}(t,\zeta)\im(t)\Big(s(t,\zeta) + \sigma(t,\zeta)v(t,\zeta)\Big){\P}(t,\zeta) \\
& - \Big(\kappaa(t,\zeta)+\muh(t,\zeta) \Big)i_a(t,\zeta){\P}(t,\zeta)\\
\Big(\frac{\partial}{\partial t}+\frac{\partial}{\partial\zeta}\Big)i_{s}(t,\zeta){\P}(t,\zeta)
= & f(t,\zeta)\beta_{h}(t,\zeta)\im(t)\Big(s(t,\zeta) + \sigma(t,\zeta)v(t,\zeta)\Big){\P}(t,\zeta) \\
& - \Big(\kappas(t,\zeta) + \delta(t,\zeta)+\muh(t,\zeta) \Big)i_s(t,\zeta){\P}(t,\zeta)\\
\Big(\frac{\partial}{\partial t}+\frac{\partial}{\partial\zeta}\Big)h(t,\zeta){\P}(t,\zeta)
= & \delta(t,\zeta)i_s(t,\zeta){\P}(t,\zeta)- \Big( \kappah(t,\zeta) + \muh(t,\zeta)+\muH(t,\zeta)\Big)h(t,\zeta){\P}(t,\zeta)\\
\Big(\frac{\partial}{\partial t}+\frac{\partial}{\partial\zeta}\Big)r(t,\zeta){\P}(t,\zeta)
= & \kappaa(t,\zeta)i_{a}(t,\zeta){\P}(t,\zeta)+\kappas(t,\zeta)i_{s}(t,\zeta){\P}(t,\zeta)+\kappah(t,\tau)h(t,\zeta){\P}(t,\zeta)\\
& -\muh(t,\zeta) r(t,\zeta){\P}(t,\zeta)\\
\frac{\d}{\d t}\Big(\sm(t) \Nm\Big)
= & \varLambda_{m}(t)-{\sm(t) \Nm} \int_{0}^{\infty}\beta_m(t,\zeta)\Pn(t,\zeta)\Big(i_s(t,\zeta)+i_a(t,\zeta)\Big)\,\d\zeta-\mum \sm(t) \Nm\\
\frac{\d}{\d t}\Big( \im(t) \Nm \Big)
= & {\sm(t) \Nm} \int_{0}^{\infty}\beta_m(t,\zeta)\Pn(t,\zeta)\Big(i_s(t,\zeta)+i_a(t,\zeta)\Big)\,\d\zeta-\mum \im(t)\Nm.
\end{split}
\end{align}
Then, by applying the derivative, we get 
\begin{align*}
& {\P}(t,\zeta)\Big(\frac{\partial s(t,\zeta)}{\partial t}+\frac{\partial s(t,\zeta)}{\partial \zeta}\Big)+s(t,\zeta)\Big(\frac{\partial \P(t,\zeta)}{\partial t}+\frac{\partial \P(t,\zeta)}{\partial \zeta}\Big)\\
& \quad =-\Big(\beta_{h}(t,\zeta)\im(t) + \alpha(t,\zeta)\Big)s(t,\zeta){\P}(t,\zeta) - \muh(t,\zeta)s(t,\zeta){\P}(t,\zeta)\\
&{\P}(t,\zeta)\Big(\frac{\partial v(t,\zeta)}{\partial t}+\frac{\partial v(t,\zeta)}{\partial \zeta}\Big)+v(t,\zeta)\Big(\frac{\partial\P(t,\zeta)}{\partial t}+\frac{\partial \P(t,\zeta)}{\partial\zeta}\Big)\\
& \quad =\alpha(t,\zeta)s(t,\zeta){\P}(t,\zeta) -\sigma(t,\zeta)\beta_{h}(t,\zeta)\im(t)v(t,\zeta){\P}(t,\zeta)-\muh(t,\zeta) v (t,\zeta){\P}(t,\zeta)\\
&{\P}(t,\zeta)\Big(\frac{\partial i_a(t,\zeta)}{\partial t}+\frac{\partial i_a(t,\zeta)}{\partial \zeta}\Big)+i_a(t,\zeta)\Big(\frac{\partial \P(t,\zeta)}{\partial t}+\frac{\partial \P(t,\zeta)}{\partial\zeta}\Big)\\
& \quad =\Big(1-f(t,\zeta)\Big)\beta_{h}(t,\zeta)\im(t)\Big(s(t,\zeta) +\sigma(t,\zeta)v(t,\zeta)\Big){\P}(t,\zeta)-\Big(\kappaa(t,\zeta)+\muh(t,\zeta) \Big)i_a(t,\zeta){\P}(t,\zeta)\\
& {\P}(t,\zeta)\Big(\frac{\partial i_s(t,\zeta)}{\partial t}+\frac{\partial i_s(t,\zeta)}{\partial \zeta}\Big)+i_s(t,\zeta)\Big(\frac{\partial \P(t,\zeta)}{\partial t}+\frac{\partial \P(t,\zeta)}{\partial\zeta}\Big)\\
& \quad =f(t,\zeta)\beta_{h}(t,\zeta)\im(t)\Big(s(t,\zeta) + \sigma(t,\zeta)v(t,\zeta)\Big){\P}(t,\zeta)-\Big(\kappas(t,\zeta)+\delta(t,\zeta)+\muh(t,\zeta)\Big)i_{s}(t,\zeta){\P}(t,\zeta)\\
& {\P}(t,\zeta)\Big(\frac{\partial h(t,\zeta)}{\partial t}+\frac{\partial h(t,\zeta)}{\partial \zeta}\Big)+h(t,\zeta)\Big(\frac{\partial \P(t,\zeta)}{\partial t}+\frac{\partial\P(t,\zeta)}{\partial\zeta}\Big)\\
&\quad =\delta(t,\zeta)i_s(t,\zeta){\P}(t,\zeta)- \Big( \kappah(t,\zeta) +\muh(t,\zeta)\Big)h(t,\zeta){\P}(t,\zeta)-\muH(t,\zeta)h(t,\zeta){\P}(t,\zeta)\\
& {\P}(t,\zeta)\Big(\frac{\partial r(t,\zeta)}{\partial t}+\frac{\partial r(t,\zeta)}{\partial  \zeta}\Big)+r(t,\zeta)\Big(\frac{\partial\P(t,\zeta)}{\partial t}+\frac{\partial\P(t,\zeta)}{\partial\zeta}\Big)\\
& \quad =\kappaa(t,\zeta)i_{a}(t,\zeta){\P}(t,\zeta)+\kappas(t,\zeta)i_{s}(t,\zeta){\P}(t,\zeta)+\kappah(t,\zeta)h(t,\zeta){\P}(r,\zeta)-\muh(t,\zeta) r(t,\zeta){\P}(t,\zeta)\\
& \Nm\frac{\d\sm(t)}{\d t}+\sm(t)\frac{\d\Nm}{\d t}
=\varLambda_{m}(t)-{\sm(t) \Nm} \int_{0}^{\infty}\beta_m(t,\zeta)\Pn(t,\zeta)\Big(i_s(t,\zeta)+i_a(t,\zeta)\Big)\,\d\zeta-\mum \sm(t) \Nm\\
& \Nm\frac{\d\im(t)}{\d t}+\im(t)\frac{\d\Nm}{\d t}
={\sm(t) \Nm} \int_{0}^{\infty}\beta_m(t,\zeta)\Pn(t,\zeta)\Big(i_s(t,\zeta)+i_a(t,\zeta)\Big)\,\d\zeta-\mum \im(t)\Nm,
\end{align*}
and by using \eqref{section1.10} and \eqref{section1.00} we can simplify these equations in the system
\begin{align}\label{section1.13}
\begin{split}
\frac{\partial s(t,\zeta)}{\partial t}+\frac{\partial s(t,\zeta)}{\partial \zeta}
&=-\Big(\beta_{h}(t,\zeta)\im(t) + \alpha(t,\zeta)\Big)s(t,\zeta)-g(t,\zeta)s(t,\zeta)\\
\frac{\partial v(t,\zeta)}{\partial t}+\frac{\partial v(t,\zeta)}{\partial\zeta}
&=\alpha(t,\zeta)s(t,\zeta) -\sigma(t,\zeta)\beta_{h}(t,\zeta)\im(t)v(t,\zeta)-g(t,\zeta)v(t,\zeta)\\
\frac{\partial i_a(t,\zeta)}{\partial t}+\frac{\partial i_a(t,\zeta)}{\partial \zeta}
&=\Big(1-f(t,\zeta)\Big)\beta_{h}(t,\zeta)\im(t)\Big(s(t,\zeta) +\sigma(t,\zeta)v(t,\zeta)\Big)-\Big(\kappaa(t,\zeta)+g(t,\zeta) \Big)i_{a}(t,\zeta)\\
\frac{\partial i_s(t,\zeta)}{\partial t}+\frac{\partial i_s(t,\zeta)}{\partial \zeta}
&=f(t,\zeta)\beta_{h}(t,\zeta)\im(t)\Big(s(t,\zeta) + \sigma(t,\zeta)v(t,\zeta)\Big)-\Big(\kappas(t,\zeta)+\delta(t,\zeta)+g(t,\zeta)\Big)i_{s}(t,\zeta)\\
\frac{\partial h(t,\zeta)}{\partial t}+\frac{\partial h(t,\zeta)}{\partial\zeta}
&=\delta(t,\zeta)i_{s}(t,\zeta)-\Big(\kappah(t,\zeta)+\muH(t,\zeta)+g(t,\zeta)\Big)h(t,\zeta),\\
\frac{\partial r(t,\zeta)}{\partial t}+\frac{\partial r(t,\zeta)}{\partial \zeta}
&=\kappaa(t,\zeta)i_{a}(t,\zeta)+\kappas(t,\zeta)i_{s}(t,\zeta)+\kappah(t,\zeta)h(t,\zeta)-g(t,\zeta) r(t,\zeta)\\
\frac{\d \sm(t)}{\d t}
&=\frac{\varLambda_{m}(t)}{\Nm}\Big(1-\sm(t)\Big)-\sm(t)\int_{0}^{\infty}\beta_m(t,\zeta)\Pn(t,\zeta)\Big(i_s(t,\zeta)+i_a(t,\zeta)\Big)\,\d\zeta\\
\frac{\d \im(t)}{\d t}
&=\sm(t)\int_{0}^{\infty}\beta_m(t,\zeta)\Pn(t,\zeta)\Big(i_s(t,\zeta)+i_a(t,\zeta)\Big)\,\d\zeta- \im(t)\frac{\varLambda_m(t)}{\Nm},
\end{split}
\end{align}
where, $g(t,\zeta)=-\muH(t,\zeta) \frac{\Hh(t,\zeta)}{\P(t,\zeta)}=-\muH(t,\zeta)h(t,\zeta)$.
For this system, the initial conditions are given by:
\begin{align*}
s_{0}(\zeta) & := \frac{{\Sh}_{0}(\zeta)}{{\P}_{0}(\zeta)}, &
v_{0}(\zeta) & := \frac{{\Vh}_{0}(\zeta)}{{\P}_{0}(\zeta)}, &
i_{a0}(\zeta) & := \frac{{\Ah}_{0}(\zeta)}{{\P}_{0}(\zeta)}, \\ 
i_{s0}(\zeta) & := \frac{{\Ih}_{0}(\zeta)}{{\P}_{0}(\zeta)}, & 
h_{0}(\zeta) & := \frac{{\Hh}_{0}(\zeta)}{{\P}_{0}(\zeta)}, & 
r_{0}(\zeta) & := \frac{{\Rh}_{0}(\zeta)}{{\P}_{0}(\zeta)}, \\ 
{\Pn}_{0}(\zeta) & := \frac{{\P}_{0}(\zeta)}{\N_{h0}}, &
\sm(0) & := \frac{{\Sm}_{0}}{\N_{m0}}, & 
\im(0) & := \frac{{\Im}_{0}}{\N_{m0}}.
\end{align*}
Here, $\N_{h0}$ denotes the total initial human population, defined as the sum of all human compartments at time zero and $\N_{m0}$ denotes the total initial mosquito population at time zero. The boundary conditions are given by 
\begin{equation*}
s(t,0)=1,\quad v(t,0)=i_{a}(t,0)=i_{s}(t,0)=h(t,0)=r(t,0)=0.
\end{equation*}

\section{Disease-free steady state}\label{sec_df}

In this section, we analyze the disease-free equilibrium (DFE) of the rescaled model, which we denote by $E^0$. 
We assume that the system starts at the DFE at time $t^{0}$ and that no transitions due to the disease occur at this time. 
This leads to the steady-state condition, where the time derivatives of all compartments are equal to zero. 
Moreover, we consider a fixed vaccination strategy at the DFE point, which implies that the vaccination rate remains time independent at the DFE.
We considered that the host proportion remains in a steady state, expressed as $s(t,\zeta) = s^0(\zeta)$, $v(t,\zeta) 
= v^0(\zeta)$, $i_a(t,\zeta) = i_a^0(\zeta)$, $i_s(t,\zeta) = i_s^0(\zeta)$,  $h(t,\zeta) = h^0(\zeta)$, and $r(t,\zeta) = r^0(\zeta)$. 

At the DFE point 
$$
E^{0}(\zeta)=\left(s^{0}(\zeta),v^{0}(\zeta),i_{a}^{0}(\zeta),i_{s}^{0}(\zeta),h^{0}(\zeta),r^{0}(\zeta),\sm^{0},\im^{0}\right)
$$
we obtain
\begin{equation*}
\frac{\d s^{0}(\zeta)}{\,\d\zeta}=-\alpha(\zeta)s^{0}(\zeta), \qquad \frac{\d v^{0}(\zeta)}{\,\d\zeta}=\alpha(\zeta)s^{0}(\zeta), \qquad \sm^{0}=1.
\end{equation*}
As a consequence we infer that 
\begin{equation*}\begin{aligned} &
s^{0}(\zeta)=e^{-\int_{0}^{\zeta}\alpha(\eta)\,\d\eta},\quad v^0(\zeta)=\int_0^\zeta \alpha(\eta) e^{-\int_0^{\eta}\alpha(\xi) d\xi}\,\d\eta=1-e^{-\int_0^\zeta \alpha(\eta)\, d\eta}=1-s^{0}(\zeta), \quad \sm^{0}=1.
\end{aligned} 
\end{equation*}
We then conclude that the DFE point exists and is uniquely given by
\begin{equation}\label{eq_DFE_sol}
E^{0}(\zeta)=\left(e^{-\int_{0}^{\zeta}\alpha(\eta)\,\d\eta},\, 1-e^{-\int_0^\zeta\alpha(\eta)\, d\eta}, 0,0,0,0,1,0\right).
\end{equation}

The following lemma provides an upper bound for the susceptible and vaccinated populations compared to the DFE state. 
This inequality will be important subsequently for showing some inequalities for the time-varying effective reproduction number.

\begin{lemma} 
Consider the system \eqref{section1.13} with $\muH=0$ and with initial conditions at the DFE, namely 
\begin{equation*}
s(t^{0},\zeta) =s^{0}(\zeta),\quad v(t^{0},\zeta) =v^{0}(\zeta),     
\end{equation*}
and boundary conditions 
\begin{equation*}
s(t,0)=1,\quad v(t,0)=0.
\end{equation*}
Then for all $(t,\zeta)\in [t^0,\infty)\times [0,\infty)$ satisfying $t-t^0<\zeta$ the following inequality holds
\begin{equation*}
s(t,\zeta)+\sigma(t,\zeta)v(t,\zeta) \leq s^{0}(t^0-t+\zeta)+\sigma(t,\zeta)v^{0}(t^0-t+\zeta).
\end{equation*}
\end{lemma}

\begin{proof}
We consider only the following susceptible and vaccinated human populations from system \eqref{section1.13} as
\begin{align*}
\frac{\partial s(t,\zeta)}{\partial t}+\frac{\partial s(t,\zeta)}{\partial\zeta}
& =-\Big(\beta_{h}(t,\zeta)\im(t)+\alpha(t,\zeta)\Big)s(t,\zeta)\\
\frac{\partial v(t,\zeta)}{\partial t}+\frac{\partial v(t,\zeta)}{\partial\zeta}
& =\alpha(t,\zeta)s(t,\zeta)-\sigma(t,\zeta)\beta_{h}(t,\zeta)\im(t)v(t,\zeta),
\end{align*}
with initial conditions and boundary conditions mentioned in the statement. 
According to Lemma \ref{App_PDE_sol}, the solution of the above system is given by
\begin{eqnarray}\label{section3.5}
s(t,\zeta)= & \begin{cases}
e^{-\int_{0}^{\zeta}[\beta_{h}(\xi+t-\zeta,\xi)\im(\xi+t-\zeta)+\alpha(\xi+t-\zeta,\xi)]\d\xi} & \text{for }t-t^0\geq\zeta,\\
s^{0}(t^0-t+\zeta)e^{-\int_{t^0}^{t}[\beta_{h}(\xi,\xi+\zeta-t)\im(\xi)+\alpha(\xi,\xi+\zeta-t)]\d\xi} & \text{for }t-t^0<\zeta.
\end{cases}
\end{eqnarray}
and 
\begin{eqnarray}\label{section3.6}
v(t,\zeta)= & \begin{cases}
\int_{0}^{\zeta}e^{-\int_{\eta}^{\zeta}\sigma(\xi+t-\zeta,\xi)\beta_{h}(\xi+t-\zeta,\xi)\im(\xi+t-\zeta)\d\xi}\alpha(\eta+t-\zeta,\eta)
s(\eta+t-\zeta,\eta)\,\d\eta & \text{for }t-t_0\geq\zeta\\
v^{0}(t^0-t+\zeta)e^{-\int_{t^0}^{t}\sigma(\xi,\xi+\zeta-t)\beta_{h}(\xi,\xi+\zeta-t)\im(\xi)\d\xi} \\
\quad +\int_{t^0}^{t}e^{-\int_{\eta}^{t}\sigma(\xi,\xi+\zeta-t)\beta_{h}(\xi,\xi+\zeta-t)\im(\xi)\d\xi}
\alpha(\eta,\eta+\zeta-t)s(\eta,\eta+\zeta-t)\,\d\eta & \text{for }t-t^0<\zeta.
\end{cases}
\end{eqnarray}

Since $\sigma(t, \zeta)\in[0, 1] $ we infer from \eqref{section3.5} and \eqref{section3.6} that for $t-t^0 \leq \zeta$
\begin{align}\label{section3.7}
\begin{split}
\sigma(t,\zeta)v(t,\zeta) 
& \leq \sigma(t,\zeta)v^{0}(t^0-t+\zeta)e^{-\int_{t^0}^{t}\sigma(\xi,\xi+\zeta-t)\beta_{h}(\xi,\xi +\zeta-t)\im(\xi)\d\xi}\\
& \quad +\int_{t^0}^{t}e^{-\int_{\eta}^{t}\sigma(\xi,\xi+\zeta-t)\beta_{h}(\xi,\xi+\zeta-t)\im(\xi)\d\xi}
\alpha(\eta,\eta+\zeta-t)s(\eta,\eta+\zeta-t)\,\d\eta,
\end{split}
\end{align}
and 
\begin{equation}\label{section3.8}
s(t,\zeta) \leq s^{0}(t^0-t+\zeta)e^{-\int_{t^0}^{t}[\sigma(\xi,\xi+\zeta-t)\beta_{h}(\xi,\xi+\zeta-t)\im(\xi)+\alpha(\xi,\xi+\zeta-t)]\d\xi}.
\end{equation}
Then, by using inequalities \eqref{section3.7} and \eqref{section3.8} we obtain  
\begin{align}\label{section3.9}
\begin{split}
s(t,\zeta)+\sigma(t,\zeta)v(t,\zeta) 
\leq &
s^{0}(t^0-t+\zeta)e^{-\int_{t^0}^{t}[\sigma(\xi,\xi+\zeta-t)\beta_{h}(\xi,\xi+\zeta-t)\im(\xi)+\alpha(\xi,\xi+\zeta-t)]\d\xi} \\
& + \sigma(t,\zeta)v^{0}(t^0-t+\zeta)e^{-\int_{t^0}^{t}\sigma(\xi,\xi+\zeta-t)\beta_{h}(\xi,\xi +\zeta-t)\im(\xi)\d\xi}\\
& +\int_{t^0}^{t}e^{-\int_{\eta}^{t}\sigma(\xi,\xi+\zeta-t)\beta_{h}(\xi,\xi+\zeta-t)\im(\xi)\d\xi}
\alpha(\eta,\eta+\zeta-t)s(\eta,\eta+\zeta-t)\,\d\eta.
\end{split}
\end{align}

To find out $s(\eta,\eta+\zeta-t)$, we get from inequality \eqref{section3.8} and for $\eta \in (t^0,t)$
\begin{equation}\label{section3.10}
s(\eta,\eta+\zeta-t)
\leq s^{0}(t^0-t+\zeta)e^{-\int_{t^0}^{\eta}[\sigma(\xi,\xi+\zeta-t)\beta_{h}(\xi,\xi+\zeta-t)\im(\xi)+\alpha(\xi,\xi+\zeta-t)]\d\xi}.
\end{equation}
By inserting \eqref{section3.10} into \eqref{section3.9}, one gets
\begin{align}\label{section3.12}
\nonumber & s(t,\zeta)+\sigma(t,\zeta)v(t,\zeta) \\
\nonumber & \leq 
s^{0}(t^0-t+\zeta)e^{-\int_{t^0}^{t}[\sigma(\xi,\xi+\zeta-t)\beta_{h}(\xi,\xi+\zeta-t)\im(\xi)+\alpha(\xi,\xi+\zeta-t)]\d\xi} 
+ \sigma(t,\zeta)v^{0}(t^0-t+\zeta)e^{-\int_{t^0}^{t}\sigma(\xi,\xi+\zeta-t)\beta_{h}(\xi,\xi +\zeta-t)\im(\xi)\d\xi}\\
\nonumber &  \quad +\int_{t^0}^{t}e^{-\int_{\eta}^{t}\sigma(\xi,\xi+\zeta-t)\beta_{h}(\xi,\xi+\zeta-t)\im(\xi)\d\xi}
\alpha(\eta,\eta+\zeta-t)s^{0}(t^0-t+\zeta)e^{-\int_{t^0}^{\eta}[\sigma(\xi,\xi+\zeta-t)\beta_{h}(\xi,\xi+\zeta-t)\im(\xi)+\alpha(\xi,\xi+\zeta-t)]\d\xi}\,\d\eta \\
\begin{split}
& = s^{0}(t^0-t+\zeta)e^{-\int_{t^0}^{t}[\sigma(\xi,\xi+\zeta-t)\beta_{h}(\xi,\xi+\zeta-t)\im(\xi)+\alpha(\xi,\xi+\zeta-t)]\d\xi} 
+ \sigma(t,\zeta)v^{0}(t^0-t+\zeta)e^{-\int_{t^0}^{t}\sigma(\xi,\xi+\zeta-t)\beta_{h}(\xi,\xi +\zeta-t)\im(\xi)\d\xi}\\
 &  \quad + s^{0}(t^0-t+\zeta) e^{-\int_{t^0}^{t}\sigma(\xi,\xi+\zeta-t)\beta_{h}(\xi,\xi+\zeta-t)\im(\xi)\d\xi}\int_{t^0}^{t}
\alpha(\eta,\eta+\zeta-t)e^{-\int_{t^0}^{\eta}\alpha(\xi,\xi+\zeta-t)\d\xi}\,\d\eta.
\end{split}
\end{align}
Then, by setting $q(\eta):=e^{-\int_{t^{0}}^{\eta}\alpha(\xi,\xi+\zeta-t)\,\d\xi}$ and
by observing that $q(t^0)=1$, one infers from the fundamental theorem of calculus applied to the function $q$ that
\begin{equation}\label{section3.13}
\int_{t^0}^{t}\alpha(\eta,\eta+\zeta-t)e^{-\int_{t^0}^{\eta}\alpha(\xi,\xi+\zeta-t)\d\xi}\,\d\eta=1-e^{-\int_{t^0}^{t}\alpha(\xi,\xi+\zeta-t)\d\xi}.
\end{equation}
Thus, by using this equality, the inequality \eqref{section3.12} becomes
\begin{align*}
& s(t,\zeta)+\sigma(t,\zeta)v(t,\zeta) \\
& \leq s^{0}(t^0-t+\zeta)e^{-\int_{t^0}^{t}[\sigma(\xi,\xi+\zeta-t)\beta_{h}(\xi,\xi+\zeta-t)\im(\xi)+\alpha(\xi,\xi+\zeta-t)]\d\xi} 
+ \sigma(t,\zeta)v^{0}(t^0-t+\zeta)e^{-\int_{t^0}^{t}\sigma(\xi,\xi+\zeta-t)\beta_{h}(\xi,\xi +\zeta-t)\im(\xi)\d\xi}\\
 &  \quad + s^{0}(t^0-t+\zeta) e^{-\int_{t^0}^{t}\sigma(\xi,\xi+\zeta-t)\beta_{h}(\xi,\xi+\zeta-t)\im(\xi)\d\xi}
\Big(1-e^{-\int_{t^0}^{t}\alpha(\xi,\xi+\zeta-t)\d\xi}\Big).
\end{align*}
By simplification, we get
\begin{align*}
s(t,\zeta)+\sigma(t,\zeta)v(t,\zeta) 
\leq & s^{0}(t^0-t+\zeta)e^{-\int_{t^0}^{t}[\sigma(\xi,\xi+\zeta-t)\beta_{h}(\xi,\xi+\zeta-t)\im(\xi)+\alpha(\xi,\xi+\zeta-t)]\d\xi} \\
& +\sigma(t,\zeta)v^{0}(t^0-t+\zeta)e^{-\int_{t^0}^{t}\sigma(\xi,\xi+\zeta-t)\beta_{h}(\xi,\xi +\zeta-t)\im(\xi)\d\xi}\\
& +s^{0}(t^0-t+\zeta) e^{-\int_{t^0}^{t}\sigma(\xi,\xi+\zeta-t)\beta_{h}(\xi,\xi+\zeta-t)\im(\xi)\d\xi}\\
& -s^{0}(t^0-t+\zeta) e^{-\int_{t^0}^{t}[\sigma(\xi,\xi+\zeta-t)\beta_{h}(\xi,\xi+\zeta-t)\im(\xi)+\alpha(\xi,\xi+\zeta-t)]\d\xi},
\end{align*}
leading to 
\begin{equation*}
s(t,\zeta)+\sigma(t,\zeta)v(t,\zeta) \leq s^{0}(t^0-t+\zeta)+\sigma(t,\zeta)v^{0}(t^0-t+\zeta),
\end{equation*}
which proves the statement.  
\end{proof}

\section{Basic effective reproduction number and basic reproduction number}\label{sec_bern}

In this section, we explore different methods for estimating the basic effective reproduction number $\Rfo$  and the basic reproduction number $R^0$ 
for the proposed age structure model. Here, $R^0$ is defined as the expected number of secondary infections generated by a single infectious individual
in a fully susceptible population (disease-free equilibrium), while $\Rfo$ denotes the expected number of secondary infections 
when the whole population is not fully susceptible, due to vaccination etc.
These numbers are crucial for understanding the spread and control of infectious diseases in populations.

\subsection{First approach for estimating the basic effective reproduction number}

In this subsection, we compute the basic effective reproduction number $\Rfo$, which is evaluated at the DFE point, 
assuming a partially susceptible population due to vaccination. 
Let us consider the initial scenario where the number of infected individuals is very small compared to the fully susceptible population.
In this case, the dynamics of the infected population can be described by the linearized equations at the DFE point, $E^{0}(\zeta)$.
 
To proceed, we introduce a change of variables in the system as follows:
\begin{align*}
s(t,\zeta) & =s^{0}(\zeta)+\per{s}(t,\zeta),\quad v(t,\zeta)=v^{0}(\zeta)+\per{v}(t,\zeta),\quad i_{a}(t,\zeta)=\per{\imath_{a}}(t,\zeta),\quad i_{s}(t,\zeta)=\per{\imath_{s}}(t,\zeta),\\
h(t,\zeta) & =\per{h}(t,\zeta),\quad r(t,\zeta)=\per{r}(t,\zeta),\quad \sm(t)=1+\per{\sm}(t),\quad \im(t)=\per{\imath_{m}}(t).
\end{align*}
Here, $\per{s}(t,\zeta)$,\, $\per{v}(t,\zeta),$\, $\per{\imath}_{a}(t,\zeta)$,\, $\per{\imath}_{s}(t,\zeta)$,\, 
$\per{r}(t,\zeta)$,\, $\per{h}(t,\zeta)$,\, $\per{\sm}(t)$ and $\per{\imath}_m(t)$ represent the perturbation terms, 
which capture small deviations from the equilibrium state. 
Similarly, the total human population of age $\zeta$ at time $t$ is given by 
$$
\Pn(t,\zeta) = \Pn^{0}(\zeta) + \per{\Pn}(t,\zeta).
$$
Then, we linearize the system \eqref{section1.13} around the DFE point and derive a new simplified system as follows
\begin{align}\label{section4.1}
\begin{split}
\frac{\partial \per{s}(t,\zeta)}{\partial t}+\frac{\partial\per{s}(t,\zeta)}{\partial\zeta}
& =-\beta_{h}(t^{0},\zeta)s^{0}(\zeta) \per{\imath_{m}}(t) -\alpha(t^{0},\zeta)\per{s}(t,\zeta)-s^{0}(\zeta)\per{g}(t,\zeta)\\
\frac{\partial\per{v}(t,\zeta)}{\partial t}+\frac{\partial\per{v}(t,\zeta)}{\partial\zeta}
& =\alpha(t^{0},\zeta)\per{s}(t,\zeta) -\sigma(t^{0},\zeta)\beta_{h}(t^{0},\zeta)v^{0}(\zeta)\per{\imath_{m}}(t)-v^{0}(\zeta)\per{g}(t,\zeta)\\
\frac{\partial\per{\imath_{a}}(t,\zeta)}{\partial t}+\frac{\partial\per{\imath_{a}}(t,\zeta)}{\partial\zeta}
& =\Big(1-f(t^{0},\zeta)\Big)\beta_{h}(t^{0},\zeta)\Big(s^{0}(\zeta)\per{\imath_{m}}(t) +\sigma(t^{0},\zeta)v^{0}(\zeta)\Big)-\kappaa(t^{0},\zeta)\per{\imath_{a}}(t,\zeta)\\
\frac{\partial\per{\imath_{s}}(t,\zeta)}{\partial t}+\frac{\partial\per{\imath_{s}}(t,\zeta)}{\partial\zeta}
& =f(t^{0},\zeta)\beta_{h}(t^{0},\zeta)\Big(s^{0}(\zeta)\per{\imath_{m}}(t) + \sigma(t^{0},\zeta)v^{0}(\zeta)\Big)-\Big(\kappas(t^{0},\zeta)+\delta(t^{0},\zeta)\Big)\per{\imath_{s}}(t,\zeta)\\
\frac{\partial\per{h}(t,\zeta)}{\partial t}+\frac{\partial\per{h}(t,\zeta)}{\partial\zeta}
& =\delta(t^{0},\zeta)\per{\imath_{s}}(t,\zeta)-\Big(\kappah(t^{0},\zeta)+\muH(t^{0},\zeta)\Big) \per{h}(t,\zeta)\\
\frac{\partial\per{r}(t,\zeta)}{\partial t}+\frac{\partial\per{r}(t,\zeta)}{\partial\zeta}
& =\kappaa(t^{0},\zeta)\per{\imath_{a}}(t,\zeta)+\kappas(t^{0},\zeta)\per{\imath_{s}}(t,\zeta)+\kappah(t^{0},\zeta)\per{h}(t,\zeta)\\
\frac{\d \per{\sm}(t)}{\d t}
& =-\frac{\varLambda_{m}(t^{0})}{\N^{0}_{m}}\per{\sm}(t)-\int_{0}^{\infty}\beta_m(t^{0},\zeta){\Pn}^{0}(\zeta)\Big(\per{\imath_s}(t,\zeta)+\per{\imath_a}(t,\zeta)\Big)\,\d\zeta\\
\frac{\d \per{\imath_{m}}(t)}{\d t}
& =\int_{0}^{\infty}\beta_m(t^{0},\zeta){\Pn}^{0}(\zeta)\Big(\per{\imath_s}(t,\zeta)+\per{\imath_a}(t,\zeta)\Big)\,\d\zeta- \frac{\varLambda_m(t^{0})}{\N^{0}_{m}}\per{\imath_{m}}(t).
\end{split}
\end{align}
where $\per{g}(t,\zeta)=-\muH(t^{0},\zeta)\per{h}(t,\zeta)$ and $\N^0_m=\N_m(t^0)$. The initial and boundary conditions are
\begin{align*}
\per{s}(t^0,0) &= 1, & \per{v}(t^0,0) &= 0, & \per{\imath_{a}}(t^0,0) &= 0,  & \per{\imath_{s}}(t^0,0) &= 0.
\end{align*}
Since we only focus on the infected classes responsible for spreading the disease, the other classes are not considered.
As a result, we concentrate on the system:
\begin{align} 
\frac{\partial\per{\imath_{a}}(t,\zeta)}{\partial t}+\frac{\partial\per{\imath_{a}}(t,\zeta)}{\partial\zeta}
& =\Big(1-f(t^{0},\zeta)\Big)\beta_{h}(t^{0},\zeta)\Big(s^{0}(\zeta) +\sigma(t^{0},\zeta)v^{0}(\zeta)\Big)\per{\imath_{m}}(t)-\kappaa(t^{0},\zeta)\per{\imath_{a}}(t,\zeta)\label{section4.2} \\
\frac{\partial\per{\imath_{s}}(t,\zeta)}{\partial t}+\frac{\partial\per{\imath_{s}}(t,\zeta)}{\partial\zeta}
& =f(t^{0},\zeta)\beta_{h}(t^{0},\zeta)\Big(s^{0}(\zeta) + \sigma(t^{0},\zeta)v^{0}(\zeta)\Big)\per{\imath_{m}}(t)-\Big(\kappas(t^{0},\zeta)+\delta(t^{0},\zeta)\Big)\per{\imath_{s}}(t,\zeta) \label{section4.3}\\
\frac{\d \per{\imath_{m}}(t)}{\d t}
& =\int_{0}^{\infty}\beta_m(t^{0},\zeta){\Pn}^{0}(\zeta)\Big(\per{\imath_s}(t,\zeta)+\per{\imath_a}(t,\zeta)\Big)\,\d\zeta- \frac{\varLambda_m(t^{0})}{\N^{0}_{m}}\per{\imath_{m}}(t). \label{section4.4} 
\end{align}
By solving the PDEs \eqref{section4.2} and \eqref{section4.3} along the characteristic line, see Lemma \ref{App_PDE_sol}, we obtain:
\begin{align}\label{section4.5}
\nonumber & \per{\imath_{a}}(t,\zeta) \\
& =  \begin{cases}
\int_{0}^{\zeta}e^{-\int_{\eta}^{\zeta}\kappaa(t^{0},\xi)\d\xi} \Big(1-f(t^{0},\eta)\Big)\beta_{h}(t^{0},\eta)\Big(s^{0}(\eta)+\sigma(t^{0},\eta)v^{0}(\eta)\Big)\per{\imath_{m}}(\eta+t-\zeta)\,\d\eta & \text{for }t-t^{0}\geq\zeta,\\
\int_{t^{0}+\zeta-t}^{\zeta}e^{-\int_{\eta}^{\zeta}\kappaa(t^{0},\xi)\d\xi} \Big(1-f(t^{0},\eta)\Big)\beta_{h}(t^{0},\eta)\Big(s^{0}(\eta)+\sigma(t^{0},\eta)v^{0}(\eta)\Big)\per{\imath_{m}}(\eta+t-\zeta)\,\d\eta & \text{for }t-t^{0}<\zeta,
\end{cases}
\end{align}
and
\begin{align}\label{section4.05}
\nonumber & \per{\imath_{s}}(t,\zeta) \\
& = \begin{cases}
\int_{0}^{\zeta}e^{-\int_{\eta}^{\zeta}[\kappas(t^{0},\xi)+\delta(t^{0},\xi)]\d\xi} f(t^{0},\eta)\beta_{h}(t^{0},\eta)\Big(s^{0}(\eta)+\sigma(t^{0},\eta)v^{0}(\eta)\Big)\per{\imath_{m}}(\eta+t-\zeta)\,\d\eta & \text{for }t-t^{0}\geq\zeta,\\ 
\int_{t^{0}+\zeta-t}^{\zeta}e^{-\int_{\eta}^{\zeta}[\kappas(t^{0},\xi)+\delta(t^{0},\xi)]\d\xi} f(t^{0},\eta)\beta_{h}(t^{0},\eta)\Big(s^{0}(\eta)+\sigma(t^{0},\eta)v^{0}(\eta)\Big)\per{\imath_{m}}(\eta+t-\zeta)\,\d\eta & \text{for }t-t^{0}<\zeta.
\end{cases}
\end{align}
Then, \eqref{section4.4} can be rewritten as
\begin{align}\label{section4.6}
\nonumber & \frac{\d\per{\imath_{m}}(t)}{\d t}+\frac{\varLambda_{m}(t^{0})}{\N_{m}^{0}}\per{\imath_{m}}(t) \\
& = \int_{0}^{t-t^{0}}\beta_{m}(t^{0},\zeta){\Pn}^{0}(\zeta)\Big(\per{\imath_{s}}(t,\zeta)+\per{\imath_{a}}(t,\zeta)\Big)\,\d\zeta+ 
\int_{t-t^{0}}^{\infty}\beta_{m}(t^{0},\zeta){\Pn}^{0}(\zeta)\Big(\per{\imath_{s}}(t,\zeta)+\per{\imath_{a}}(t,\zeta)\Big)\,\d\zeta.
\end{align}
Thus, by summing the solutions for $\per{\imath}_{a}(t,\zeta)$ and $\per{\imath}_{s}(t,\zeta)$ when $t-t^{0}\geq\zeta$, we get
\begin{equation}\label{section4.7}
\per{\imath_{a}}(t,\zeta)+\per{\imath_{s}}(t,\zeta)\\
=\int_{0}^{\zeta}\beta_{h}(t^{0},\eta)\Big(s^{0}(\eta)+\sigma(t^{0},\eta)v^{0}(\eta)\Big)
F(\eta,\zeta)\per{\imath_{m}}(\eta+t-\zeta)\,\d\eta.
\end{equation}
with the new factor 
\begin{equation}\label{eq_def_of_F}
F(\eta,\zeta):=e^{-\int_{\eta}^{\zeta}\kappaa(t^{0},\xi)\d\xi}\Big(1-f(t^{0},\eta)\Big)+e^{-\int_{\eta}^{\zeta}
[\kappas(t^{0},\xi)+\delta(t^{0},\xi)]\d\xi}f(t^{0},\eta)
\end{equation}
which takes into account the recovery of the symptomatic and of the asymptomatic individuals.
From the above equation \eqref{section4.7}, we infer the following integral equation
\begin{align}\label{section4.8}
\nonumber &\int_{0}^{t-t^{0}}\beta_{m}(t^{0},\zeta){\Pn}_{0}(\zeta)\Big(\per{\imath_{s}}(t,\zeta)+\per{\imath_{a}}(t,\zeta)\Big)\,\d\zeta \\
& =\int_{0}^{t-t^{0}}\beta_{m}(t^{0},\zeta){\Pn}^{0}(\zeta)\int_{0}^{\zeta}\beta_{h}(t^{0},\eta)\Big(s^{0}(\eta)+\sigma(t^{0},\eta)v^{0}(\eta)\Big)
F(r,\eta,\zeta) \per{\imath_{m}}(\eta+t-\zeta) \,\d\eta\,\d\zeta.
\end{align}
Similarly, by adding the solutions for $\per{\imath}_{a}(t,\zeta)$ and $\per{\imath}_{s}(t,\zeta)$ from 
\eqref{section4.5} and \eqref{section4.05} for $t-t^{0}<\zeta$, we obtain
\begin{equation}\label{section4.9}
\per{\imath_{a}}(t,\zeta)+\per{\imath_{s}}(t,\zeta)
 =\int_{t^{0}+\zeta-t}^{\zeta} \beta_{h}(t^{0},\eta)
\Bigl(s^{0}(\eta)+\sigma(t^{0},\eta)v^{0}(\eta)\Bigr) F(\eta,\zeta)\per{\imath_{m}}(\eta+t-\zeta)\,\d\eta.
\end{equation}
From \eqref{section4.9}, we then obtain the following integral form
\begin{align}\label{section4.10}
\nonumber &\int_{t-t^{0}}^{\infty}\beta_{m}(t^{0},\zeta){\Pn^{0}}(\zeta)\Big(\per{\imath_{s}}(t,\zeta)+\per{\imath_{a}}(t,\zeta)\Big)\,\d\zeta\\
& =\int_{t-t^{0}}^{\infty}\beta_{m}(t^{0},\zeta){\Pn}^{0}(\zeta)\int_{t^{0}+\zeta-t}^{\zeta}\beta_{h}(t^{0},\eta)\Big(s^{0}(\eta)+\sigma(t^{0},\eta)v^{0}(\eta)\Big)
F(\eta,\zeta)\per{\imath_{m}}(\eta+t-\zeta)\,\d\eta\,\d\zeta.
\end{align}
Thus, by adding the content of \eqref{section4.8} and \eqref{section4.10}, we finally get
\begin{align}\label{section4.11}
\nonumber & \int_{0}^{\infty}\beta_{m}(t^{0},\zeta){\Pn}^{0}(\zeta)\Big(\per{\imath_{s}}(t,\zeta)+\per{\imath_{a}}(t,\zeta)\Big)\,\d\zeta \\
\nonumber & =\int_{0}^{t-t^{0}}\beta_{m}(t^{0},\zeta){\Pn}^{0}(\zeta)\int_{0}^{\zeta}\beta_{h}(t^{0},\eta)\Big(s^{0}(\eta)+\sigma(t^{0},\eta)v^{0}(\eta)\Big)
F(\eta,\zeta)\per{\imath_{m}}(\eta+t-\zeta)\,\d\eta\,\d\zeta\\
\nonumber &\quad  +\int_{t-t^{0}}^{\infty}\beta_{m}(t^{0},\zeta){\Pn}^{0}(\zeta)\int_{t^{0}+\zeta-t}^{\zeta}\beta_{h}(t^{0},\eta)\Big(s^{0}(\eta)+\sigma(t^{0},\eta)v^{0}(\eta)\Big)
F(\eta,\zeta)\per{\imath_{m}}(\eta+t-\zeta)\,\d\eta\,\d\zeta \\
& = \int_{0}^{\infty}\beta_{m}(t^{0},\zeta){\Pn}^{0}(\zeta)\int_{\max\{0,t^{0}+\zeta-t\}}^{\zeta}\beta_{h}(t^{0},\eta)\Big(s^{0}(\eta)+\sigma(t^{0},\eta)v^{0}(\eta)\Big)
F(\eta,\zeta)\per{\imath_{m}}(\eta+t-\zeta)\,\d\eta\,\d\zeta.
\end{align}
By substituting \eqref{section4.11} into \eqref{section4.6} this yields to the renewal equation:
\begin{align}\label{section4.13}
\begin{split}
&\frac{\d\per{\imath_{m}}(t)}{\d t}+\frac{\varLambda_{m}(t^{0})}{\N_{m}^{0}}\per{\imath_{m}}(t) \\
& = \int_{0}^{\infty}\beta_{m}(t^{0},\zeta){\Pn}^{0}(\zeta)\int_{\max\{0,t^{0}+\zeta-t\}}^{\zeta}\beta_{h}(t^{0},\eta)\Big(s^{0}(\eta)+\sigma(t^{0},\eta)v^{0}(\eta)\Big)
F(\eta,\zeta)\per{\imath_{m}}(\eta+t-\zeta)\,\d\eta\,\d\zeta.
\end{split}
\end{align}

We seek a solution in the exponential form by assuming that the incidence follows $e^{\lambda (t-t^0)}$ 
with a fixed growth rate $\lambda$ during the initial phase of the outbreak. 
By following the approach provided for example in  \cite[Sec.~2]{Anazawa2025}, \cite[Thm.~1]{Cai2013}, or in \cite[Sec.~2.1]{Nishiura2009}, 
the number of new infections $\per{\imath_{m}}(t)$ can be expressed by 
\begin{equation}\label{section4.14}
\frac{\d \per{\imath_{m}}(t)}{\d t}=\lambda \per{\imath_{m}}(t),
\end{equation}
or equivalently by the relation 
\begin{equation}\label{section4.15}
\per{\imath_{m}}(t-\epsilon) = \per{\imath_{m}}(t) e^{-\lambda \epsilon}.
\end{equation}
By using \eqref{section4.14} and \eqref{section4.15}, the renewal equation \eqref{section4.13} becomes 
\begin{align*}
&\Big(\lambda +\frac{\varLambda_m(t^{0})}{\N^{0}_{m}} \Big)\per{\imath_{m}}(t) \\
& = \per{\imath_{m}}(t) \int_{0}^{\infty}\beta_{m}(t^{0},\zeta){\Pn}^{0}(\zeta)\int_{\max\{0,\,t^{0}+\zeta-t\|}^{\zeta}\beta_{h}(t^{0},\eta)\Big(s^{0}(\eta)+\sigma(t^{0},\eta)v^{0}(\eta)\Big) F(\eta,\zeta)e^{-\lambda(\zeta-\eta) }\,\d\eta\,\d\zeta.
\end{align*}
with $F(\eta,\zeta)$ defined in \eqref{eq_def_of_F}.
This equation can be rewritten as 
\begin{equation}\label{lotka_equation}
\per{\imath_{m}}(t) 
 = \frac{\per{\imath_{m}}(t)\N^{0}_{m}}{\lambda\N^{0}_{m}+\varLambda_{m}(t^{0})} \int_{0}^{\infty}\beta_{m}(t^{0},\zeta){\Pn}^{0}(\zeta)\int_{\max\{0,\,t^{0}+\zeta-t\}}^{\zeta}\beta_{h}(t^{0},\eta)\Big(s^{0}(\eta)+\sigma(t^{0},\eta)v^{0}(\eta)\Big) F(\eta,\zeta)e^{-\lambda(\zeta-\eta) }\,\d\eta\,\d\zeta,
\end{equation}
which is known as the Lotka equation (or also as the renewal equation). 
Then there exists a nonzero solution $\per{\imath_{m}}(t)$ to \eqref{lotka_equation} if and only if there exists a real or complex number $\lambda$ such that
\begin{equation}\label{section4.18}
1 = \frac{\N^{0}_{m}}{\lambda\N^{0}_{m}+\varLambda_{m}(t^{0})} \int_{0}^{\infty}\beta_{m}(t^{0},\zeta){\Pn}^{0}(\zeta)\int_{\max\{0,\,t^{0}+\zeta-t\}}^{\zeta}\beta_{h}(t^{0},\eta)\Big(s^{0}(\eta)+\sigma(t^{0},\eta)v^{0}(\eta)\Big) F(\eta,\zeta)e^{-\lambda(\zeta-\eta) }\,\d\eta\,\d\zeta.
\end{equation}

The next idea is to take the limit as $t\to\infty$ for estimating the basic reproduction number $\Rfo$
by keeping all other parameters fixed at $t^0$. This allows us to understand how the system behaves in the long term, 
with the assumed exponential growth rate $\lambda$. 
Therefore, by considering the limit as $t\to\infty$, we can observe how the disease spreads with the growth rate $\lambda$. 
The dynamics at large $t$ allows us to observe the impact of $\Rfo$ on the infection spread, 
assuming no major interventions in the system. 
Therefore, the steady-state behavior of the epidemic can be characterized by the values of $\Rfo$, namely
if $\Rfo > 1$ or if $\Rfo < 1$, as discussed in Section \ref{sec_stability}.

For these investigations, let us already set
\begin{equation}\label{section4.19}
L(\lambda) 
:=\frac{\N^{0}_{m}}{\lambda\N^{0}_{m}+\varLambda_{m}(t^{0})} \int_{0}^{\infty}\beta_{m}(t^{0},\zeta){\Pn}^{0}(\zeta)
\int_{0}^{\zeta}\beta_{h}(t^{0},\eta)\Big(s^{0}(\eta)+\sigma(t^{0},\eta)v^{0}(\eta)\Big) 
F(\eta,\zeta)e^{-\lambda(\zeta-\eta) }\,\d\eta\,\d\zeta.
\end{equation}
The equation $L(\lambda)=1$ represents the Lotka characteristic equation. 
Following \cite[Sec.~9.5.1]{Li2008} and \cite[Sec.~3]{Sanchez2019}, we define the basic effective reproduction number as $L(0)$, namely
\begin{equation}\label{section4.20}
\Rfo := \frac{\N^{0}_{m}}{\varLambda_{m}(t^{0})} \int_{0}^{\infty}\beta_{m}(t^{0},\zeta){\Pn}^{0}(\zeta)\int_{0}^{\zeta}\beta_{h}(t^{0},\eta) \Big(s^{0}(\eta)+\sigma(t^{0},\eta)v^{0}(\eta)\Big) F(\eta,\zeta)\,\d\eta\,\d\zeta.
\end{equation}

\subsection{Second approach for estimating the basic effective reproduction number}

In this subsection, we compute the basic effective reproduction number $\Rfo$ using a slightly different approach. 
For this, we consider the number of infected classes, which are linearized at the DFE point and given in \eqref{section4.2}, \eqref{section4.3}, 
and \eqref{section4.4}. For convenience we recall this system:
\begin{align}\label{eq_i_m_only} 
\nonumber \frac{\partial\per{\imath_{a}}(t,\zeta)}{\partial t}+\frac{\partial\per{\imath_{a}}(t,\zeta)}{\partial\zeta}
& =\Big(1-f(t^{0},\zeta)\Big)\beta_{h}(t^{0},\zeta)\Big(s^{0}(\zeta) +\sigma(t^{0},\zeta)v^{0}(\zeta)\Big)\per{\imath_{m}}(t)-\kappaa(t^{0},\zeta)\per{\imath_{a}}(t,\zeta) \\
\nonumber \frac{\partial\per{\imath_{s}}(t,\zeta)}{\partial t}+\frac{\partial\per{\imath_{s}}(t,\zeta)}{\partial\zeta}
\nonumber & =f(t^{0},\zeta)\beta_{h}(t^{0},\zeta)\Big(s^{0}(\zeta) + \sigma(t^{0},\zeta)v^{0}(\zeta)\Big)\per{\imath_{m}}(t)-\Big(\kappas(t^{0},\zeta)+\delta(t^{0},\zeta)\Big)\per{\imath_{s}}(t,\zeta)\\ 
\frac{\d \per{\imath_{m}}(t)}{\d t}
& =\int_{0}^{\infty}\beta_m(t^{0},\zeta){\Pn}^{0}(\zeta)\Big(\per{\imath_s}(t,\zeta)+\per{\imath_a}(t,\zeta)\Big)\,\d\zeta- \frac{\varLambda_m(t^{0})}{\N^{0}_{m}}\per{\imath_{m}}(t),
\end{align}
with boundary conditions $\per{\imath_{a}}(t,0)=0$ and $\per{\imath_{s}}(t,0)=0$.
As before the solutions of the first two equations read
\begin{align}\label{section4.24}
\nonumber & \per{\imath_{a}}(t,\zeta) \\
& =  \begin{cases}
\int_{0}^{\zeta}e^{-\int_{\eta}^{\zeta}\kappaa(t^{0},\xi)\d\xi} \Big(1-f(t^{0},\eta)\Big)\beta_{h}(t^{0},\eta)\Big(s^{0}(\eta)+\sigma(t^{0},\eta)v^{0}(\eta)\Big)\per{\imath_{m}}(\eta+t-\zeta)\,\d\eta & \text{for }t-t^{0}\geq\zeta,\\
\int_{t^{0}+\zeta-t}^{\zeta}e^{-\int_{\eta}^{\zeta}\kappaa(t^{0},\xi)\d\xi} \Big(1-f(t^{0},\eta)\Big)\beta_{h}(t^{0},\eta)\Big(s^{0}(\eta)+\sigma(t^{0},\eta)v^{0}(\eta)\Big)\per{\imath_{m}}(\eta+t-\zeta)\,\d\eta & \text{for }t-t^{0}<\zeta,
\end{cases}
\end{align}
and
\begin{align}\label{section4.26}
\nonumber & \per{\imath_{s}}(t,\zeta) \\
& = \begin{cases}
\int_{0}^{\zeta}e^{-\int_{\eta}^{\zeta}[\kappas(t^{0},\xi)+\delta(t^{0},\xi)]\d\xi} f(t^{0},\eta)\beta_{h}(t^{0},\eta)\Big(s^{0}(\eta)+\sigma(t^{0},\eta)v^{0}(\eta)\Big)\per{\imath_{m}}(\eta+t-\zeta)\,\d\eta & \text{for }t-t^{0}\geq\zeta,\\ 
\int_{t^{0}+\zeta-t}^{\zeta}e^{-\int_{\eta}^{\zeta}[\kappas(t^{0},\xi)+\delta(t^{0},\xi)]\d\xi} f(t^{0},\eta)\beta_{h}(t^{0},\eta)\Big(s^{0}(\eta)+\sigma(t^{0},\eta)v^{0}(\eta)\Big)\per{\imath_{m}}(\eta+t-\zeta)\,\d\eta & \text{for }t-t^{0}<\zeta.
\end{cases}
\end{align}
The equations \eqref{section4.24} and \eqref{section4.26} describe the number of asymptomatic individuals $\per{\imath_{a}}(t,\zeta)$ 
and symptomatic individuals $\per{\imath_{s}}(t,\zeta)$ at a given time $t$ and age $\zeta$. 
To compute the basic effective reproduction number, $\Rfo$, we want to see the long term behavior of the population. Thus, by assuming $t$ large enough one has
\begin{equation}\label{section4.26a}
\per{\imath_{a}}(t,\zeta)= 
\int_{0}^{\zeta}e^{-\int_{\eta}^{\zeta}\kappaa(t^{0},\xi)\d\xi} \Big(1-f(t^{0},\eta)\Big)\beta_{h}(t^{0},\eta)\Big(s^{0}(\eta)+\sigma(t^{0},\eta)v^{0}(\eta)\Big)\per{\imath_{m}}(\eta+t-\zeta)\,\d\eta.
\end{equation}
and 
\begin{equation}\label{section4.27}
\per{\imath_{s}}(t,\zeta)= \int_{0}^{\zeta}e^{-\int_{\eta}^{\zeta}[\kappas(t^{0},\xi)+\delta(t^{0},\xi)]\d\xi} f(t^{0},\eta)\beta_{h}(t^{0},\eta)\Big(s^{0}(\eta)+\sigma(t^{0},\eta)v^{0}(\eta)\Big)\per{\imath_{m}}(\eta+t-\zeta)\,\d\eta.
\end{equation}
By inserting \eqref{section4.26a} and \eqref{section4.27} into \eqref{eq_i_m_only} one gets the new renewal equation
\begin{equation}\label{section4.28}
\frac{\d\per{\imath_{m}}(t)}{\d t}+\frac{\varLambda_{m}(t^{0})}{\N_{m}^{0}}\per{\imath_{m}}(t)  =\int_{0}^{\infty}\beta_{m}(t^{0},\zeta){\Pn}^{0}(\zeta)\int_{0}^{\zeta}\beta_{h}(t^{0},\eta)\Big(s^{0}(\eta)+\sigma(t^{0},\eta)v^{0}(\eta)\Big)
F(\eta,\zeta)\per{\imath_{m}}(t+\eta-\zeta)\,\d\eta\,\d\zeta.
\end{equation}

As in the previous subsection, we assume that the incidence follows an exponential increase of the form $e^{\lambda (t-t^0)}$ 
with a fixed growth rate $\lambda$ during the initial phase of the outbreak. 
Accordingly, we can use \eqref{section4.14} and \eqref{section4.15} for $\per{\imath_{m}}(t)$, and we get 
\begin{equation*}
\per{\imath_{m}}(t) = \frac{\per{\imath_{m}}(t)\N^{0}_{m}}{\lambda\N^{0}_{m}+\varLambda_{m}(t^{0})} \int_{0}^{\infty}\beta_{m}(t^{0},\zeta){\Pn}^{0}(\zeta)\int_{0}^{\zeta}\beta_{h}(t^{0},\eta)\Big(s^{0}(\eta)+\sigma(t^{0},\eta)v^{0}(\eta)\Big)F(\eta,\zeta)e^{-\lambda(\zeta-\eta) }\,\d\eta\,\d\zeta.
\end{equation*}
This equation is again the Lotka equation or the renewal equation. 
Since $\per{\imath_{m}}(t)$ is nonzero for all $t \geq 0$, we can cancel it from both sides, 
yielding the same basic effective reproduction number as in equation \eqref{section4.20}.

\subsection{Third approach for estimating the basic effective reproduction number}

In this subsection, we compute the basic effective reproduction number $\Rfo $ by using a third approach. 
We consider again the infected classes satisfying \eqref{section4.2}, \eqref{section4.3} and \eqref{section4.4}
together with the boundary condition $\per{\imath_{a}}(t,0)=0$ and $\per{\imath_{s}}(t,0)=0$.
We then assume that the solutions have an exponential form given by
\begin{equation}\label{section4.35}
\per{\imath_{a}}(t,\zeta)=\per{\imath_{a}}(\zeta)e^{\lambda (t-t^0)}, 
\quad \per{\imath_{s}}(t,\zeta)=\per{\imath_{s}}(\zeta)e^{\lambda (t-t^0)},
\quad \per{\imath_{m}}(t)=\per{\imath}_{m0}e^{\lambda (t-t^0)}.
\end{equation}
By using \eqref{section4.35} we transform the system of PDEs into a system of ODEs as
\begin{align}
\frac{\per{\imath_{a}}(\zeta)}{\d\zeta}+\Big(\kappaa(t^{0},\zeta)+\lambda\Big)\per{\imath_{a}}(\zeta)
& =\per{\imath}_{m0}\Big(1-f(t^{0},\zeta)\Big)\beta_{h}(t^{0},\zeta)\Big(s^{0}(\zeta) +\sigma(t^{0},\zeta)v^{0}(\zeta)\Big),\label{section4.36}\\
\frac{\per{\imath_{s}}(\zeta)}{\d\zeta}+\Big(\kappas(t^{0},\zeta)+\delta(t^{0},\zeta)+\lambda\Big)\per{\imath_{s}}(\zeta)
& =\per{\imath}_{m0}f(t^{0},\zeta)\beta_{h}(t^{0},\zeta)\Big(s^{0}(\zeta) + \sigma(t^{0},\zeta)v^{0}(\zeta)\Big), \label{section4.37}\\
\lambda \per{\imath}_{m0} + \per{\imath}_{m0}\frac{\varLambda_m(t^{0})}{\N^{0}_{m}}
& =\int_{0}^{\infty}\beta_m(t^{0},\zeta){\Pn}^{0}(\zeta)\Big(\per{\imath_s}(\zeta)+\per{\imath_a}(\zeta)\Big)\,\d\zeta \label{section4.38} .
\end{align}
The solutions of the first two equations of the above system are
\begin{equation}\label{section4.39}
\per{\imath_{a}}(\zeta)
=  \per{\imath}_{m0} \int_{0}^{\zeta} e^{-\int_{\eta}^{\zeta}\kappaa(t^{0},\xi)\d\xi} \Big(1-f(t^{0},\eta)\Big)\beta_{h}(t^{0},\eta)
\Big(s^{0}(\eta)+\sigma(t^{0},\eta)v^{0}(\eta)\Big)e^{-\lambda(\zeta-\eta)} \,\d\eta
\end{equation}
and 
\begin{equation}\label{section4.40}
\per{\imath_{s}}(\zeta)
= \per{\imath}_{m0} \int_{0}^{\zeta}e^{-\int_{\eta}^{\zeta}[\kappas(t^{0},\xi)+\delta(t^{0},\xi)]\d\xi}f(t^{0},\eta)\beta_{h}(t^{0},\eta) 
\Big(s^{0}(\eta)+\sigma(t^{0},\eta)v^{0}(\eta)\Big)e^{-\lambda(\zeta-\eta)}\,\d\eta.
\end{equation}
By substituting \eqref{section4.39} and \eqref{section4.40} into \eqref{section4.38} one gets a new Lotka equation or renewal equation, namely
\begin{equation}\label{section4.41}
\per{\imath}_{m0} = \frac{\per{\imath}_{m0}\N^{0}_{m}}{\lambda\N^{0}_{m}+\varLambda_{m}(t^{0})} \int_{0}^{\infty}\beta_{m}(t^{0},\zeta){\Pn}^{0}(\zeta)\int_{0}^{\zeta}\beta_{h}(t^{0},\eta)\Big(s^{0}(\eta)+\sigma(t^{0},\eta)v^{0}(\eta)\Big)
F(\eta,\zeta) e^{-\lambda(\zeta-\eta) }\,\d\eta\,\d\zeta.
\end{equation}
where $F(\eta,\zeta)$ has been defined in \eqref{eq_def_of_F}.
By canceling $\per{\imath}_{m0}$ from both sides, we obtain the same equation as in~\eqref{section4.18}.

\subsection{Interpretation of the expression obtained}

The estimation of $\Rfo$ is helpful for determining whether the disease spreads or declines for different values of $\lambda$. 
We will discuss the behavior of $\lambda$ in detail in Section \ref{sec_stability}.  
For now, let us recall the formula obtained for the basic effective reproduction number in \eqref{section4.20}
and let us interpret its different factors. One has
\begin{equation}\label{eq_Rfo}
\Rfo := \frac{\N^{0}_{m}}{\varLambda_{m}(t^{0})} \int_{0}^{\infty}\beta_{m}(t^{0},\zeta){\Pn}^{0}(\zeta)\int_{0}^{\zeta}\beta_{h}(t^{0},\eta) \Big(s^{0}(\eta)+\sigma(t^{0},\eta)v^{0}(\eta)\Big) F(\eta,\zeta)\,\d\eta\,\d\zeta.
\end{equation}
In \eqref{eq_Rfo}, $\beta_{m}(t^{0},\zeta)$ and $\beta_{h}(t^{0},\eta)$ represent the transmission rates from vector to host and host to vector,
at the disease free equilibrium point.
The factors $\frac{\N^{0}_{m}}{\varLambda_{m}(t^{0})}$ and $s^{0}(\eta)+\sigma(t^{0},\eta)v^{0}(\eta)$ 
correspond respectively to the susceptible mosquitoes and humans, also at the DFE. 
Note that the vaccination efficiency rate $\sigma(t^{0},\eta)$ indicates the effectiveness of the vaccination.
The factor $F(\eta,\zeta)$, namely
\begin{equation*}
e^{-\int_{\eta}^{\zeta}\kappaa(t^{0},\xi)\d\xi}\Big(1-f(t^{0},\eta)\Big)+e^{-\int_{\eta}^{\zeta}
[\kappas(t^{0},\xi)+\delta(t^{0},\xi)]\d\xi}f(t^{0},\eta)
\end{equation*}
represent the total probability that an individual infected at age $\eta$ remains infectious at age $\zeta$, 
both asymptomatic and symptomatic. 
Finally the factor $p^0$ provides the distribution of the human population at time $t^0$.
By putting the solutions for $s^0(\zeta)$ and $v^{0}(\zeta)$ obtained at the DFE in \eqref{eq_DFE_sol} one gets
\begin{equation}\label{section4.44}
\Rfo = \frac{\N^{0}_{m}}{\varLambda_{m}(t^{0})} \int_{0}^{\infty}\beta_{m}(t^{0},\zeta){\Pn}^{0}(\zeta)\int_{0}^{\zeta}\beta_{h}(t^{0},\eta) \Big(e^{-\int_{0}^{\eta}\alpha(\xi)\,\d\xi} + \sigma(t^{0},\eta) (1-e^{-\int_0^\eta\alpha(\xi)\d\xi}) \Big) F(\eta,\zeta)\,\d\eta\,\d\zeta.
\end{equation}

In summary, the basic effective reproduction number for our model can be understood from different perspectives:
\begin{equation}
\begin{aligned} 
\Rfo =&  \underbrace{\frac{\N^{0}_{m}}{\varLambda_{m}(t^{0})}}_{\text{Number of mosquitoes}} 
\int_{0}^{\infty}  
\underbrace{\beta_{m}(t^{0},\zeta)}_{\text{Human-to-mosquito transmission rate}} 
\underbrace{{\Pn}^{0}(\zeta)}_{\text{Human population distribution}} \\[6pt]
&\times \int_{0}^{\zeta}  
\underbrace{\beta_{h}(t^{0},\eta)}_{\text{Mosquito-to-human transmission rate}}  
\underbrace{\left(s^{0}(\eta)+\sigma(t^{0},\eta)v^{0}(\eta)\right)}_{\text{Susceptible humans (including vaccination failures)}} \\[6pt]
&\times \bigg[ 
\underbrace{e^{-\int_{\eta}^{\zeta}\kappaa(t^{0},\xi)\,\d\xi}\Big(1-f(t^{0},\eta)\Big)}_{\text{Probability of infection without symptoms}} 
+ \underbrace{e^{-\int_{\eta}^{\zeta}[\kappas(t^{0},\xi)+\delta(t^{0},\xi)]\,\d\xi}f(t^{0},\eta)}_{\text{Probability of infection with symptoms}} 
\bigg] \, \d\eta \, \d\zeta .
\label{section4.45}
\end{aligned}
\end{equation}
Alternatively, since $\beta_{h}(t^{0},\eta)$ is the transmission rate from infected mosquitoes to humans, 
the total number of humans infected by a single mosquito is 
$\beta_{h}(t^{0},\eta)\frac{\N^{0}_{m}}{\varLambda_{m}(t^{0})}{\Pn}^{0}(\zeta)\left(s^{0}(\eta)+\sigma(t^{0},\eta)v^{0}(\eta)\right)$. 
The infected mosquito spreads the disease for its entire life because there is no recovery rate for mosquitoes.
The basic effective reproduction number can then be understood as
\begin{equation}
\begin{aligned} 
\Rfo = &\int_{0}^{\infty}\int_{0}^{\zeta}\underbrace{\beta_{h}(t^{0},\eta) \frac{\N^{0}_{m}}{\varLambda_{m}(t^{0})}  {\Pn}^{0}(\zeta)\left(s^{0}(\eta)+\sigma(t^{0},\eta)v^{0}(\eta)\right) }_{\text{Number of human infections caused by a single infected mosquito over its lifetime }}
 \\[2.0mm] 
& \times \underbrace{\beta_{m}(t^{0},\zeta) \left(e^{-\int_{\eta}^{\zeta}\kappaa(t^{0},\xi)\d\xi}f(t^{0},\eta)+e^{-\int_{\eta}^{\zeta}(\kappas(t^{0},\xi)+\delta(t^{0},\xi))\d\xi}f(t^{0},\eta) \right)}_{\text{Number of mosquito infections caused by a single infected human during their infectious period}} \,\d\eta \, \d\zeta.
\label{section4.47}
\end{aligned}
\end{equation}

\begin{remark}[Basic reproduction number]
If the vaccination rate in \eqref{section4.44} is zero, namely if $\alpha(t^{0},\zeta)=0$, then the entire population is susceptible.
This defines the basic reproduction number $R^0$ of our age structured model. 
$R^0$ represent the average number of secondary infections caused by a single infected individual in a completely susceptible population.
In our model, it is given by
\begin{equation}\label{section4.48}
R^0 = \frac{\N^{0}_{m}}{\varLambda_{m}(t^{0})} \int_{0}^{\infty}\beta_{m}(t^{0},\zeta){\Pn}^{0}(\zeta)\int_{0}^{\zeta}\beta_{h}(t^{0},\eta) 
F(\eta,\zeta)\,\d\eta\,\d\zeta.
\end{equation}
Since the vaccination failure rate always satisfies $0\leq\sigma(t^{0},\eta)\leq 1$,
it follows that  
$$
e^{-\int_{0}^{\eta}\alpha(\xi)\,\d\xi} + \sigma(t^{0},\eta) \big(1-e^{-\int_0^\eta\alpha(\xi)\d\xi}\big) \leq 1,
$$
and as a consequence $\Rfo \leq R^0$. In other words, 
the basic effective reproduction number is always smaller than or equal to the basic reproduction number.
Clearly, this confirms that vaccination reduces the disease transmission. 
\end{remark}

\section{Comparison of basic reproduction number obtained with the Lotka equation 
and with the next generation matrix approach}\label{sec_comparison}

In this section, we compare the basic reproduction number obtained in the previous section and using the Euler-Lotka characteristic equation
with the reproduction number derived using the Next Generation Matrix (NGM) approach. 
To avoid confusion in notation, we denote the reproduction number obtained via the Euler–Lotka characteristic equation as 
$\RL0$.
On the other hand, the reproduction number obtained directly from the ODE model using the next generation matrix is going to be denoted by $\RN0$.
Note however that latter approach is rather rudimentary, since it is based on ODEs instead of PDEs, 
and since the parameters are assumed to be time-independent. For that reason, the expression borrowed from the previous
section will have to be simplified a lot.

\subsection{Derivation of the basic reproduction number using the Euler-Lotka approach}

To derive $\RL0$ for the ODE model, which assumes that the disease transmission is not dependent on the age structure of the human population.
We also assume that the parameters are independent of age $\zeta$ and time $t$. We shall therefore consider the following time independent parameters:
\begin{align}\label{section5.1}
\begin{split}
& \beta_{h}(t,\zeta) \equiv \beta_{h}, \quad \beta_{m}(t,\zeta) \equiv \beta_{m}, \quad \kappa_{a}(t,\zeta) \equiv \kappa_{a}, 
\quad \kappa_{s}(t,\zeta) \equiv \kappa_{s},\quad \delta(t,\zeta) \equiv \delta, \\
&\kappa_h(t,\zeta) \equiv \kappa_h, \quad \muh(t,\zeta) \equiv \mu_h, \quad f(t,\zeta) \equiv f,\quad \varLambda_{m}(t) \equiv \varLambda_{m}.
\end{split}
\end{align}
By making these assumptions and using \eqref{section4.48}, we define
\begin{align*}
\RL0 & =\frac{\N_{m}^{0}\beta_{h}\beta_{m}}{\varLambda_{m}}\int_{0}^{\infty}{\Pn}^{0}(\zeta)\int_{0}^{\zeta}
\bigg[e^{-\int_{\eta}^{\zeta}\kappaa\d\xi}(1-f)+e^{-\int_{\eta}^{\zeta}(\kappas+\delta)\d\xi}f\bigg]\,\d\eta\,\d\zeta\\
 & =\frac{\N_{m}^{0}\beta_{m}\beta_{h}}{\varLambda_{m}}\int_{0}^{\infty}{\Pn}^{0}(\zeta)\left(\int_{0}^{\zeta}e^{-\kappaa(\zeta-\eta)}(1-f)\,\d\eta+\int_{0}^{\zeta}e^{-(\kappas+\delta)(\zeta-\eta)}f\,\d\eta\right)\,\d\zeta
\end{align*}
which can be further simplified as
\begin{align*}
\RL0 & =\frac{\N_{m}^{0}\beta_{m}\beta_{h}}{\varLambda_{m}}\int_{0}^{\infty}{\Pn}^{0}(\zeta)\left(\int_{0}^{\zeta}e^{-\kappaa(\zeta-\eta)}(1-f)\,\d\eta+\int_{0}^{\zeta}e^{-(\kappas+\delta)(\zeta-\eta)}f\,\d\eta\right)\,\d\zeta\\
& =\frac{\N_{m}^{0}\beta_{m}\beta_{h}}{\varLambda_{m}}\int_{0}^{\infty}{\Pn}^{0}(\zeta)\bigg[\frac{(1-f)}{\kappaa}\left(1-e^{-\kappaa\zeta}\right)+\frac{f}{(\kappas+\delta)}\left(1-e^{-(\kappas+\delta)\zeta}\right)\bigg]\,\d\zeta\\
& =\frac{\N_{m}^{0}\beta_{m}\beta_{h}}{\varLambda_{m}}\left[\frac{(1-f)}{\kappaa}\left(1-\int_{0}^{\infty}{\Pn}^{0}(\zeta)e^{-\kappaa\zeta}\,\d\zeta\right)+\frac{f}{(\kappas+\delta)}\left(1-\int_{0}^{\infty}{\Pn}^{0}(\zeta)e^{-(\kappas+\delta)\zeta}\right)\,\d\zeta\right].
\end{align*}
 
As seen in Figure \ref{fig3} the population distribution denoted as ${\Pn}^{0}$ can vary significantly between regions. 
In the two examples of Figure \ref{fig3} it is clear that Pakistan has a younger population with a high birth rate, 
while Japan has an old population with declining birth rates. Thus, in any concrete situation, the exact distribution $p^0$
should be implemented, in order to obtain the basic reproduction number $\RL0$ for a given human population.
Quite interestingly, a crude estimate shows that density $p^0$ concentrated near $0$ would lead to a basic reproduction number $\RL0$
smaller compared to a density $p^0$ having a large extension towards large ages.
By keeping the example of Pakistan and Japan, the former is likely to have a smaller $\RL0$ compared to the 
one computed with the Japanese population distribution. 

In order to produce a basic reproduction number independent of any concrete population, let us use the idealistic
population derived in Remark \ref{rem_ideal_pop}.
Namely, we consider the age distribution given by $\Pn^{0}(\zeta) = \muh e^{-\muh \zeta}$. 
For this population, we can continue the above computation and get: 
\begin{align*}
\RL0 & =\frac{\N_{m}^{0}\beta_{m}\beta_{h}}{\varLambda_{m}}\muh\int_{0}^{\infty}e^{-\muh\zeta}\bigg[\frac{(1-f)}{\kappaa}\left(1-e^{-\kappaa\zeta}\right)+\frac{f}{(\kappas+\delta)}\left(1-e^{-(\kappas+\delta)\zeta}\right)\bigg]\,\d\zeta,\\
& =\frac{\N_{m}^{0}\beta_{m}\beta_{h}}{\varLambda_{m}}\muh\left[\frac{(1-f)}{\kappaa}\int_{0}^{\infty}\left(e^{-\muh\zeta}-e^{-(\kappaa+\muh)\zeta}\right)\,\d\zeta+\frac{f}{(\kappas+\delta)\zeta}\int_{0}^{\infty}\left(e^{-\muh\zeta}-e^{-(\kappas+\delta+\muh)\zeta}\right)\,\d\zeta\right],\\
& =\frac{\N_{m}^{0}}{\varLambda_{m}}\beta_{m}\beta_{h}\muh\bigg[\frac{(1-f)}{\kappaa}\bigg(\frac{-1}{-\muh}-\frac{-1}{-(\kappaa+\muh)}\bigg)+\frac{f}{(\kappas+\delta)}\bigg(\frac{-1}{-\muh}-\frac{-1}{-(\kappas+\delta+\muh)}\bigg)\bigg],\\
& =\frac{\N_{m}^{0}}{\varLambda_{m}}\beta_{m}\beta_{h}\muh\bigg[\frac{(1-f)}{\muh(\kappaa+\muh)}+\frac{f}{\muh(\kappas+\delta+\muh)}\bigg].
\end{align*}
Thus, the basic reproduction number $\RL0$ for the idealistic population is given by:
\begin{equation}\label{section5.6}
\RL0= \frac{\beta_{m} \beta_{h}\N_{m}^{0}}{\varLambda_{m}} \bigg( \frac{1-f}{\kappa_{a}+\muh} + \frac{f}{\kappa_{s}+\delta+\muh} \bigg).
\end{equation}

\subsection{Derivation of the basic reproduction number using the next generation matrix approach}

In order to use the approach of the next generation matrix, the first task is to get an ODE system from our original PDE system.
Starting from the initial system given in \eqref{section1.2}--\eqref{section1.9}, 
and using the simplifying assumptions \eqref{section5.1}, we can derive the following ODE model:
\begin{align}\label{section5.5}
\begin{split}
\frac{\d \Sh(t)}{\d t} & =-\Big(\beta_{h}\frac{\Im(t)}{\Nm}+\alpha\Big)\Sh(t)-\muh \Sh(t),\\
\frac{\d \Vh(t)}{\d t} & =\alpha \Sh(t)-\sigma\beta_{h}\frac{\Im(t)}{\Nm}\Vh(t)-\muh \Vh(t),\\
\frac{\d \Ah(t)}{\d t} & =\Big(1-f\Big)\beta_{h}\frac{\Im(t)}{\Nm}\Big(\Sh(t)+\sigma \Vh(t)\Big)-\Big(\kappaa+\muh\Big)\Ah(t),\\
\frac{\d \Ih(t)}{\d t} & =f\beta_{h}\frac{\Im(t)}{\Nm}\Big(\Sh(t)+\sigma \Vh(t)\Big)-\Big(\kappas+\delta+\muh\Big)\Ih(t),\\
\frac{\d \Hh(t)}{\d t} & =\delta \Ih(t)-\Big(\kappah+\muH+\muh\Big)\Hh(t),\\
\frac{\d \Rh(t)}{\d t} & =\kappaa \Ah(t)+\kappas \Ih(t)+\kappah \Hh(t)-\muh \Rh(t),\\
\frac{\d{\Sm}(t)}{{\d}t} & =\varLambda_{m}-\beta_m\frac{{\Ih}(t,\zeta)+{\Ah}(t)}{{\Nht}}{\Sm}(t)-\mum{\Sm}(t),\\
\frac{\d{\Im}(t)}{{\d}t} & =\beta_m\frac{{\Ih}(t)+{\Ah}(t)}{{\Nht}}{\Sm}(t)-\mum{\Im}(t).
\end{split}
\end{align}
with the initial conditions 
\begin{eqnarray*}
 & \Sh(0)={\Sh}_{0}\geq0,\quad \Vh(0)={\Vh_{0}}\geq0,\quad \Ah(0)={\Ah}_{0}>0,\quad \Ih(0)={\Ih}_{0}>0,\\
 & \Hh(0)={\Hh}_{0}\geq0,\quad \Rh(0)={\Rh}_{0}\geq0,\quad \Sm(0)={\Sm}_{0}\geq0,\quad \Im(0)=I_{m\,0}>0.
\end{eqnarray*}

To compute the basic reproduction number of the age independent model \eqref{section5.5}, we use the next generation matrix approach. 
This approach consists in finding the dominant eigenvalue of the next generation matrix. 
The next generation matrix is given by $G=FV^{-1},$ where $F=\frac{\partial \mathcal{F}_{i}}{\partial x_{j}}(E^{0})$ 
is the Jacobian matrix of the new infection vector-valued function $\mathcal{F}$, 
and $V=\frac{\partial\mathcal{V}_{i}}{\partial x_{j}}(E^{0})$  is the Jacobian matrix of the transition vector-valued function $\mathcal{V}$. 
Here $x_{i}$ denotes the infected classes and $E^{0}$ corresponds to the disease free equilibrium point. 
Therefore, we define $FV^{-1}$ as the next-generation vector for the model and set 
$\RN0:=\rho(FV^{-1}),$ where $\rho(A)$ denotes the spectral radius of $A$. We also refer to \cite{Driessche2017} for more details.

In the model \eqref{section5.5}, $\Ah,\Ih,$ and $\Im$ are the infected classes. 
We do not consider the class $\Hh$ because it does not interact with susceptible mosquitoes and does not spread the disease. 
As for relation \eqref{section5.6}, we assume that $\sigma=0$.
Therefore, we have the functions $\mathcal{F}$ and $\mathcal{V}$ as 
\begin{equation*}
\mathcal{F}=  \left(\begin{array}{c}
\Big(1-f\Big)\beta_{h}\frac{\Im(t)}{\Nm}\Sh(t)\\
f\beta_{h}\frac{\Im(t)}{\Nm}\Sh(t)\\
\beta_m\frac{{\Ih}(t)+{\Ah}(t)}{{\Nht}}{\Sm}(t)
\end{array}\right),
\qquad \mathcal{V}=\left(\begin{array}{c}
\Big(\kappaa+\muh\Big)\Ah(t)\\
\Big(\kappas+\delta+\muh\Big)\Ih(t)\\
\mum\Im(t) \end{array}\right).
\end{equation*}
For the Jacobian matrices $\mathcal{F}$ and $\mathcal{V}$ at the DFE point we get
\begin{equation*}
F= \left(\begin{array}{ccc}
0 & 0 & \frac{\left(1-f\right)\beta_{h}\Sh^{0}}{\N^{0}_{m}}\\
0 & 0 & \frac{f\beta_{h}\Sh^{0}}{\N^{0}_{m}}\\
\frac{\beta_m \Sm^{0}}{\N^{0}_{h}} & \frac{\beta_m \Sm^{0}}{\N^{0}_{h}} & 0
\end{array}\right), \qquad 
V=\left(\begin{array}{ccc}
\kappaa+\muh & 0 & 0\\
0 & \kappas+\delta+\muh & 0\\
0 & 0 & \mum
\end{array}\right).
\end{equation*}
As a consequence, the next generation matrix $G$ is given by
\begin{equation}\label{eq_def_G}
G=FV^{-1}=  \left(\begin{array}{ccc}
0 & 0 & \frac{\left(1-f\right)\beta_{h}\Sh^{0}}{\N^{0}_{m}\mum}\\
0 & 0 & \frac{f\beta_{h}\Sh^{0}}{\N^{0}_{m}\mum}\\
\frac{\beta_{m}\Sm^{0}}{(\kappaa+\muh)\N^{0}_{h}} & \frac{\beta_{m}\Sm^{0}}{(\kappas+\delta+\muh)\N^{0}_{h}} & 0
\end{array}\right).
\end{equation}
The analysis of this matrix is provided in the next lemma

\begin{lemma} \label{lemaappend}
The eigenvalues of the next generation matrix $G$ defined in \eqref{eq_def_G} are 
\[
\lambda_1 = 0, \quad 
\lambda_2 = -\sqrt{\frac{\beta_m\beta_h \Sm^{0}\Sh^0 }{\mum\N^{0}_m\N^{0}_h} \left( \frac{f}{\kappa_s + \delta + \mu_h} + \frac{1 - f}{\kappa_a + \mu_h} \right)}, \quad \lambda_3 = \sqrt{\frac{\beta_m\beta_h \Sm^{0}\Sh^0 }{\mum\N^{0}_m\N^{0}_h} \left( \frac{f}{\kappa_s + \delta + \mu_h} + \frac{1 - f}{\kappa_a + \mu_h} \right)}.
\]
\end{lemma}

\begin{proof}
The eigenvalues $\lambda$ of the matrix $G$ are the solutions to the characteristic equation $\det(G-\lambda I)=0$, 
where $I$ is the identity matrix. Thus we have 
\begin{equation*}
\det\left(G-\lambda I\right)=\left|\begin{array}{ccc}
-\lambda & 0 & \frac{\left(1-f\right)\beta_{h}S^{0}}{\N_{m}^{0}\mu_{m}}\\
0 & -\lambda & \frac{f\beta_{h}S^{0}}{\N_{m}^{0}\mu_{m}}\\
\frac{\beta_{m}S_{m}^{0}}{(\kappa_{a}+\mu_{h})\N_{h}^{0}} & \frac{\beta_{m}S_{m}^{0}}{(\kappa_{s}+\delta+\mu_{h})\N_{h}^{0}} & -\lambda
\end{array}\right|=0.
\end{equation*}
By computing the determinant we get
\begin{equation*}
-\lambda\left(\lambda^{2}-\frac{f\beta_{h}S^{0}}{\N_{m}^{0}\mu_{m}}\frac{\beta_{m}S_{m}^{0}}{(\kappa_{s}+\delta+\mu_{h})\N_{h}^{0}}\right)+\lambda\frac{\left(1-f\right)\beta_{h}S^{0}}{\N_{m}^{0}\mu_{m}}\frac{\beta_{m}S_{m}^{0}}{(\kappa_{a}+\mu_{h})\N_{h}^{0}}=0.
\end{equation*}
As a consequence, we infer that $\lambda_{1}=0$, and the remaining equation becomes
\begin{equation}
-\lambda^{2}+\frac{f\beta_{h}S^{0}}{\N_{m}^{0}\mu_{m}}\frac{\beta_{m}S_{m}^{0}}{(\kappa_{s}+\delta+\mu_{h})\N_{h}^{0}}+\frac{\left(1-f\right)\beta_{h}S^{0}}{\N_{m}^{0}\mu_{m}}\frac{\beta_{m}S_{m}^{0}}{(\kappa_{a}+\mu_{h})\N_{h}^{0}}=0.\label{appendixb4}
\end{equation}
After simplification, we directly infer the other two eigenvalues mentioned in the statement. 
\end{proof}

Let us recall that at the DFE point, we have $\Sh^0=\N^{0}_{h}$ and $\Sm^0=\N^{0}_{m}$.
In addition, the death rate for mosquitoes satisfies the relation $\mu_m=\frac{\varLambda_m}{\N_m^0}$, 
as a consequence of \eqref{section1.00}.
Thus, from the previous statement and from these relations one sets
\begin{equation*}
\RN0:=\sqrt{\frac{\beta_{h}\beta_{m}\N_{m}^{0}}{\varLambda_{m}}\left(\frac{(1-f)}{\kappa_a+\mu_h}+\frac{f}{(\delta+\kappa_s+\mu_h)}\right)}.
\end{equation*}

It immediately appears that the following relation holds:
$$
\RN0=\left(\RL0\right)^{2}.
$$
A natural question arises about the square root difference. 
The presence of the square root in $\RN0$ is due to the fundamental concept of vector to host and host to vector transmission dynamics. 
This square root formulation accounts for the two step transmission process, 
where the infection cycle involves both vector to host and host to vector transitions, rather than direct host to host transmission. 
In \cite[Sec.~2.2]{Heffernan2005} the difference between the Lotka approach and the next generation matrix approach is further explained. 
Lotka's method gives the total infectivity in a class, assuming that each infected individual contributes to new infections within the same class. 
In contrast, the next generation matrix calculates the average number of new infections per infected individual, 
considering that infections occur across multiple classes in the infection cycle. 
Therefore, the square root in $\RN0$ arises naturally from the multi-step transmission dynamics present in vector borne diseases.

\section{Stability analysis}\label{sec_stability}

In this section, we investigate the local behavior of the basic reproduction number at the disease-free equilibrium point $E^{0}$. 
To establish the local stability of the equilibrium solutions of the model, 
we apply analytical techniques presented in \cite[Sec.~9.5.1]{Li2008}.

\begin{thm}
The DFE point $E^{0}$ is locally asymptotically stable if $\Rfo<1$ and unstable if $\Rfo>1$. 
\end{thm}

\begin{figure}[htbp]
    \centering
    \subfloat[${\Rfo=1 \text{ and } \lambda^*=0}$]{\includegraphics[width=0.33\textwidth]{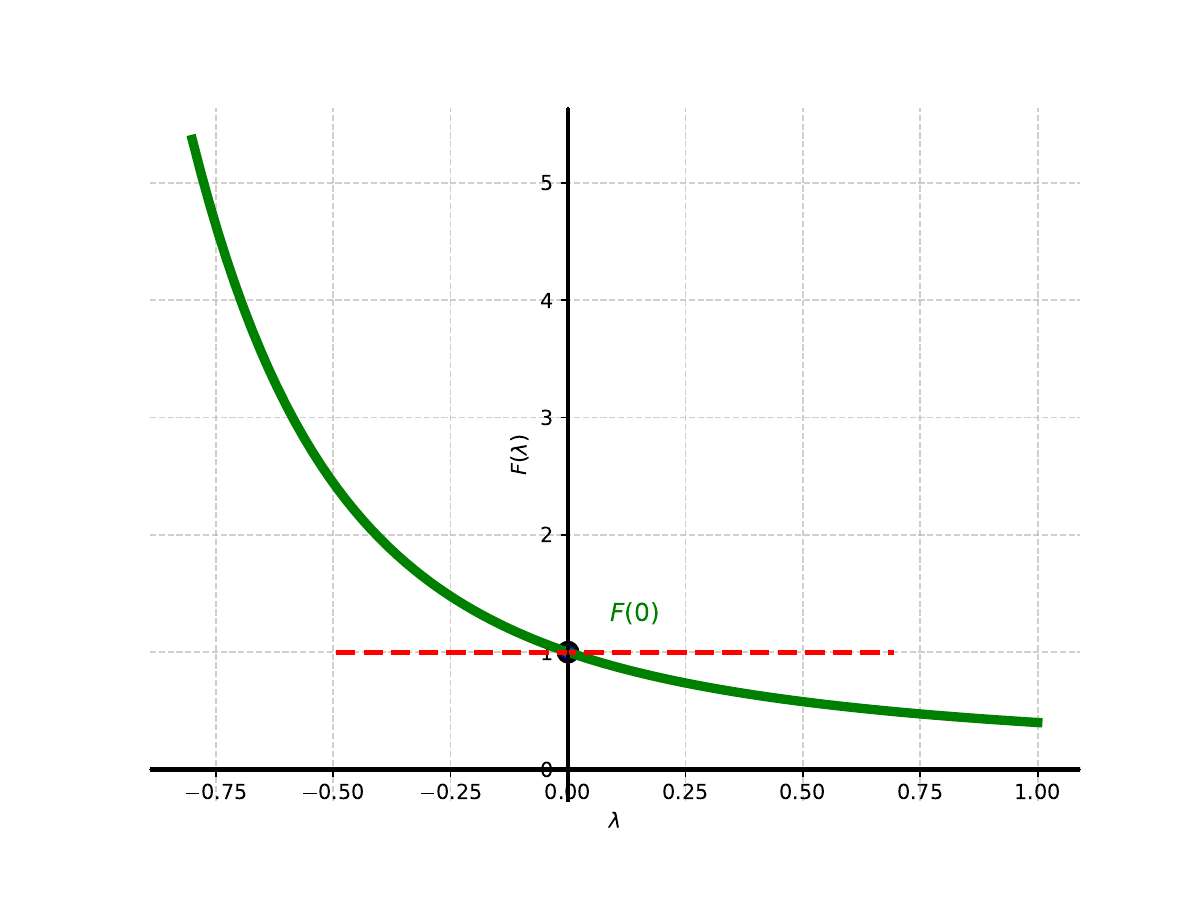}}
    \hfill
    \subfloat[${\Rfo<1 \text{ and } \lambda^*<0}$]{\includegraphics[width=0.33\textwidth]{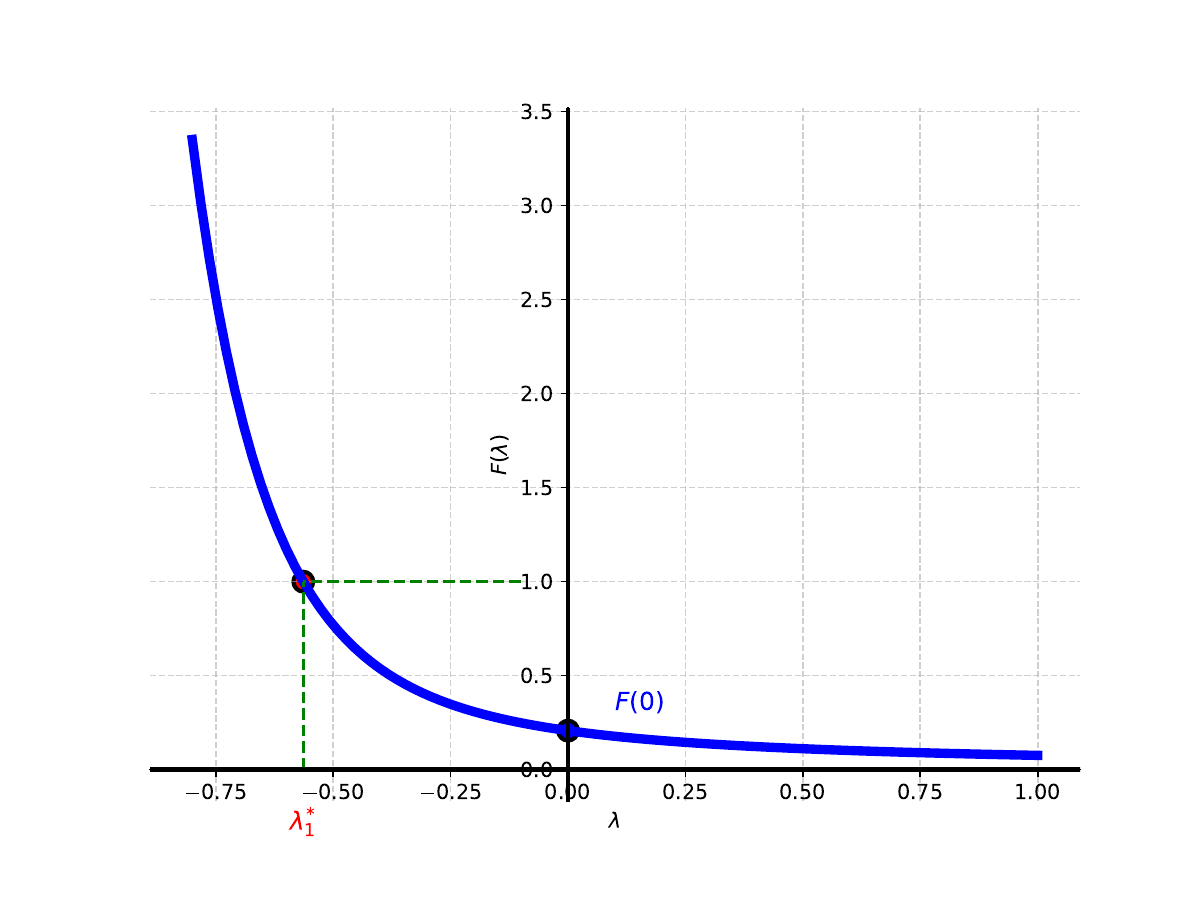}}
    \hfill
    \subfloat[${\Rfo>1 \text{ and } \lambda^*>0}$]{\includegraphics[width=0.33\textwidth]{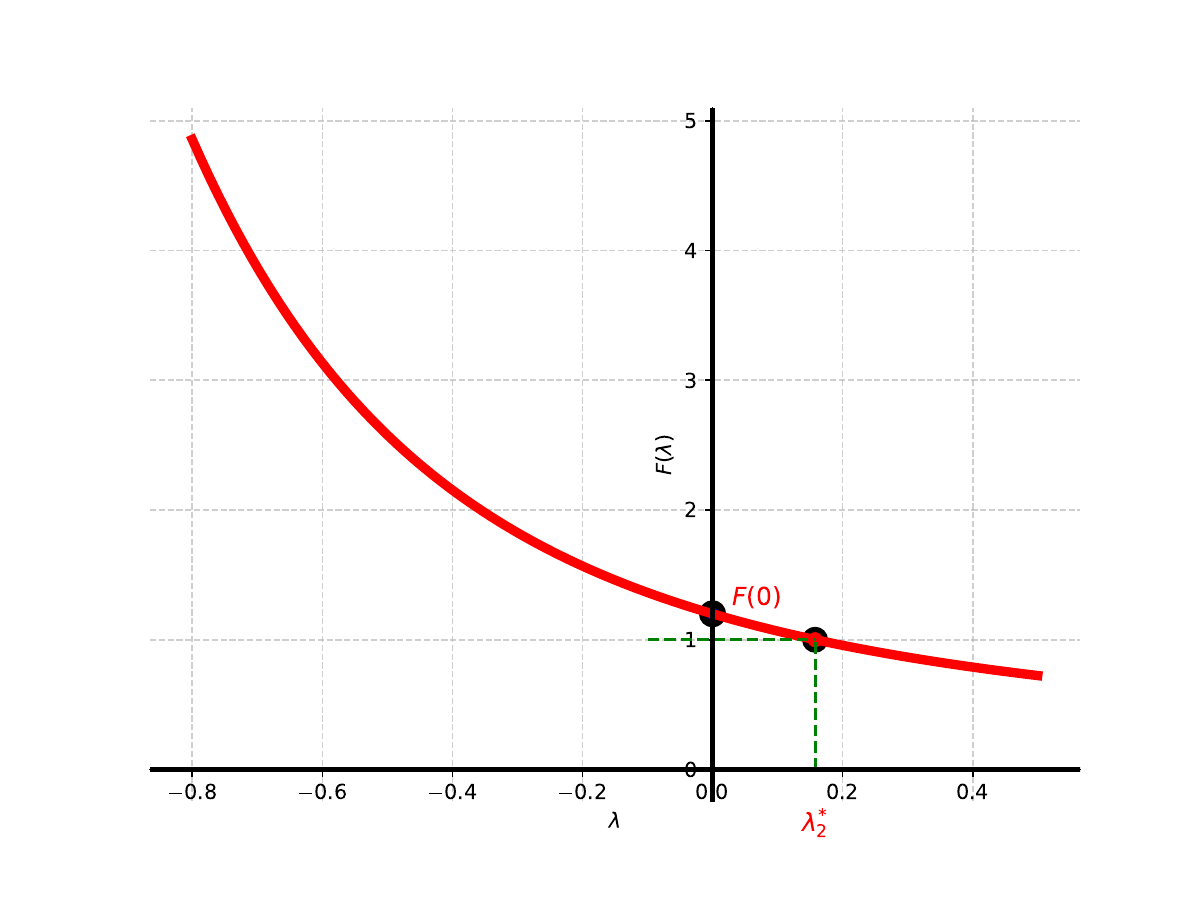}}
    
    \caption{Illustration of the three possible relations between 
    basic effective reproduction number $\Rfo$ and growth rate $\lambda^*$.}
    \label{fig4}
\end{figure}

\begin{proof}
Let us recall that the function $\lambda \mapsto L(\lambda)$ has been introduced in \eqref{section4.19} 
and reads
\begin{equation*}
L(\lambda) 
:=\frac{\N^{0}_{m}}{\lambda\N^{0}_{m}+\varLambda_{m}(t^{0})} \int_{0}^{\infty}\beta_{m}(t^{0},\zeta){\Pn}^{0}(\zeta)
\int_{0}^{\zeta}\beta_{h}(t^{0},\eta)\Big(s^{0}(\eta)+\sigma(t^{0},\eta)v^{0}(\eta)\Big) 
F(\eta,\zeta)e^{-\lambda(\zeta-\eta) }\,\d\eta\,\d\zeta.
\end{equation*}
for $\lambda \in \R$. By differentiating with respect to $\lambda$ one gets 
\begin{align}\label{section6.1}
\nonumber &\frac{\d L}{\d \lambda}(\lambda)  \\
\nonumber &= - \frac{(\N^{0}_{m})^2}{\left(\lambda\N^{0}_{m} + \varLambda_{m}(t^{0})\right)^2} \int_{0}^{\infty} \beta_{m}(t^{0},\zeta) {\Pn}^{0}(\zeta) 
\int_{0}^{\zeta}\beta_{h}(t^{0},\eta)\Big(s^{0}(\eta)+\sigma(t^{0},\eta)v^{0}(\eta)\Big) 
F(\eta,\zeta)e^{-\lambda(\zeta-\eta) }\,\d\eta\,\d\zeta  \\
&\quad-\frac{\N^{0}_{m}}{\lambda\N^{0}_{m} + \varLambda_{m}(t^{0})} \int_{0}^{\infty} \beta_{m}(t^{0},\zeta) {\Pn}^{0}(\zeta) \int_{0}^{\zeta} (\zeta-\eta) \beta_{h}(t^{0},\eta) \left(s^{0}(\eta)+\sigma(t^{0},\eta)v^{0}(\eta)\right)
F(\eta,\zeta)e^{-\lambda(\zeta-\eta)} \,\d\eta\,\d\zeta.
\end{align}
From \eqref{section4.18} and \eqref{section6.1}, we deduce that $L$ is a strictly decreasing function. 
It also satisfies the following properties:
$$
\lim_{\lambda\searrow \lambda_{s}}L(\lambda)=\infty, \qquad \lim_{\lambda\nearrow\infty}L(\lambda)=0,
$$
where $\lambda_{s}=-\frac{\varLambda_{m}(t^{0})}{\N^{0}_{m}}$.

Recall now that $\Rfo:=L(0)$, with its precise expression provided in \eqref{section4.20}.
We also define the growth rate $\lambda^*\in \R$ by the equation $L(\lambda^*)=1$.
Clearly, if $\Rfo=1$, then $\lambda^*=0$, as illustrated in Figure \ref{fig4}.(a).
On the other hand, if $\Rfo<1$, then by monotonicity of $L$ one has $\lambda^*<0$, see Figure \ref{fig4}.(b).
Finally, if $\Rfo>1$, then we must have $\lambda^*>0$, as shown in Figure \ref{fig4}.(c)
These relation between $\Rfo$ and $\lambda^*$ lead directly to the statement.
\end{proof}

\section{Time-varying effective reproduction number}\label{sec_tvern}

During an epidemic, the number of susceptible individuals in the host $\Big(s(t,\zeta)+\sigma(t,\zeta)v(t,\zeta)\Big)$
and in the vector populations $\sm(t)$ gradually decrease as they become infected and are eventually removed from the disease transmission cycle.  
As time progresses, the basic effective reproduction number, which helps measure how the disease spreads in the early infection stage, 
is no longer a reliable indicator. 
Additionally, due to vaccination the values of $s(t,\zeta)$ and $v(t,\zeta)$ also change over time. 
This makes it necessary to estimate the time-varying effective reproduction number $\Rf(t)$.
Therefore, our next aim is to better understand how the disease spreads over time. 

For this, we consider the non-linear normalized system \eqref{section1.13} as
\begin{align}
\frac{\partial s(t,\zeta)}{\partial t}+\frac{\partial s(t,\zeta)}{\partial\zeta}
& =-\Big(\beta_{h}(t,\zeta)\im(t)+\alpha(t,\zeta)\Big)s(t,\zeta)\\
\frac{\partial v(t,\zeta)}{\partial t}+\frac{\partial v(t,\zeta)}{\partial\zeta}
& =\alpha(t,\zeta)s(t,\zeta)-\sigma(t,\zeta)\beta_{h}(t,\zeta)\im(t)v(t,\zeta),\\
\label{section8.1} \frac{\partial i_a(t,\zeta)}{\partial t}+\frac{\partial i_a(t,\zeta)}{\partial \zeta}
&=\Big(1-f(t,\zeta)\Big)\beta_{h}(t,\zeta)\im(t)\Big(s(t,\zeta) +\sigma(t,\zeta)v(t,\zeta)\Big)-\Big(\kappaa(t,\zeta)+g(t,\zeta) \Big)i_{a}(t,\zeta)\\
\label{section8.2} \frac{\partial i_s(t,\zeta)}{\partial t}+\frac{\partial i_s(t,\zeta)}{\partial \zeta}
&=f(t,\zeta)\beta_{h}(t,\zeta)\im(t)\Big(s(t,\zeta) + \sigma(t,\zeta)v(t,\zeta)\Big)-\Big(\kappas(t,\zeta)+\delta(t,\zeta)+g(t,\zeta)\Big)i_{s}(t,\zeta)\\
\label{section8.3}\frac{\d \im(t)}{\d t}
&=\sm(t)\int_{0}^{\infty}\beta_m(t,\zeta)\Pn(t,\zeta)\Big(i_s(t,\zeta)+i_a(t,\zeta)\Big)\,\d\zeta- \im(t)\frac{\varLambda_m(t)}{\Nm},
\end{align}
with initial condition at $t^0$
\begin{equation*}
s(t^{0},\zeta) =s^{0}(\zeta),\quad v(t^{0},\zeta) =v^{0}(\zeta), \quad i_{a}(t^{0},\zeta)=0,\quad i_{s}(t^{0},\zeta)=0, \quad \im(t^0)=i_{m0},
\end{equation*}
and boundary condition
\begin{equation*}
s(t,0) =1,\quad v(t,0) =0, \quad i_{a}(t,0)=0,\quad i_{s}(t,0)=0.
\end{equation*}
By solving \eqref{section8.1} and \eqref{section8.2} along the characteristic line, see Lemma \ref{App_PDE_sol}, we get 
\begin{eqnarray}\label{section8.4}
i_{a}(t,\zeta)= & \begin{cases}
\int_{0}^{\zeta}e^{-\int_{\eta}^{\zeta}[\kappaa(\xi+t-\zeta,\xi)+g(\xi+t-\zeta,\xi)]\d\xi}\Big(1-f(\eta+t-\zeta,\eta)\Big)\beta_{h}(\eta+t-\zeta,\eta)\\
\quad \times \ {\im}(\eta+t-\zeta)\Big(s(\eta+t-\zeta,\eta)+\sigma(\eta+t-\zeta,\eta)v(\eta+t-\zeta,\eta)\Big)\,\d\eta, & \text{for }t-t^{0}\geq\zeta,\\
\int_{t^{0}+\zeta-t}^{\zeta}e^{-\int_{\eta}^{\zeta}[\kappaa(\xi+t-\zeta,\xi)+g(\xi+t-\zeta,\xi)]\d\xi}\Big(1-f(\eta+t-\zeta,\eta)\Big)\beta_{h}(\eta+t-\zeta,\eta)\\
\quad \times \ {\im}(\eta+t-\zeta)\Big(s(\eta+t-\zeta,\eta)+\sigma(\eta+t-\zeta,\eta)v(\eta+t-\zeta,\eta)\Big)\,\d\eta, & \text{for } t-t^{0}<\zeta,
\end{cases}
\end{eqnarray}
and 
\begin{eqnarray}\label{section8.5}
i_{s}(t,\zeta)= & \begin{cases}
\int_{0}^{\zeta}e^{-\int_{\eta}^{\zeta}[\kappas(\xi+t-\zeta,\xi)+\delta(\xi+t-\zeta,\xi)+g(\xi+t-\zeta,\xi)]\d\xi}f(\eta+t-\zeta,\eta)\beta_{h}(\eta+t-\zeta,\eta)\\
\quad \times\ {\im}(\eta+t-\zeta)\Big(s(\eta+t-\zeta,\eta)+\sigma(\eta+t-\zeta,\eta)v(\eta+t-\zeta,\eta)\Big)\,\d\eta, & \text{for }t-t^{0}\geq\zeta,\\
\int_{t^{0}+\zeta-t}^{\zeta}e^{-\int_{\eta}^{\zeta}[\kappas(\xi+t-\zeta,\xi)+\delta(\xi+t-\zeta,\xi)+g(\xi+t-\zeta,\xi)]\d\xi}f(\eta+t-\zeta,\eta)\beta_{h}(\eta+t-\zeta,\eta)\\
\quad \times\ {\im}(\eta+t-\zeta)\Big(s(\eta+t-\zeta,\eta)+\sigma(\eta+t-\zeta,\eta)v(\eta+t-\zeta,\eta)\Big)\,\d\eta, & \text{for }t-t^{0}<\zeta.
\end{cases}
\end{eqnarray}
Equation \eqref{section8.3} can also be rewritten as 
\begin{align}\label{section8.6}
\nonumber &\frac{\d\im(t)}{\d t}+\im(t)\frac{\varLambda_{m}(t)}{\Nm} \\
& =\int_{0}^{t-t^{0}}\sm(t)\beta_{m}(t,\zeta)\Pn(t,\zeta)\Big(i_{s}(t,\zeta)+i_{a}(t,\zeta)\Big)\,\d\zeta+\int_{t-t^{0}}^{\infty}\sm(t)\beta_{m}(t,\zeta)\Pn(t,\zeta)\Big(i_{s}(t,\zeta)+i_{a}(t,\zeta)\Big)\,\d\zeta.
\end{align}
By summing \eqref{section8.4} and \eqref{section8.5} for  $t-t^{0}\geq\zeta$ and substituting the result into the integral expression, we have
\begin{align}\label{section8.7}
\nonumber &\int_{0}^{t-t^{0}}\beta_{m}(t,\zeta){\Pn}(t,\zeta)\Big(i_{s}(t,\zeta)+i_{a}(t,\zeta)\Big)\,\d\zeta \\
&= \int_{0}^{t-t^{0}}\beta_{m}(t,\zeta){\Pn}(t,\zeta)\int_{0}^{\zeta}\beta_{h}(\eta+t-\zeta,\eta){\im}(\eta+t-\zeta)
\big[s+\sigma v\big](\eta+t-\zeta,\eta)  F(t,\eta,\zeta)\,\d\eta\,\d\zeta,
\end{align}
with the new factor 
\begin{align}\label{eq_new_F}
\begin{split}
&F(t,\eta, \zeta) \\
&:=e^{-\int_{\eta}^{\zeta}[\kappaa(\xi+t-\zeta,\xi)+g(\xi+t-\zeta,\xi)]\d\xi}\Big(1-f(\eta+t-\zeta,\eta)\Big)
+e^{-\int_{\eta}^{\zeta}[\kappas(\xi+t-\zeta,\xi)+\delta(\xi+t-\zeta,\xi)+g(\xi+t-\zeta,\xi)]\d\xi}f(\eta+t-\zeta,\eta).
\end{split}
\end{align}
By combining \eqref{section8.4} and \eqref{section8.5} for $t-t^{0}<\zeta$, we also get
\begin{align}\label{section8.8}
\nonumber &\int_{t-t^{0}}^{\infty}\beta_{m}(t,\zeta){\Pn}(t,\zeta)\Big(i_{s}(t,\zeta)+i_{a}(t,\zeta)\Big)\,\d\zeta \\
& =\int_{t-t^{0}}^{\infty}\beta_{m}(t,\zeta){\Pn}(t,\zeta)\int_{t^{0}+\zeta-t}^{\zeta}\beta_{h}(\eta+t-\zeta,\eta){\im}(\eta+t-\zeta) 
\big[s+\sigma v\big](\eta+t-\zeta,\eta)  F(t,\eta,\zeta)\,\d\eta\,\d\zeta.
\end{align}
By adding \eqref{section8.7} and \eqref{section8.8} we then infer that
\begin{align}\label{section8.9}
\nonumber & \int_{0}^{\infty}\beta_{m}(t,\zeta){\Pn}(t,\zeta)\Big(i_{s}(t,\zeta)+i_{a}(t,\zeta)\Big)\,\d\zeta \\
& =\int_{0}^{\infty}\beta_{m}(t,\zeta){\Pn}(t,\zeta)\int_{\max\{0,t^{0}+\zeta-t\}}^{\zeta}\beta_{h}(\eta+t-\zeta,\eta){\im}(\eta+t-\zeta)
\big[s+\sigma v\big](\eta+t-\zeta,\eta)  F(t,\eta,\zeta)\,\d\eta\,\d\zeta.
\end{align}
And finally, by substituting \eqref{section8.9} into \eqref{section8.6}, we get the following Lotka-Euler equation
\begin{equation}\label{section8.10}
\begin{aligned}
& \frac{\d\im(t)}{\d t}+\im(t)\frac{\varLambda_{m}(t)}{\Nm}\\
&= \int_{0}^{\infty}\sm(t)\beta_{m}(t,\zeta){\Pn}(t,\zeta)\int_{\max\{0,t^{0}+\zeta-t\}}^{\zeta}\beta_{h}(\eta+t-\zeta,\eta){\im}(\eta+t-\zeta)
\big[s+\sigma v\big](\eta+t-\zeta,\eta)  F(t,\eta,\zeta)\,\d\eta\,\d\zeta,
\end{aligned}
\end{equation}
with $F(t,\eta,\zeta)$ defined in \eqref{eq_new_F}.

For the definition of the time-varying effective reproduction number $\Rf(t)$, we take the decay in time of susceptible individuals 
into account. Therefore, we can not assume an exponential growth, which was applicable only during the very early phase of an epidemic.
For $\Rf(t)$, we use the notion of instantaneous growth rate, which can be defined as 
\begin{equation}\label{section8.11}
r_t := \frac{\d \log(\im(t))}{\d t}. 
\end{equation}
We refer to \cite[Sec.~2.1]{Parag2022} for additional information.
Then, equation \eqref{section8.10} can be written as 
\begin{align}\label{section8.010}
\nonumber &\im(t)\Bigg(\frac{\d(\log \im(t))}{\d t}+\frac{\varLambda_{m}(t)}{\Nm}\Bigg)\\
& =\int_{0}^{\infty}\sm(t)\beta_{m}(t,\zeta){\Pn}(t,\zeta)\int_{\max\{0,t^{0}+\zeta-t\}}^{\zeta}\beta_{h}(\eta+t-\zeta,\eta){\im}(\eta+t-\zeta)
\big[s+\sigma v\big](\eta+t-\zeta,\eta)  F(t,\eta,\zeta)\,\d\eta\,\d\zeta.
\end{align}
By using this definition of the instantaneous growth rate in \eqref{section8.11}, we have
\begin{equation*}
\im(t)={\im}_{0} e^{\int_{t^{0}}^{t} r_{\tau}\d \tau},
\end{equation*}
which implies that 
\begin{equation}\label{section8.0011}
\im(\eta+t-\zeta)={\im}_{0} e^{\int_{t^{0}}^{\eta+t-\zeta} r_{\tau}\d \tau}
={\im}_{0} e^{\int_{t^{0}}^{t} r_{\tau}\d \tau-\int_{\eta+t-\zeta}^{t} r_{\tau}\d \tau}
=\im(t)e^{-\int_{\eta+t-\zeta}^{t} r_{\tau}\d \tau}.
\end{equation}
By using the definition of the instantaneous growth rate and \eqref{section8.0011}, 
\eqref{section8.010} can be written as 
\begin{align*}
& \im(t)\left( r_{t} + \frac{\varLambda_{m}(t)}{\Nm} \right)\\
& = \sm(t)\int_{0}^{\infty} \beta_{m}(t, \zeta) \, \Pn(t, \zeta) \int_{\max\{0, t^{0}+\zeta - t\}}^{\zeta} \beta_{h}(\eta + t - \zeta, \eta) 
[s + \sigma v](\eta + t - \zeta, \eta)F(t,\eta,\zeta)\im(t)e^{-\int_{\eta+t-\zeta}^{t} r_{\tau} \, \d\tau} \, \d\eta \, \d\zeta.
\end{align*}
By canceling ${\im}(t)$ from both sides, we obtain the equation $1= \L(t,r_t)$ with 
\begin{align*}
&\L(t,r_t) :=\frac{\Nm\sm(t)}{r_{t}\Nm+\varLambda_{m}(t)} \\
& \qquad\quad \times \ \int_{0}^{\infty}\beta_{m}(t,\zeta){\Pn}(t,\zeta)\int_{\max\{0,t^{0}+\zeta-t\}}^{\zeta}\beta_{h}(\eta+t-\zeta,\eta) 
\big[s+\sigma v\big](\eta+t-\zeta,\eta)  F(t,\eta,\zeta) e^{-\int_{\eta+t-\zeta}^{t}  r_{\tau} \, \d\tau} \,\d\eta\,\d\zeta.
\end{align*}

The equation $\L(t, r_{t})=1$ represents the Lotka characteristic equation. 
Following \cite[Eq.~(24C)]{Mills2024} and \cite[Sec.~2.2]{Nishiura2009}, 
we define the time-varying effective reproduction number by 
\begin{equation}\label{section8.0012}
\Rf(t) :=
 \frac{\Nm\sm(t)}{\varLambda_{m}(t)} \int_{0}^{\infty} \beta_{m}(t,\zeta){\Pn}(t,\zeta) \int_{\max\{0,t^{0}+\zeta-t\}}^{\zeta} \beta_{h}(\eta+t-\zeta,\eta)
 \big[s+\sigma v\big](\eta+t-\zeta,\eta)  F(t,\eta,\zeta) \,\d\eta\,\d\zeta.
\end{equation}
The reproduction number $\Rf(t)$ represents the average number of secondary infections caused by a single infected individual at time $t$. 
The expected number of secondary cases generated by an individual is determined by the cumulative infectivity 
during the period from when the individual is infected (at age $\eta$) until the current age $\zeta$, 
with this infectivity being diminished by recovery or death. 

\section{Relation between instantaneous growth rate and time-varying effective reproduction number}\label{sec_relations}

In this section, we derive the relation between the instantaneous growth rate $r_t$ and the time-varying effective reproduction number $\Rf(t)$,
introduced in \eqref{section8.0012}.
The interpretation for the time-varying effective reproduction number $\Rf(t)$ is rather straightforward. 
Suppose a disease has a generation time of 7 days, which is defined as the average interval of time between the time 
an individual becomes infected and the time at which this individual transmits the infection to others. 
If $\Rf(t)=3$, this implies that $7$ days later, namely on day $t+7$, the number of new cases will be three times the number 
of new cases on day $t$. Note that in this example we consider $t$ as a measure of the days. 
Therefore, whenever $\Rf(t)>1$, the number of cases in the future will be larger than the current number. 
Conversely, if $\Rf(t)<1$, the number of new cases will be smaller and the incidence is decreasing. 
If $\Rf(t)=1$, the incidence is stable. 
Thus, one observes that $\Rf(t)$ is dimensionless and it cannot provide any information about time, 
or more precisely about the rapidity of the spread of the disease.

For example, consider two diseases, $\D_{1}$ and $\D_{2}$, both with $\Rf(t)=3$ but with different generation time:
$7$ days for the disease $\D_{1}$ and $2$ months for the disease $\D_{2}$. 
For the disease $\D_{1}$, $\Rf(t)=3$
means that the number of new cases will triple every $7$ days, while for the disease $\D_{2}$ 
it means that the number of new cases will triple every $2$ months. 
Clearly, $\Rf(t)$ gives no information about how quickly the number of cases is increasing or decreasing.

An alternative measure is the instantaneous growth rate $r_t$, as introduced in \eqref{section8.11}, 
and which is less used than $\Rf(t)$. 
An instantaneous growth rate $r_t$ greater than $0$ means that the incidence is increasing, while if less than $0$ means that the incidence is declining.
For example, if $r_{t}=0.5$/week it means that the number of cases on day $t+7$ will be 50\% higher than the number of cases on day $t$. 
If the growth rate is $0$, then the number of cases will remain the same for all time $t$. 
Therefore, the epidemic growth rate has a few advantages over $\Rf(t)$: it is attached with a notion of time.

In order to better understand this, we analyze the dynamics of the infected classes in system \eqref{section1.13} 
and examine relation between the time-varying effective reproduction number $\Rf(t)$ and the instantaneous growth rate $r_t$.
Let us consider the system:
\begin{align}
\label{section9.1} \frac{\partial i_a(t,\zeta)}{\partial t}+\frac{\partial i_a(t,\zeta)}{\partial \zeta}
&=\Big(1-f(t,\zeta)\Big)\beta_{h}(t,\zeta)\im(t)\Big(s(t,\zeta) +\sigma(t,\zeta)v(t,\zeta)\Big)-\Big(\kappaa(t,\zeta)+g(t,\zeta) \Big)i_{a}(t,\zeta)\\
\label{section9.2} \frac{\partial i_s(t,\zeta)}{\partial t}+\frac{\partial i_s(t,\zeta)}{\partial \zeta}
&=f(t,\zeta)\beta_{h}(t,\zeta)\im(t)\Big(s(t,\zeta) + \sigma(t,\zeta)v(t,\zeta)\Big)-\Big(\kappas(t,\zeta)+\delta(t,\zeta)+g(t,\zeta)\Big)i_{s}(t,\zeta)\\
\label{section9.3}\frac{\d \im(t)}{\d t}
&=\sm(t)\int_{0}^{\infty}\beta_m(t,\zeta)\Pn(t,\zeta)\Big(i_s(t,\zeta)+i_a(t,\zeta)\Big)\,\d\zeta- \im(t)\frac{\varLambda_m(t)}{\Nm},
\end{align}
with initial condition at $t^0$
\begin{equation*}
s(t^{0},\zeta) =s^{0}(\zeta),\quad v(t^{0},\zeta) =v^{0}(\zeta), \quad i_{a}(t^{0},\zeta)=0,\quad i_{s}(t^{0},\zeta)=0, \quad \im(t^0)=i_{m0},
\end{equation*}
and boundary condition
\begin{equation*}
s(t,0) =1,\quad v(t,0) =0, \quad i_{a}(t,0)=0,\quad i_{s}(t,0)=0.
\end{equation*}
By using the solution of equations \eqref{section9.1} and \eqref{section9.2} 
given respectively in \eqref{section8.4} and \eqref{section8.5}, we can substitute these solutions into \eqref{section9.3} and obtain
\begin{align*}
& \frac{\d\im(t)}{\d t}+\im(t)\frac{\varLambda_{m}(t)}{\Nm} \\
&= \int_{0}^{\infty}\sm(t)\beta_{m}(t,\zeta){\Pn}(t,\zeta)\int_{\max\{0,t^{0}+\zeta-t\}}^{\zeta}\beta_{h}(\eta+t-\zeta,\eta){\im}(\eta+t-\zeta)
\big[s+\sigma v\big](\eta+t-\zeta,\eta)  F(t,\eta,\zeta)\,\d\eta\,\d\zeta,
\end{align*}
with $F(t,\eta,\zeta)$ defined in \eqref{eq_new_F}.
By dividing both sides by $\im(t)$, we then get 
\begin{equation}\label{section9.5}
\begin{aligned}
& \frac{1}{\im(t)}\frac{\d\im(t)}{\d t}+\frac{\varLambda_{m}(t)}{\Nm}\\
&= \int_{0}^{\infty}\sm(t)\beta_{m}(t,\zeta){\Pn}(t,\zeta)\int_{\max\{0,t^{0}+\zeta-t\}}^{\zeta}\beta_{h}(\eta+t-\zeta,\eta)\frac{{\im}(\eta+t-\zeta)}{\im(t)}
\big[s+\sigma v\big](\eta+t-\zeta,\eta)  F(t,\eta,\zeta)\,\d\eta\,\d\zeta.
\end{aligned}
\end{equation}
or equivalently
\begin{equation}\label{section9.6}
\begin{aligned}
\frac{1}{\im(t)}\frac{\d\im(t)}{\d t} 
= & \frac{\varLambda_{m}(t)}{\Nm}\Bigg[\frac{\Nm}{\varLambda_{m}(t)}\int_{0}^{\infty}\sm(t)\beta_{m}(t,\zeta){\Pn}(t,\zeta)\int_{\max\{0,t^{0}+\zeta-t\}}^{\zeta}\beta_{h}(\eta+t-\zeta,\eta)\frac{{\im}(t-(\zeta-\eta))}{\im(t)}  \\
& \qquad\qquad \times \big[s+\sigma v\big](\eta+t-\zeta,\eta)  F(t,\eta,\zeta) \,\d\eta\,\d\zeta-1\Bigg].
\end{aligned}
\end{equation}

Let us now approximate the term $\im(t-(\zeta-\eta))$ by using a first-order Taylor expansion. 
Since $\zeta-\eta$ represents the infectivity age of the infected individuals, 
which is relatively small, and since the age of the mosquitoes is also very small, 
we have $\frac{i_{m}(t-(\zeta-\eta)}{i_{m}(t)}\approx1-\frac{\frac{\d}{\d t} i_m(t) (\zeta - \eta)}{i_m(t)}$.
This leads to 
\begin{equation}\label{section9.7}
\begin{aligned}
\frac{1}{\im(t)}\frac{\d\im(t)}{\d t} 
= & \frac{\varLambda_{m}(t)}{\Nm}\Bigg[\frac{\Nm}{\varLambda_{m}(t)}\int_{0}^{\infty}\sm(t)\beta_{m}(t,\zeta){\Pn}(t,\zeta)\int_{\max\{0,t^{0}+\zeta-t\}}^{\zeta}\beta_{h}(\eta+t-\zeta,\eta)\bigg(1-\frac{\frac{\d}{\d t} i_m(t)(\zeta - \eta)}{i_m(t)}\bigg)  \\
& \qquad\qquad \times \big[s+\sigma v\big](\eta+t-\zeta,\eta)  F(t,\eta,\zeta) \,\d\eta\,\d\zeta-1\Bigg].
\end{aligned}
\end{equation}

By using the definition of the instantaneous growth rate $r_{t}$ introduced in \eqref{section8.11} and explained in \cite[Sec.~2.1]{Parag2022},
the previous equation reads
\begin{equation}\label{section9.7a}
\begin{aligned}
r_t 
= & \frac{\varLambda_{m}(t)}{\Nm}\Bigg[\frac{\Nm}{\varLambda_{m}(t)}\int_{0}^{\infty}\sm(t)\beta_{m}(t,\zeta){\Pn}(t,\zeta)\int_{\max\{0,t^{0}+\zeta-t\}}^{\zeta}\beta_{h}(\eta+t-\zeta,\eta) \big(1-r_{t}(\zeta - \eta)\big)  \\
& \qquad\qquad \times \big[s+\sigma v\big](\eta+t-\zeta,\eta)  F(t,\eta,\zeta) \,\d\eta\,\d\zeta-1\Bigg].
\end{aligned}
\end{equation}
For the final step, by using the expression for the time-varying effective reproduction number $\Rf(t)$
obtained in \eqref{section8.0012}, one infers that
\begin{equation}\label{section9.8}
r_{t} =\frac{\varLambda_{m}(t)}{\Nm}\Big(\Rf(t)-1\Big)-r_{t}\K_{t}
\end{equation}
with
\begin{equation}\label{section9.9}
\K_{t} 
=\int_{0}^{\infty}\sm(t)\beta_{m}(t,\zeta){\Pn}(t,\zeta)\int_{\max\{0,t^{0}+\zeta-t\}}^{\zeta}\beta_{h}(\eta+t-\zeta,\eta) (\zeta - \eta)  
\big[s+\sigma v\big](\eta+t-\zeta,\eta)  F(t,\eta,\zeta) \,\d\eta\,\d\zeta.
\end{equation}
Equivalently, the relation between $r_{t}$ and $\Rf(t)$ can be rewritten as 
\begin{equation}\label{section9.10}
r_{t} =\frac{\varLambda_{m}(t)}{\Nm\Big(1+\K_{t}\Big)}\Big(\Rf(t)-1\Big).
\end{equation}

A similar relation between the instantaneous growth rate $r_t$ and the time-varying effective reproduction number 
was derived in \cite[Eq.~(3)]{Arroyo2021} for a discrete-time model, namely
\begin{equation}\label{section9.11}
r_{t} =\gamma(t)\Big(\Rf(t)-1\Big).
\end{equation}
For our model, one has $\gamma(t)=\frac{\varLambda_{m}(t)}{\Nm\big(1+\K_{t}\big)}$. 
Equation \eqref{section9.11} describes the relation between the instantaneous growth rate $r_{t}$ 
and the time-varying effective reproduction number $\Rf(t)$. 
In particular, it shows that when $\Rf(t) = 1$, the instantaneous growth rate $r_t$ is $0$, 
meaning that the number of cases is neither increasing nor decreasing at time~$t$. 
If $\Rf(t) > 1$, the instantaneous growth rate $r_t$ is positive, indicating that the incidence is increasing. 
In contrast, if $\Rf(t) < 1$, the instantaneous growth rate $r_t$ is negative, 
implying that the incidence is declining at time~$t$.
If $\Rf(t)=0,$ the number of cases is decreasing, as the disease is no longer spreading. 
Note that \eqref{section9.11} can also be rewritten as
\begin{equation}\label{section9.12}
\Rf(t)=\frac{r_{t}}{\gamma(t)}+1.
\end{equation}
This presentation emphasizes that if $r_t=0$, then $\Rf(t)=1$, which corresponds to a stable incidence.

\part{Applications to Bazil}

In this second part we use the model developed so far for studying the epidemic of dengue in Brazil, 
data from from the first week of 2021 to the last week of 2024 can be downloaded from the website of the Health Ministry of Brazil,
see \cite{Brazil_source}.
The reported cases include confirmed symptomatic cases, hospitalized patients, and deaths among hospitalized patients, 
and these numbers are reported on a weekly basis. 
The data are subdivided into four age groups: child (age $0$ to $19$), adult (age $20$ to $39$), 
middle-aged adult (ages $40$ to $59$), and elderly (ages $60$ and above).
\section{Medical parameters' estimation}\label{sec_med_pre}

In this first section, we estimate some medical parameters of our the model 
using the weekly dengue incidence data from Brazil mentioned above.
Since the data are provided on a weekly basis, it means that our time parameter $t$ is going to denote
weeks. We treat the parameters of interest as unknowns and estimate them using the observed data.
A particle filter technique is implemented for estimating the uncertainties, see \ref{sec_particle} for more details.

\subsection{Death rate}

We use a discrete time approach for firstly estimating the death rate $\muH$.
The weekly death rate among hospitalized dengue cases is calculated by dividing the number of reported deaths 
in a specific age group during the week by the number of hospitalized cases in that same age group during the same week. 
This approach is similar to the one used in \cite{Kim2020} and \cite{Sun2023} to which we refer additional information.

\begin{figure}[htbp]
    \centering
    \includegraphics[scale=0.55]{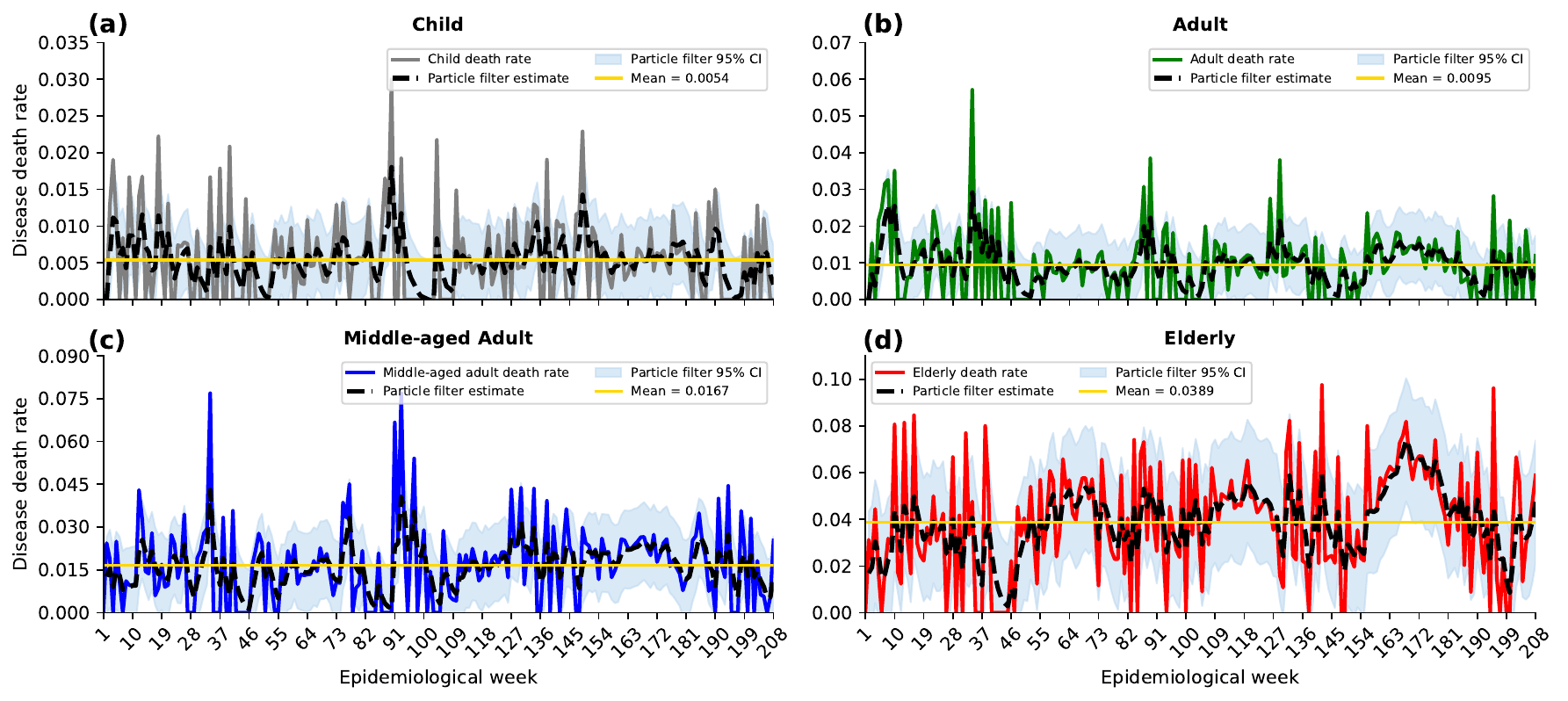}
    \caption[Time-varying death rate due to dengue in Brazil]{\textbf{Time varying death rate 
    due to dengue in Brazil.} 
    The Subfigure present the death rate due to dengue per week for child susceptible (a), for adult susceptible (b), for middle-aged adult susceptible(c), and for elderly susceptible (d) in Brazil from 2021 to 2024.  The solid gold line represents the mean value of the different age groups, while the black dashed line shows the values estimated by the particle filter.  The sky blue colored region indicates the 95\% confidence interval of these estimated values.}
    \label{fig11}
\end{figure}

The weekly death rate caused by dengue is shown in Figure \ref{fig11}. 
The average death rate computed over $4$ years for dengue is $0.0054$ for child, $0.0095$ for adult, $0.0167$ for middle-aged adult,
and $0.0389$ for the elderly. This matches findings from a study performed in Singapore.
The study found that elderly individuals stayed in the hospital longer and were more likely to develop serious symptoms
as compared to other age group, see \cite{Rowe2014} for the details.

\subsection{Recovery rates}

Two recovery rates can be evaluated: the recovery rate $\kappas$ of symptomatic individuals, and the recovery rate $\kappah$
of hospitalized individuals.

\begin{figure}[htbp]
    \centering
    \includegraphics[scale=0.55]{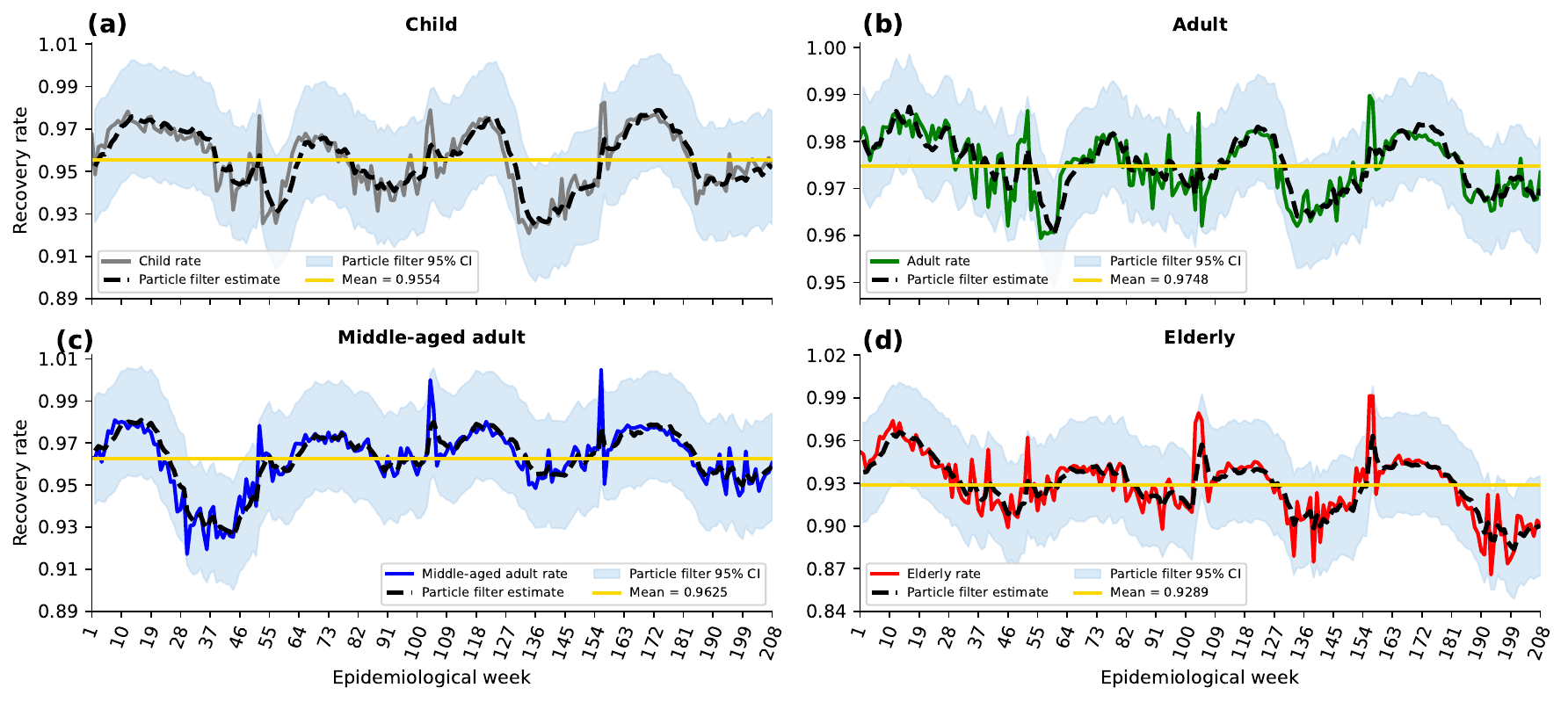}

    \caption[Time-varying recovery rate of symptomatic dengue patients in Brazil]{\small
    \textbf{Time varying recovery rate of symptomatic dengue patients in Brazil}. Recovery rate for symptomatic dengue patients in Brazil from 2021 to 2024. Subfigures (a), (b), (c), and (d) show recovery rate for child, adult, middle-aged adult, and the elderly, respectively.
    Particle filter estimate is marked by dashed black lines, and sky blue colored regions represent the 95\% confidence interval of the estimates.}
    \label{fig_kappas}
\end{figure}

To compute $\kappas(t,\zeta)$ for week $t$ and age $\zeta$, we divide the number of symptomatic recoveries 
in the age group recorded during this week by the total number of symptomatic cases during the same week:
$$
\kappas(t,\zeta)=\frac{\text{symptomatic recoveries}\;\!(t,\zeta)}{\text{total symptomatic cases}\;\!(t,\zeta)}.
$$
The recovery rates for the four age groups are reported in Figure \ref{fig_kappas}, and the average on the period of 
four years is also computed.
We observe that on average the highest recovery rates are for adult and middle-aged adult, followed by the rate for child and then for the
elderly. Elderly people have the lowest recovery rate since these individuals tend to have a higher death rates and a longer recovery time.

\begin{figure}[htbp]
    \centering
    \includegraphics[scale=0.55]{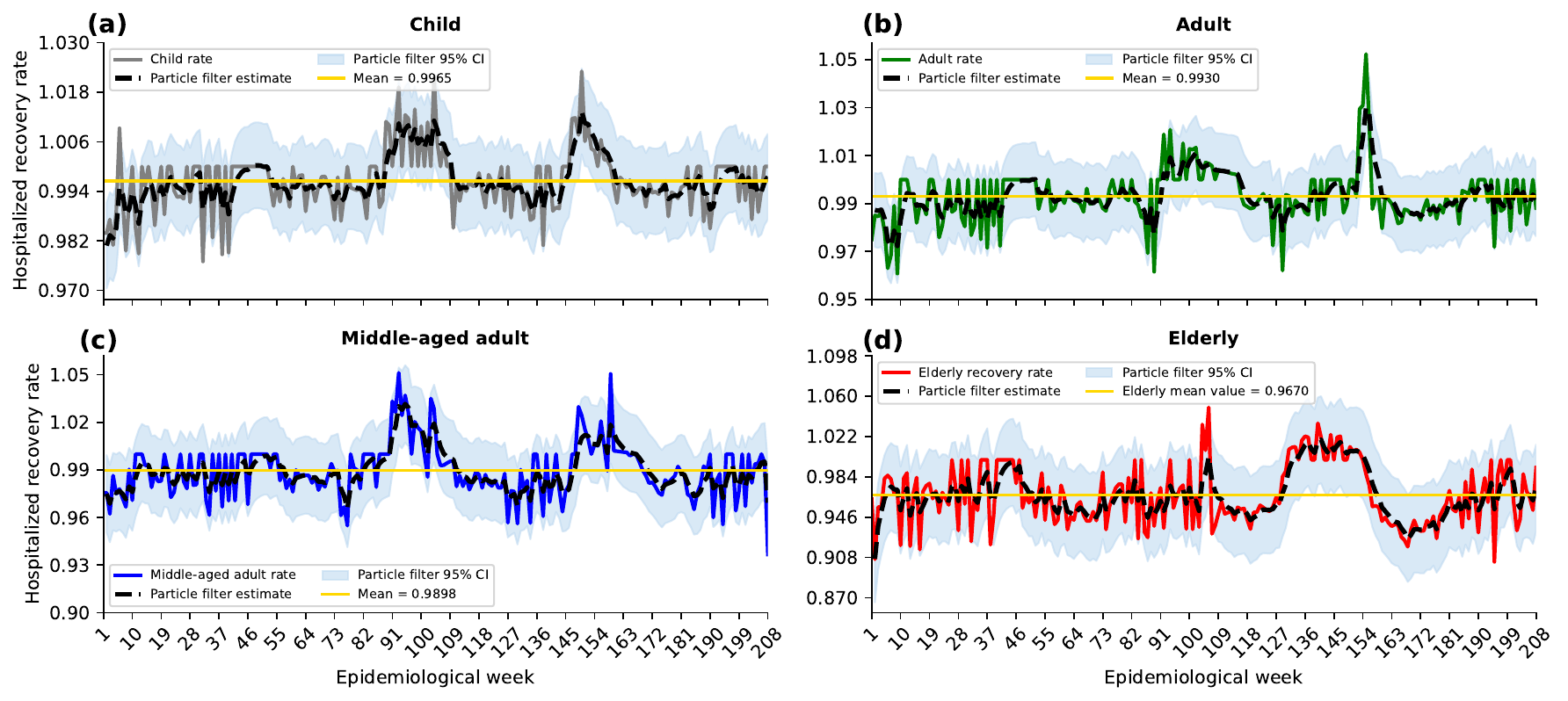}
    \caption[Time-varying recovery rate for hospitalized dengue patients in Brazil]{\small
    \textbf{Time-varying recovery rate for hospitalized dengue patients in Brazil}. Weekly estimation of the recovery rate for hospitalized dengue patients in Brazil from week 1 of 2021 to the last week of 2024. Subfigures (a), (b), (c), and (d) show recovery rates for child, adult, middle-aged adult, and the elderly, respectively.
    Particle filter estimate is marked by dashed black lines, and sky blue colored regions represent the 95\% confidence interval of the estimates.}
    \label{fig13}
\end{figure}

Similarly, the estimate of the recovery rate $\kappah$ of hospitalized individuals using actual data from Brazil for four age groups
is represented in Figure \ref{fig13}. The mean recovery rate over the four years is also represented for each age group. 
On average, the highest recovery rate from hospital is for the child group, while the lowest one is again for the elderly. 
Clearly, these individuals face a high hospitalization rates and lower recovery rates compared to the other age groups due to underlying health conditions and age-related physiological changes.  

\subsection{Hospitalization rate}

\begin{figure}[htbp]
    \centering
    \includegraphics[scale=0.55]{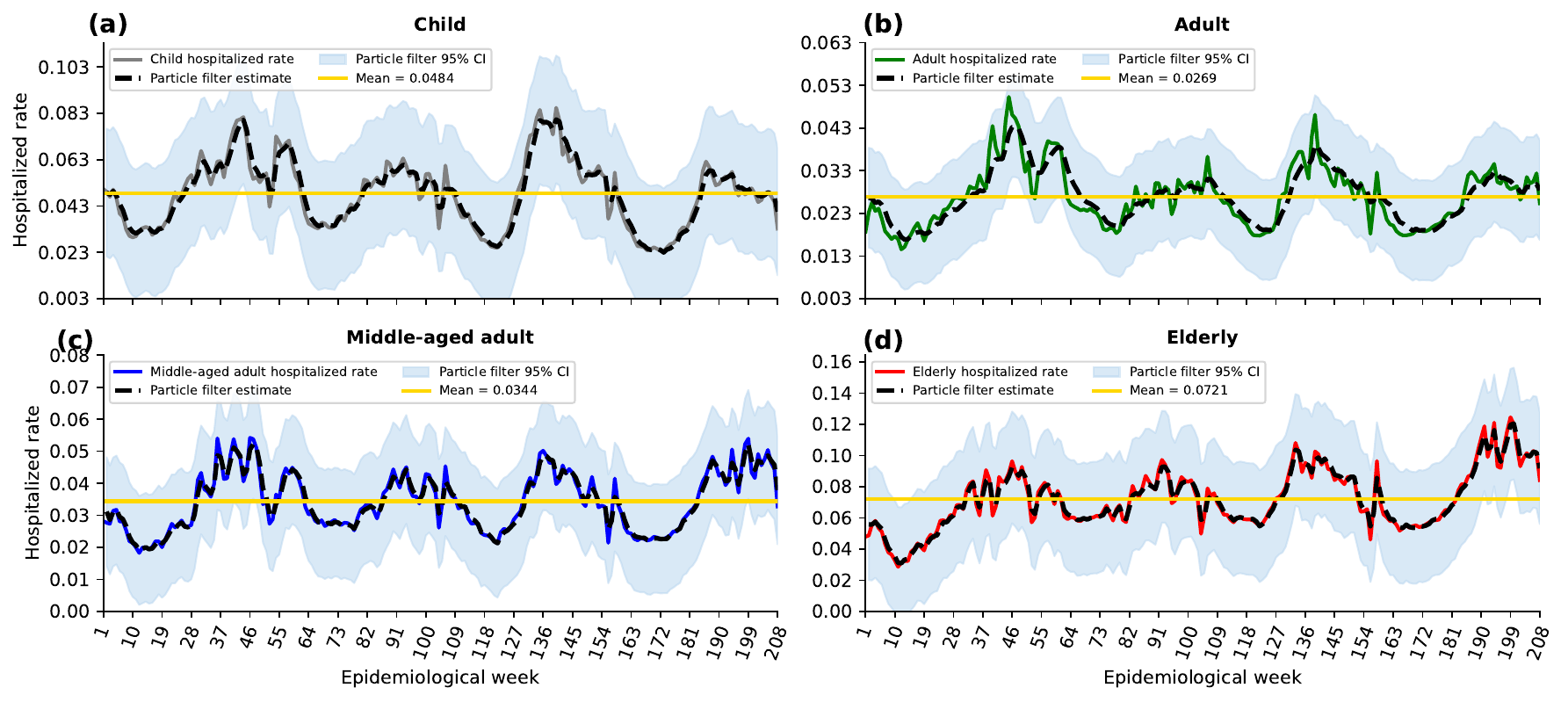}
    \caption[Time-varying hospitalized rate of dengue patients in Brazil]{\small
    \textbf{Time varying hospitalized rate of dengue patients in Brazil}. Hospitalized rate for dengue patients in Brazil from 2021 to 2024. Subfigures (a), (b), (c), and (d) show recovery rates for child, adult, middle-aged adult, and the elderly, respectively.
    Particle filter estimate is marked by dashed black lines, and sky blue colored regions represent the 95\% confidence interval of the estimates.}
    \label{fig12}
\end{figure}

To estimate the hospitalization rate $\delta$ of symptomatic individuals for week $t$, 
we divide the number of hospitalized cases in age group $\zeta$ recorded that week by the total number of infected cases 
in the same age group that week:
$$
\delta(t,\zeta)=\frac{\text{hospitalized}\;\!(t,\zeta)}{\text{infected}\;\!(t,\zeta)}.
$$
This time-varying parameter is reported in Figure \ref{fig12} and the mean value for the period of four years is also provided.
On average, the elderly have clearly the highest hospitalization rate while the adult group has the lowest one.

In Table \ref{table2} we provide the mean value obtained for the medical parameters discussed above and for each age group.
Since no information for the recovery rate of asymptomatic individuals is available, we shall simply use the value obtained
for the symptomatic individuals of the same age group.
Additional medical parameters provided by the literature are also indicated, with the corresponding reference.
Note that the parameter $\alpha$ has been evaluated based on reference \cite{Xue2021}.
The mean values of these parameters will be used for the computations in the subsequent sections.

\begin{table}[htbp]
\centering
\caption{Parameter values, ranges, and sources used in the proposed model.}
\scalebox{0.74}{
\begin{tabular}{lllll}
\hline
\hline
\textbf{Parameter} & \textbf{Value} & \textbf{Range} & \textbf{Units} & \textbf{Source} \\ 
\hline 
\hline
$\beta_{h}(t,\zeta)$ & 0.44 & - & - & \cite{Churcher2015} \\ 
$\mu_H(t,\zeta)$ (Child) & 0.0054 & 0–0.057 & week$^{-1}$ & Estimated \\ 
$\mu_H(t,\zeta)$ (Adult) & 0.0095 & 0–0.057 & week$^{-1}$ & Estimated \\ 
$\mu_H(t,\zeta)$ (Middle-Aged Adult) & 0.0167 & 0–0.0769 & week$^{-1}$ & Estimated \\ 
$\mu_H(t,\zeta)$ (Elderly) & 0.0389 & 0–0.0976 & week$^{-1}$ & Estimated \\ 
$\sigma(t,\zeta)$ & 0.6 & 0–1 & - & \cite{Xue2021}   \\
$\kappa_a(t,\zeta)$ & $\kappa_s(t,\zeta)$ & - & week$^{-1}$ & Estimated \\ 
$\kappa_s(t,\zeta)$ (Child) & 0.9554 & 0.9245 - 0.9813 & week$^{-1}$ & Estimated \\ 
$\kappa_s(t,\zeta)$ (Adult) & 0.9748 & 0.9652-0.9857 & week$^{-1}$ & Estimated \\ 
$\kappa_s(t,\zeta)$ (Middle-Aged Adult) & 0.9825 & 0.9172-1.0 & week$^{-1}$ & Estimated \\ 
$\kappa_s(t,\zeta)$ (Elderly) & 0.9289 & 0.8989-0.96877 & week$^{-1}$ & Estimated \\ 
$\varLambda_{m}(t)$ & 0.2  & 0.15–0.2 & week$^{-1}$ & \cite{Yi2021} \\ 
$\kappa_H(t,\zeta)$ (Child) & 0.9965 & 0.9795-1.0125 & week$^{-1}$ & Estimated \\ 
$\kappa_H(t,\zeta)$ (Adult) & 0.9930 & 0.9684-1.0125 & week$^{-1}$ & Estimated \\ 
$\kappa_H(t,\zeta)$ (Middle-Aged Adult) & 0.9898 & 0.9561-1.0156 & week$^{-1}$ & Estimated \\ 
$\kappa_H(t,\zeta)$ (Elderly) & 0.9670 & 0.9038-1.0099 & week$^{-1}$ & Estimated \\ 
$\alpha(t,\zeta)$ & 0.0121 & 0–1 & week$^{-1}$ & Evaluated \\ 
$\delta(t,\zeta)$ (Child) & 0.0484 & 0.0237-0.0854 & week$^{-1}$ & Estimated \\ 
$\delta(t,\zeta)$ (Adult) & 0.0269 & 0.0167-0.0542 & week$^{-1}$ & Estimated \\ 
$\delta(t,\zeta)$ (Middle-Aged Adult) & 0.0344 & 0.0097-0.0609 & week$^{-1}$ & Estimated \\ 
$\delta(t,\zeta)$ (Elderly) & 0.0721 & 0.0275-0.0999 & week$^{-1}$ & Estimated \\ 
$f(t,\zeta)$ & 0.4074 & 0–1 & - & \cite{Asish2023} \\ 
$\mu_m$ & 0.5  & - & week$^{-1}$ & \cite{Yi2021} \\ 
\hline
\end{tabular}
}
\label{table2}
\end{table}

\section{Transmission rates}\label{sec_transmission_r}

We now move to a more important and more involved parameter's determination: the transmission rates.
The most important parameters in the model are the transmission rates: the mosquito-to-human transmission rate $\beta_{h}(t)$
and the human-to-mosquito transmission rate $\beta_{m}(t)$. These rates govern the spread of the disease between mosquitoes and humans. 
Estimating these parameters is very difficult, and getting them with their dependence to the age variable $\zeta$
seems out of reach. Nevertheless, for our investigations we shall use and compare different values for these parameters
obtained by different approaches. These approaches will be explained subsequently, but for clarity let us already 
list the different notations we shall use: 
\begin{enumerate}
\item[i)] $\beta_h$ indicated in Table \ref{table2} and borrowed from the literature,
\item[ii)] $\beta^{D}_{m}(t)$ estimated from the data,
\item[iii)] $\beta^{C}_{m}(t)$ and $\beta^{C}_{h}(t)$  estimated from the temperature and the humidity (the index $C$ stands for climate).
\end{enumerate}

Let us directly stress that the choice of the method used for getting the transmission rates will 
directly impact the estimation of the time-varying effective reproduction number.
Indeed, two main approaches will be used for determining this essential parameter: the data-based
approach and the model-based approach. For the latter, 
the transmission rates are involved in the computation.
Therefore, the outcome will depend on the method used for getting the transmission rate, 
and accordingly various notations for the time-varying effective reproduction number have to be introduced. 
In the sequel we shall use:
\begin{enumerate}
\item[i)] $\RD(t)$, for $\Rf(t)$ based on data only, 
\item[ii)] $\RMC(t)$, for $\Rf(t)$  based on the model and using $\beta^{C}_{m}(t)$ and $\beta^{C}_{h}(t)$,
\item[iii)] $\RMD(t)$, for $\Rf(t)$  based on the model and using $\beta^{D}_{m}(t)$ and $\beta_{h}$,
\item[iv)] $\RMDC(t)$, for $\Rf(t)$  based on the model and using $\beta^{D}_{m}(t)$ and $\beta^{C}_{h}(t)$.
\end{enumerate}
Once again, additional information will be provided subsequently. 

\subsection{Data-based transmission rate}\label{sec_trans_rate_data}

Firstly, to estimate the human–to–mosquito transmission rate $\beta^{D}_m(t)$ from observed data, 
we shall assume that this parameter is age-independent. We can then use its relations to the 
time-varying effective reproduction number provided in \eqref{section8.0012}. Thus, the transmission rate can be written as:
\begin{equation}\label{section10.1}
\beta^{D}_{m}(t)=\frac{\RD(t)\varLambda_{m}(t)}{\Nm\sm(t)\Big(\Psi_{\scalebox{0.4}{\text{Child}}}(t)+\Psi_{\scalebox{0.4}{\text{Adult}}}(t)+\Psi_{\scalebox{0.4}{\text{Middle-Aged Adult}}}(t)+\Psi_{\scalebox{0.4}{\text{Elderly}}}(t)\Big)},
\end{equation}
where 
\begin{align*}
\Psi_{\scalebox{0.4}{\text{Child}}}(t) 
& =\int_{0}^{20}{\Pn}(t,\zeta)\int_{\max\{0,t^{0}+\zeta-t\}}^{\zeta}\beta_{h}\,
\big[s+\sigma v\big](\eta+t-\zeta,\eta)  F(t,\eta,\zeta)\,\d\eta\,\d\zeta, \\
\Psi_{\scalebox{0.4}{\text{Adult}}}(t) 
& = \int_{20}^{40}{\Pn}(t,\zeta)\int_{\max\{0,t^{0}+\zeta-t\}}^{\zeta}\beta_{h}\, \big[s+\sigma v\big](\eta+t-\zeta,\eta)  F(t,\eta,\zeta)\,\d\eta\,\d\zeta,  \\
\Psi_{\scalebox{0.4}{\text{Middle-Aged Adult}}}(t)
& =\int_{40}^{60}{\Pn}(t,\zeta)\int_{\max\{0,t^{0}+\zeta-t\}}^{\zeta}\beta_{h}\, \big[s+\sigma v\big](\eta+t-\zeta,\eta)  F(t,\eta,\zeta)\,\d\eta\,\d\zeta, \\ 
\Psi_{\scalebox{0.4}{\text{Elderly}}}(t) 
& =\int_{60}^{90}{\Pn}(t,\zeta)\int_{\max\{0,t^{0}+\zeta-t\}}^{\zeta}\beta_{h}\, \big[s+\sigma v\big](\eta+t-\zeta,\eta)  F(t,\eta,\zeta)\,\d\eta\,\d\zeta. 
\end{align*}
The different age groups have been separated since the parameters involved in the integrands are different for each of them.

\begin{figure}[H]
    \centering
    \includegraphics[scale=0.5]{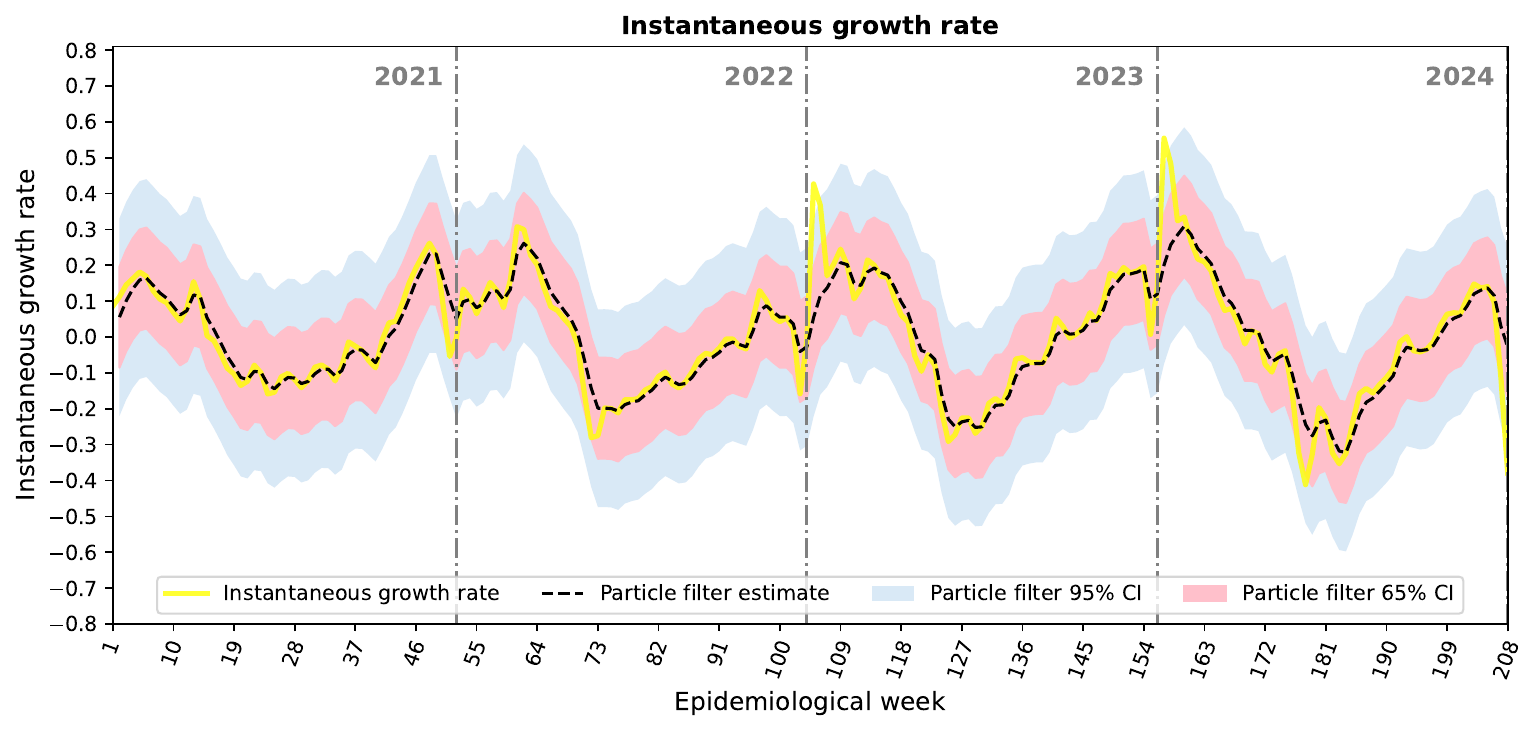}
    \caption[Instantaneous growth rate of dengue in Brazil]{\small
    \textbf{Instantaneous growth rate of dengue in Brazil}. Instantaneous growth rate $r_t$ of dengue in Brazil from week~1 of 2021 to the subsequent 208 weeks. The yellow line represents the estimated values of $r_t$, while the dashed black line shows the estimates obtained from the particle filter. The 95\% confidence interval is shown as a sky blue colored region, and the 65\% confidence interval as a pink colored region.}
    \label{fig5}
\end{figure}

Our aim now is to estimate the r.h.s.~of \eqref{section10.1} based on the data.
For the time-varying effective reproduction number on the numerator, we shall use \eqref{section9.12}.
From the weekly data given in Figure \ref{fig1} we firstly estimate the instantaneous growth rate $r_t$ by using
the symmetric discrete differentiation
\begin{equation}\label{growth_rate_plot}
r_t = \frac{ \log(I_{t+1}) -  \log(I_{t-1})}{2}, \qquad \hbox{ for } t = 2,\dots,T-1,
\end{equation} 
where $I_t$ denotes the total number of infected cases at time $t$.
At the endpoints, we use the one-sided differences given by 
$$
r_1 =\log(I_2) - \log(I_1),
\qquad
r_T = \log(I_T) - \log(I_{T-1}).
$$
Based on these formulas, the instantaneous growth rate $r_{t}$ is reported in Figure \ref{fig5}.

\begin{remark}\label{rem_outliers}
It is visible in Figure \ref{fig5} and in all subsequent figures deduced from it that
an artificial phenomenon appears on weeks $53$, $105$, and $157$. By looking more carefully
at the data, it clearly appears that the numbers of infected individuals are not properly 
recorded on these specific weeks (corresponding to changes of year). We decided
to keep the data provided untouched, but to stress that any value computed on these specific weeks has to
be taken with a grain of salt.
\end{remark}

\begin{figure}[htbp]
    \centering
    \includegraphics[scale=0.5]{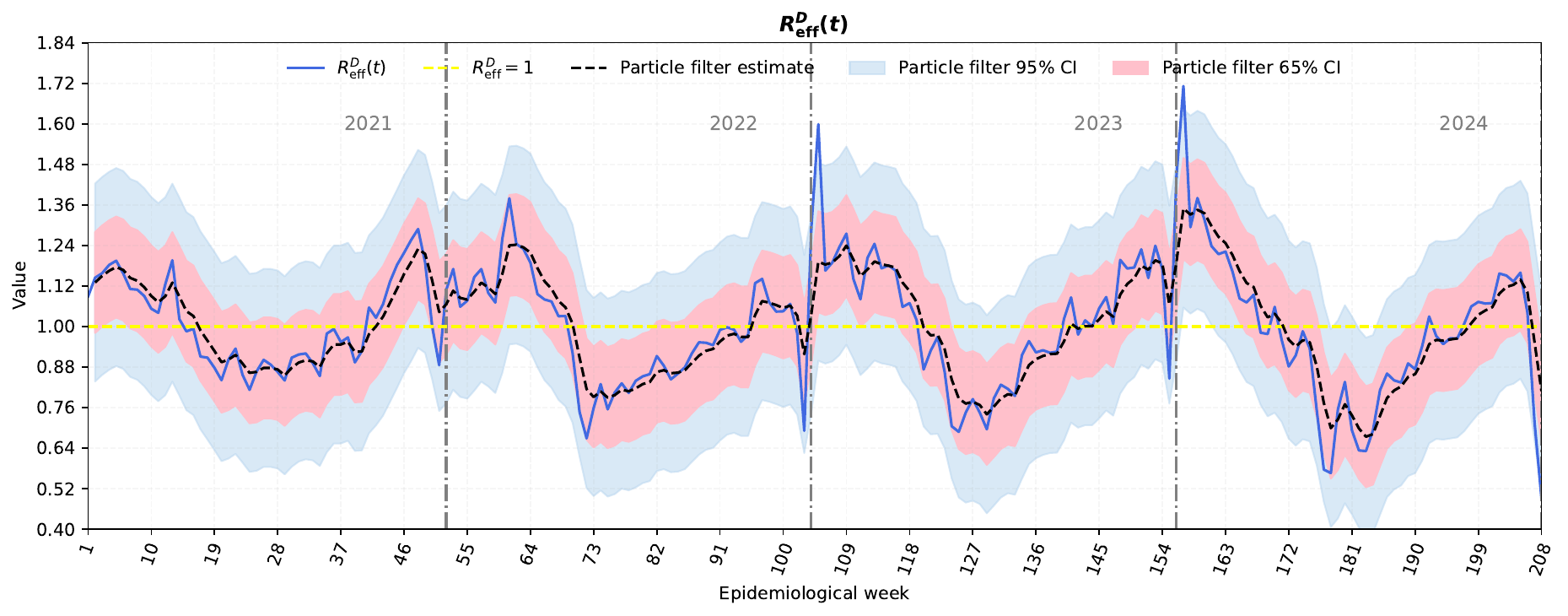}
    \caption[Time-varying effective reproduction number $\RD(t)$ for dengue in Brazil]{\small
    \textbf{$\boldsymbol{\RD(t)}$ for dengue in Brazil}. $\RD(t)$
    for dengue in Brazil from week 1 of 2021 to the last week of 2024 estimated on real data.  The sky blue and pink colored regions representing the 95\% and 90\% confidence intervals of the  estimated values. The gold solid line indicates the threshold value~$1$.}
    \label{fig6}
\end{figure}

In order to infer $\RD(t)$ from Equation~\eqref{section9.12}, we should evaluate the factor $\gamma(t)$ which appears 
in the relation.
However, since we have no access to this parameter, we shall use a fixed $\gamma$ corresponding to the recovery/removal rate, 
by assuming that the mean infectious period is approximately constant. This approach is based on the substitution approach 
of \cite[Eq.~(3)]{Arroyo2021}. 
Based on the computation of the previous section, the mean recovery rate across all ages for infected individuals 
is given by $\gamma = 0.9604$.
The resulting value for $\RD(t)$ is reported in Figure \ref{fig6}.

\begin{figure}[htbp]
    \centering
    \includegraphics[scale=0.5]{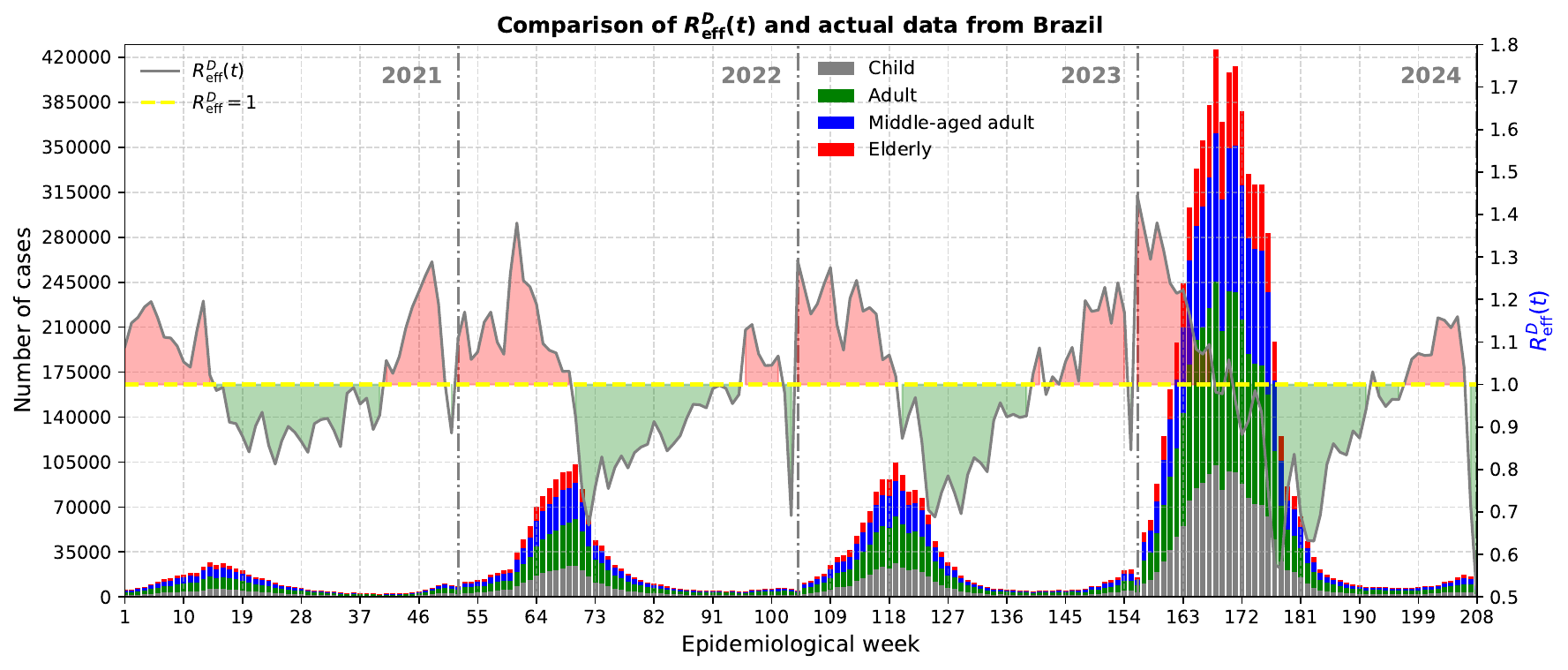}
    \caption[Time-varying effective reproduction number $\RD(t)$ with data from Brazil]{\small
    \textbf{$\boldsymbol{\RD(t)}$ and actual data from Brazil}. The figure compares $\RD(t)$ and time-series data from Brazil from 2021 to 2024. In the $\RD(t)$ curve the green region represent $\RD(t)<1$, the red region indicates $\RD(t)>1$ and the gold  line shows when $\RD(t)=1$.
}
    \label{fig7}
\end{figure}

For convenience, we provide in Figure \ref{fig7} simultaneously the information about the symptomatic infected individuals and the $\RD(t)$ computed from this data. The phenomenon explained in Remark \ref{rem_outliers}
which takes place on weeks $53$, $105$, and $157$ is clearly visible on these figures.

\begin{figure}[htbp]
    \centering
    \includegraphics[scale=0.55]{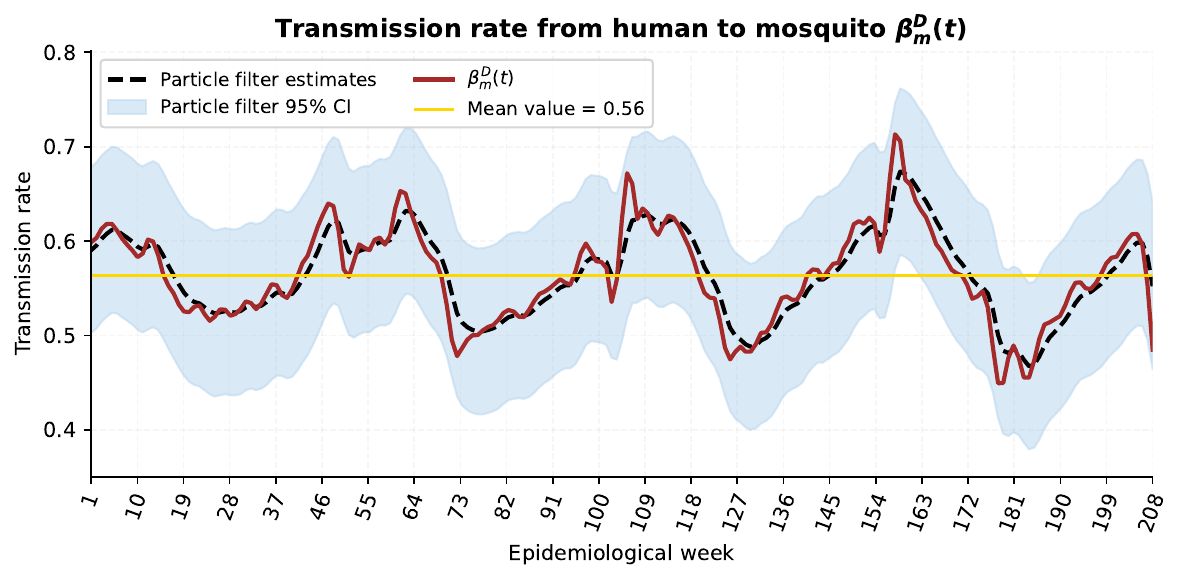}
    \caption[Transmission rate from humans to mosquitoes]{\small
    \textbf{Transmission rate from humans to mosquitoes}. Estimates are based on time-series data using the relation in \eqref{section10.1}. The brown curve represents the estimated values, the dashed black line is the particle filter estimate with the 95\% confidence interval shown in sky blue.}
\label{fig8}
\end{figure}

The final step for evaluating the transmission rate $\beta^{D}_m(t)$ is to compute the denominator on the r.h.s.~of \eqref{section10.1}.
This can be performed with a lengthy computation which takes into account the data and all medical parameters of Table \ref{table2}.
As a final result, we get the value of $\beta^{D}_m(t)$ reported in Figure \ref{fig8}.
This figure shows the fluctuations in the transmission rate due to changes in the mosquito biting rate. 
During the high-incidence period, as for example around week $160$, the rate reached to peaked at nearly $0.7$.
The mean value of $\beta^{D}_m(t)$ is also provided for the period of four years.

\subsection{Temperature and humidity based transmission rates}\label{sec_thbased}
\begin{figure}[htbp]
    \centering
    \includegraphics[scale=0.55]{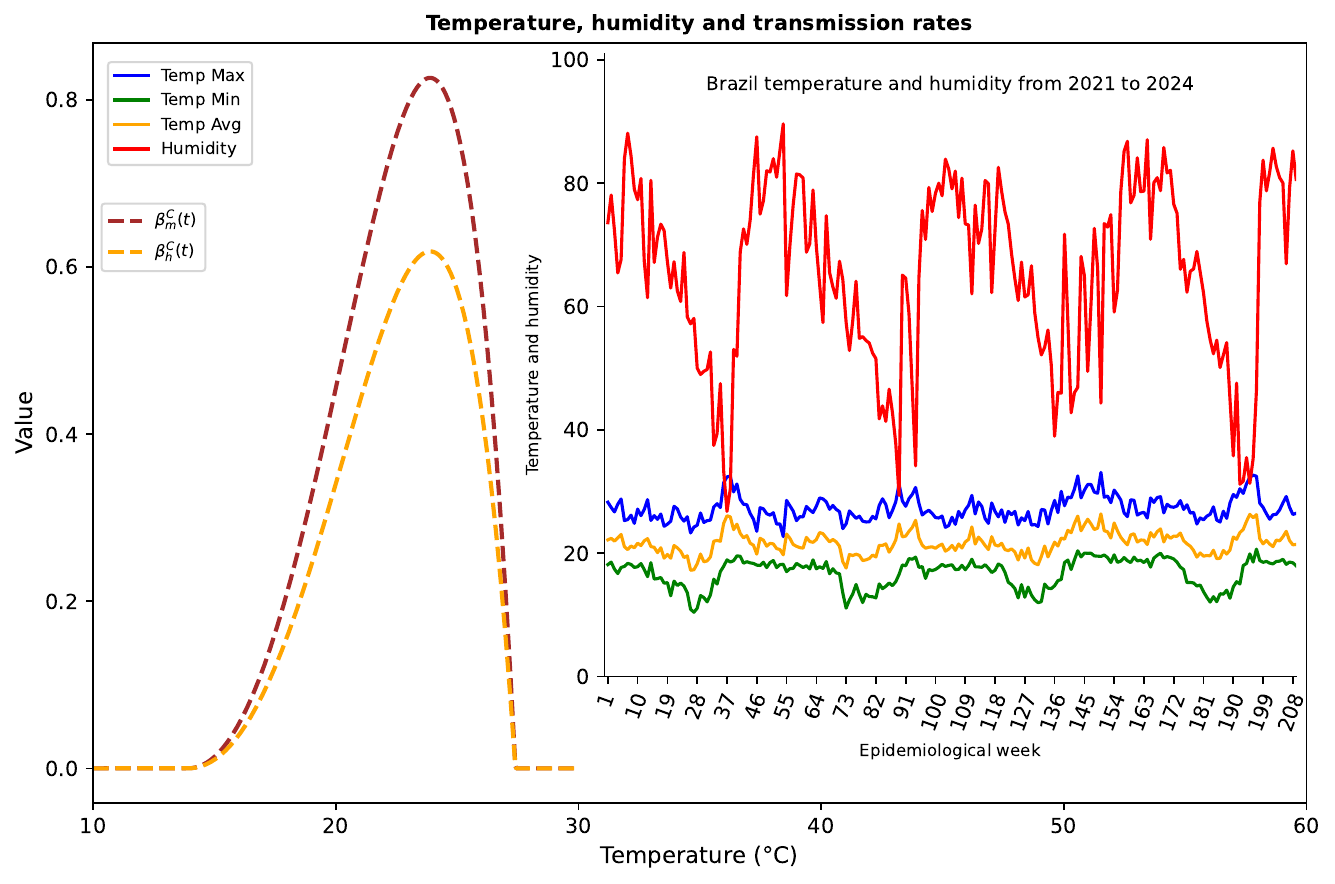}
    \caption[Behavior of the parameters $\beta^{C}_{m}$ and $\beta^{C}_{h}$ / Temperature and humidity between 2021 and 2024]{\small
    \textbf{Behavior of the parameters}. 
    Behavior of the parameters $\beta^{C}_{m}(t)$ and $\beta^{C}_{h}(t)$ based on relations \eqref{section10.6} and  \eqref{section10.7} with temperature $T$ ranging from $10^\circ \text{C}$ to $30^\circ \text{C}$. The factor
    $E(H)$ is assumed to be equal to $1$.The subplot shows the weekly temperature and humidity from Brazil between 2021 and 2024.}
    \label{fig9}
\end{figure}

In this subsection, we estimate the transmissions rates $\beta^{C}_m(t)$ and $\beta^{C}_h(t)$ by using weekly temperature and humidity data from Brazil. We recall that the index $C$ stands for climate.
This is possible since dengue is transmitted through the bite of adult female mosquitoes and their population is significantly 
influenced by climate conditions such as precipitation, humidity, and temperature. Modeling mosquito populations would require 
regional environmental data to estimate the local abundance of female mosquitoes, and such data are not easily accessible. 
To capture the effects of temperature and humidity, we employ another method based on earlier researches. Indeed, in \cite{Kim2020} it is 
directly shown that changes in temperature and humidity are closely related to an increased incidence of dengue cases, since
these parameters are crucial for the development and survival of Aedes mosquitoes. 
Maximum temperature, minimum temperature, and average temperature are shown on the right top corner in Figure \ref{fig9}, together with
the average humidity. These data are obtained from \cite{Brazil_weather}.

Let us now introduce the temperature and humidity based the mosquito-to-human transmission rate $\beta^{C}_{h}(t)$
and the human-to-mosquito transmission rate $\beta^{C}_{m}(t)$. 
Thus, the new parameters are defined by 
\begin{equation}\label{section10.6}
\beta^{C}_m(T,H) := b(T)P_{m}(T)E(H),
\end{equation}
and
\begin{equation}\label{section10.7}
\beta^{C}_h(T,H) := b(T)P_h(T)E(H),
\end{equation}
where $b(T)$ represents the mosquitoes time dependent biting rate (the number of bites per mosquito at temperature $T$), 
$P_{m}(T)$ is the probability of transmission from an infected human 
to a susceptible mosquito during a single bite,
and $P_{h}(T)$ is probability of transmission from an infected mosquito 
to a susceptible human during a single bite.
The factors $b$, $P_m$, and $P_h$ depend on the temperature, as indicated below. 
On the other hand, the new factor $E(H)$ is a correction factor which takes the 
effect of humidity $H$ on dengue transmission into account.
Subsequently, these parameters will depend on the time parameter $t$ since $T$ and $H$
are functions of time.

Following \cite{Kim2020} and based on the earlier paper \cite{Mordecai2017}, the biting rates
and the probability of transmission are described by a Bri\`ere function of the form
$$
T\mapsto 
\begin{cases} 
\mathrm{c} T (T - T_1)\sqrt{T_2 - T} & \text{ if } T_1 \leq T \leq T_2, \\
0 & \text{ otherwise.}
\end{cases}
$$
with $\mathrm{c}$ a numerical factor, and with $T_1, T_2$ the minimal and the maximal empirical value
for the phenomenon to be described, see also \cite{Jin2022} for a generalization of this function.
Based on this idea and on the data from Brazil, 
the empirical biting rate $b$ of mosquitoes depending on the temperature is given by the relation
\begin{equation}\label{eq_b1}
b\big(T(t)\big) = 
\begin{cases} 
{\mathrm{c}_1}T(t) \Big( T(t) - 13.9 \Big)\sqrt{ 27.4 - T(t) }, & \text{if } 13.9 \leq T(t) \leq 27.4, \\
0, & \text{otherwise.}
\end{cases}
\end{equation}
where $T(t)$ is the average weekly temperature in degrees Celsius $^\circ \text{C}$.
Similarly, the probability $P_m$ of transmission from infected humans to susceptible mosquitoes is given by
\begin{equation}\label{eq_b2}
P_{m}\big(T(t)\big) = 
\begin{cases} 
{\mathrm{c}_2}T(t) \Big( T(t) - 13.9 \Big)\sqrt{ 27.4 - T(t) }, & \text{if } 13.9 \leq T(t) \leq 27.4, \\
0, & \text{otherwise,}
\end{cases}
\end{equation}
while the probability $P_h$ of transmission from infected mosquitoes to susceptible hosts is given by
\begin{equation}\label{eq_b3}
P_h\big(T(t)\big) = 
\begin{cases} 
{\mathrm{c}_3}T(t) \Big( T(t) - 13.9 \Big)\sqrt{ 27.4 - T(t) }, & \text{if } 13.9 \leq T(t) \leq 27.4, \\
0, & \text{otherwise.}
\end{cases}
\end{equation}
In these three relations, the parameters ${\mathrm{c}_1}$, ${\mathrm{c}_2}$, and ${\mathrm{c}_3}$ are scaling parameters 
which have been tuned.
More precisely, we used the values ${\mathrm{c}_1}=0.000202$, ${\mathrm{c}_2}=0.0204609$ and ${\mathrm{c}_3}=0.0153192$
by fitting  the model with the four years of dengue cases in Brazil.

\begin{figure}[htbp]
    \centering
    \includegraphics[scale=0.65]{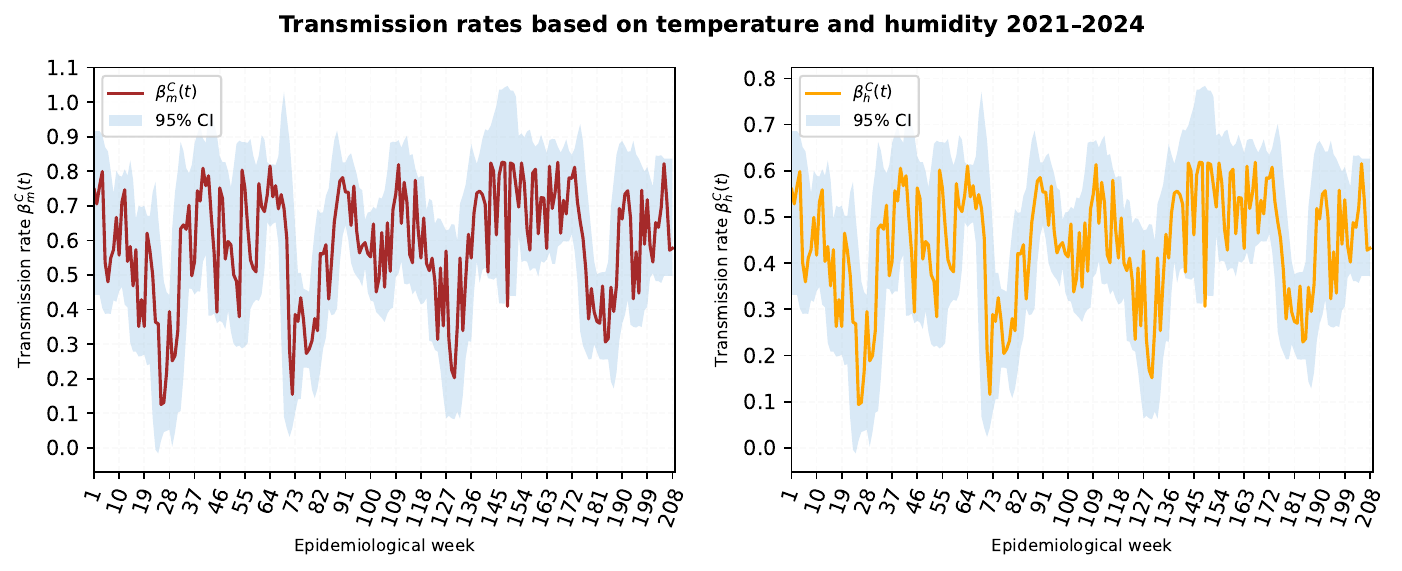}
    \caption[Transmission rates from mosquitoes to humans and from humans to mosquitoes]{\small
\textbf{Transmission rates $\boldsymbol{\beta^{C}_{m}(t)}$ and $\boldsymbol{\beta^{C}_{h}(t)}$.}
The left subplot shows the transmission rate from mosquitoes to humans $\beta^{C}_{m}(t)$ in brown color as defined by \eqref{section10.6}. The right subplot shows the transmission rate from humans to mosquitoes $\beta^{C}_{h}(t)$ in orange color, as defined by \eqref{section10.7}. The 95\% confidence interval is shown in sky blue.}
    \label{fig10}
\end{figure}

Finally, the factor $E$ depending on humidity and introduced in relations \eqref{section10.6} and \eqref{section10.7}
is given by 
\begin{equation*}
E(H(t)) = 
\begin{cases} 
1.0, & \text{if } 60 \leq H(t) \leq 80, \\
0.90, & \text{otherwise,}
\end{cases}
\end{equation*}
where $H(t)$ is the average weekly humidity.
This additional empirical factor represents the effect of humidity on dengue transmission. The resulting formulas
for $\beta^{C}_{m}(t)$ and $\beta^{C}_{h}(t)$ extend the previous ones by incorporating humidity 
in order to obtain more accurate estimates of the transmission rates for dengue.
By using these time series data, Figure~\ref{fig10} presents the corresponding estimated empirical transmission rates for Brazil
for the period 2021--2024.

\section{Numerical simulations 2021--2024}\label{sec_21_24}

In this section, we simulate the propagation of dengue in Brazil over the period 2021--2024. This will allow us to assess
our model and to discuss the role of some parameters. Let us emphasize that these simulations are not completely independent
from the data since some medical parameters have been estimated based on them. Test simulations independent of the data will be
performed in the next section. The simulations are based on the time-dependent model \eqref{section1.13} that we integrate
over the $4$ years.

\subsection{Transmission rates and estimation of compartments}\label{subsec_trans_r}

In this subsection, we want to compare the two transmission rates introduced and discussed in Section \ref{sec_transmission_r}:
the data-based transmission rate as shown in Figure \ref{fig8}, and the temperature-humidity-dependent transmission rate as 
represented in Figure \ref{fig10}.
This analysis can be performed by simulating the propagation of dengue in Brazil over the four years and by comparing some compartments  with the real data.
For that purpose, we also use the mean values of the medical parameters, as provided in Table \ref{table2}.
The initial conditions (corresponding to the $1^{\rm st}$ week of 2021) for the different age groups are provided in Table
\ref{table:initial_conditions} and have been estimated on the real data of this period.

\begin{table}[htbp]
\centering
\caption{Initial conditions for different age groups.}
\scalebox{0.6}{
\begin{tabular}{lcccccccc}
\hline
\hline
\textbf{Age Group} & \textbf{Susceptible} & \textbf{Vaccinated} & \textbf{Asymptomatic } & \textbf{Symptomatic} & \textbf{Hospitalized} & \textbf{Recovered} & \textbf{ Susceptible Mosquitoes} & \textbf{Infected Mosquitoes} \\ \hline \hline
\textbf{Child}              & 0.69093   & 0.28  & 0.0000608    & 0.0000251  & 0.0000013  & 0.0289828  & 0.99912  & 0.00088  \\ 
\textbf{Adults}             & 0.7199153 & 0.28  & 0.00006072   & 0.0000227  & 0.0000013  & 0.00       & 0.99912  & 0.00088  \\ 
\textbf{Middle-Aged Adults} & 0.7199153 & 0.28  & 0.00006072   & 0.0000227  & 0.0000013  & 0.00       & 0.99912  & 0.00088  \\ 
\textbf{Elderly}            & 0.69193   & 0.28  & 0.00006021   & 0.0000238  & 0.0000013  & 0.02798469  & 0.99912  & 0.00088  \\ \hline
\end{tabular}
}
\label{table:initial_conditions}
\end{table}

\begin{figure}[htbp]
    \centering
    \includegraphics[scale=0.55]{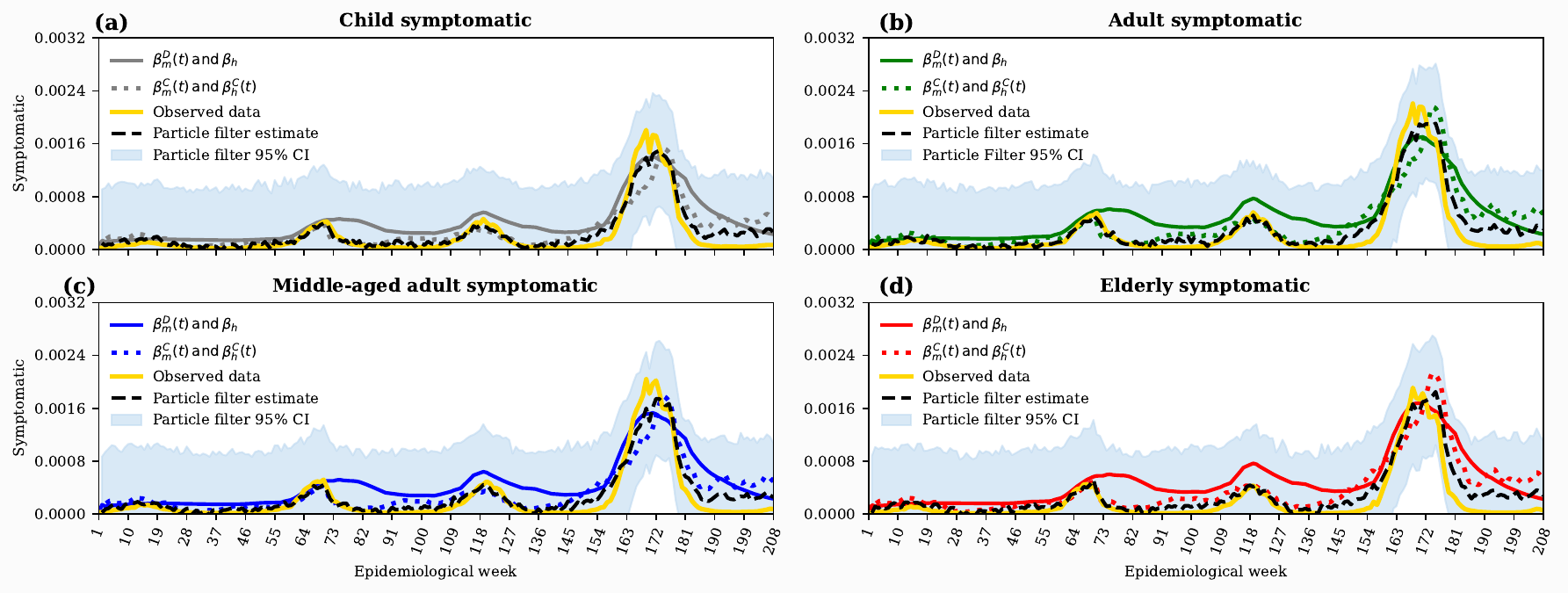}
    \caption[Dynamics of symptomatic]{\small
    \textbf{Dynamics of symptomatic}. Dynamics of normalized symptomatic incidence among child, adult, middle-aged adult and the elderly, as represented by model \eqref{section1.13} under varying transmission rates. The solid gold line shows actual data from Brazil (2021–2024). The dashed black line shows the particle filter estimate with the 95\% confidence interval shaded in sky blue.}
    \label{fig17}
\end{figure}

\begin{figure}[htbp]
    \centering
    \includegraphics[scale=0.55]{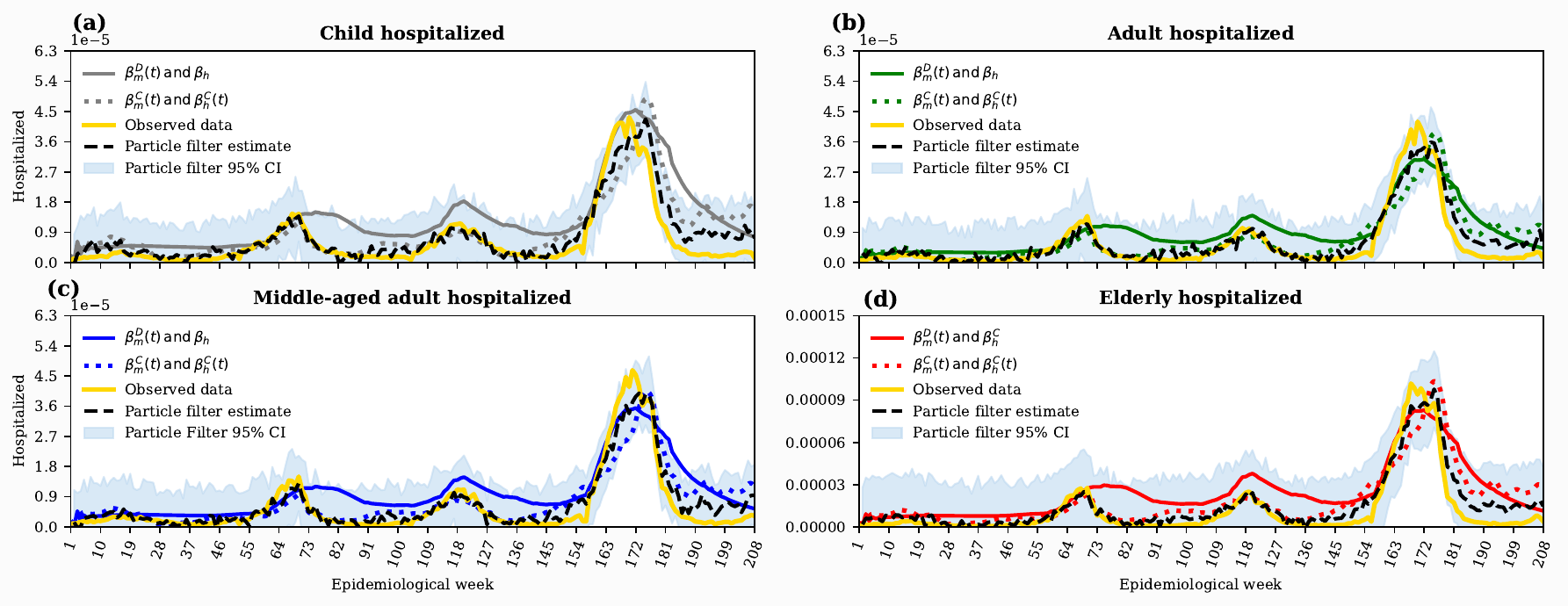}
    \caption[Dynamics of hospitalized]{\small
    \textbf{Dynamics of hospitalized}. Dynamics of normalized hospitalized incidence among child, adult, middle-aged adult and the elderly, as represented by model \eqref{section1.13} under varying transmission rates. The solid gold line shows actual data from Brazil (2021–2024). The dashed black line shows the particle filter estimate with the 95\% confidence interval shaded in sky blue.}
    \label{fig18}
\end{figure}

We firstly provide a comparison of the symptomatic individuals, since data about this compartment are available. The four age
groups are studied separately, and the outcome is shown in Figure \ref{fig17}. Similarly, for hospitalized individuals a 
comparison can be provided between the simulations obtained with the two transmission rates and the real data, 
see Figure \ref{fig18}.  We can observe that the model fed with both transmission rate fit well with the reported dengue cases.
In these two figures, the yellow line represents the observed data, while the dotted lines represent the model estimation 
using the temperature-humidity dependent transmission rates. The gray line for child, green for adults, bright blue line for middle-aged adults, and red line for the elderly, represent the model estimations based on data based transmission rates. 
The black dashed line represents the particle filter estimation with a 95\% confidence interval.
We can observe that our model fits well with the observed symptomatic and hospitalized dengue case data from Brazil by capturing the seasonality and the growth trend of dengue.
In particular, the model correctly describes the outbreaks by using temperature-humidity based transmission rates 
while there is a slightly delay and an overshoot in the 2024 outbreak. These estimations also fit well with the model simulation with the estimation of data based transmission rates but slight increases are seen in 2022 and 2023.
About the hospitalized individuals, the curves show more variation in the elderly and middle-aged adults, 
as expected and as visible in Figure \ref{fig18}. 

\begin{figure}[htbp]
    \centering
    \includegraphics[scale=0.55]{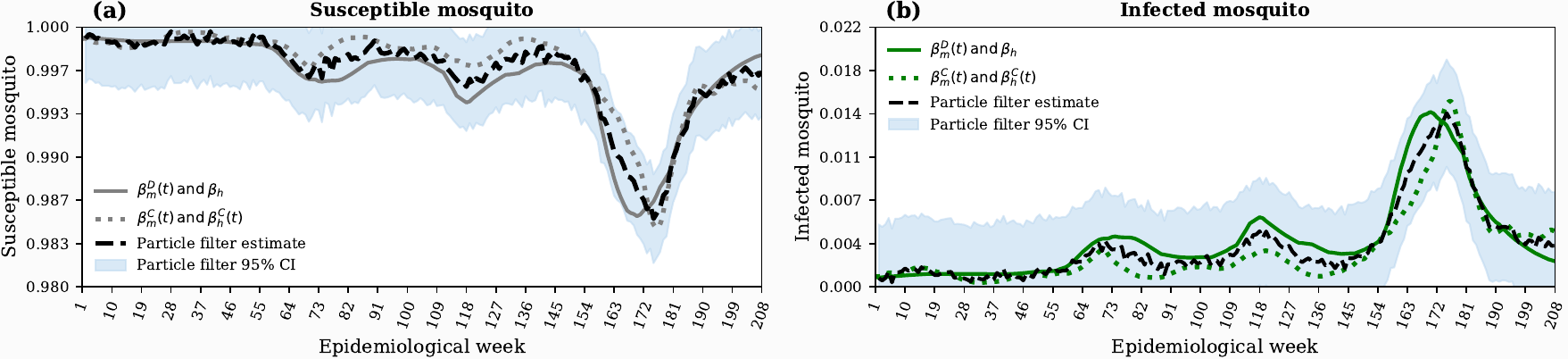}
    \caption[Dynamics of mosquitoes in the model]{\small
    \textbf{Dynamics of mosquitoes}. Dynamics of normalized susceptible and infected mosquitoes, as represented by model \eqref{section1.13} under different transmission rates. The dashed black line shows the particle filter estimate with the 95\% confidence interval shaded in sky blue.}
    \label{fig20}
\end{figure}

Quite interestingly, we also provide in Figure \ref{fig20} the fraction of susceptible mosquitoes and the fraction
of infected mosquitoes. There is no data supporting these quantities, but these graphs show that our model provides
takes the evolution of these population into account and provide meaningful information about them. 
More precisely, Figure \ref{fig20} illustrates the dynamics of mosquitoes in the model. 
We observe a stable trend for infected mosquitoes, with a slight increase over time, which is influenced by the transmission rates. 
Our simulation displays that the mosquito population follows an annual cycle with peaks and decreases. 
This suggests that a model with temperature as an external forcing function can replicate the observed patterns, 
with the fourth peak being larger than the first and slightly delayed. 
The effect of temperature and humidity on mosquito dynamics in Brazil is sufficient to drive the observed dengue seasonality even in the absence of observable data.

\subsection{Sensitivity of parameters}

\begin{figure}[htbp]
    \centering
    \includegraphics[scale=0.5]{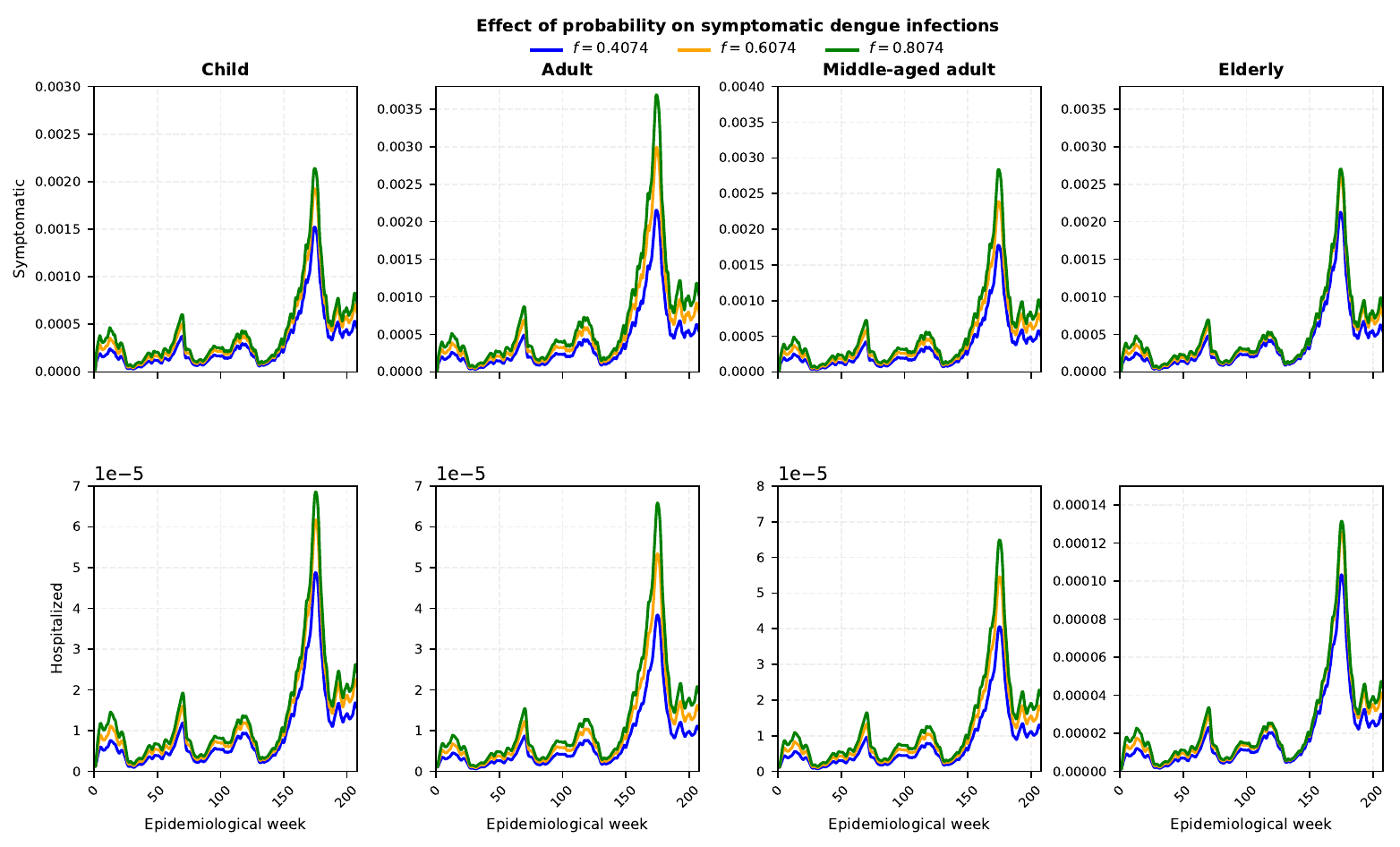}
    \caption[Effect of the probability that dengue infected individuals become symptomatic]{\small
    \textbf{Effect of the probability that dengue infected individuals become symptomatic}. Effects of different probabilities that dengue infected individuals become symptomatic on the normalized incidence of symptomatic cases and hospitalized in the model}
    \label{fig21}
\end{figure}

\begin{figure}[htbp]
    \centering
    \includegraphics[scale=0.5]{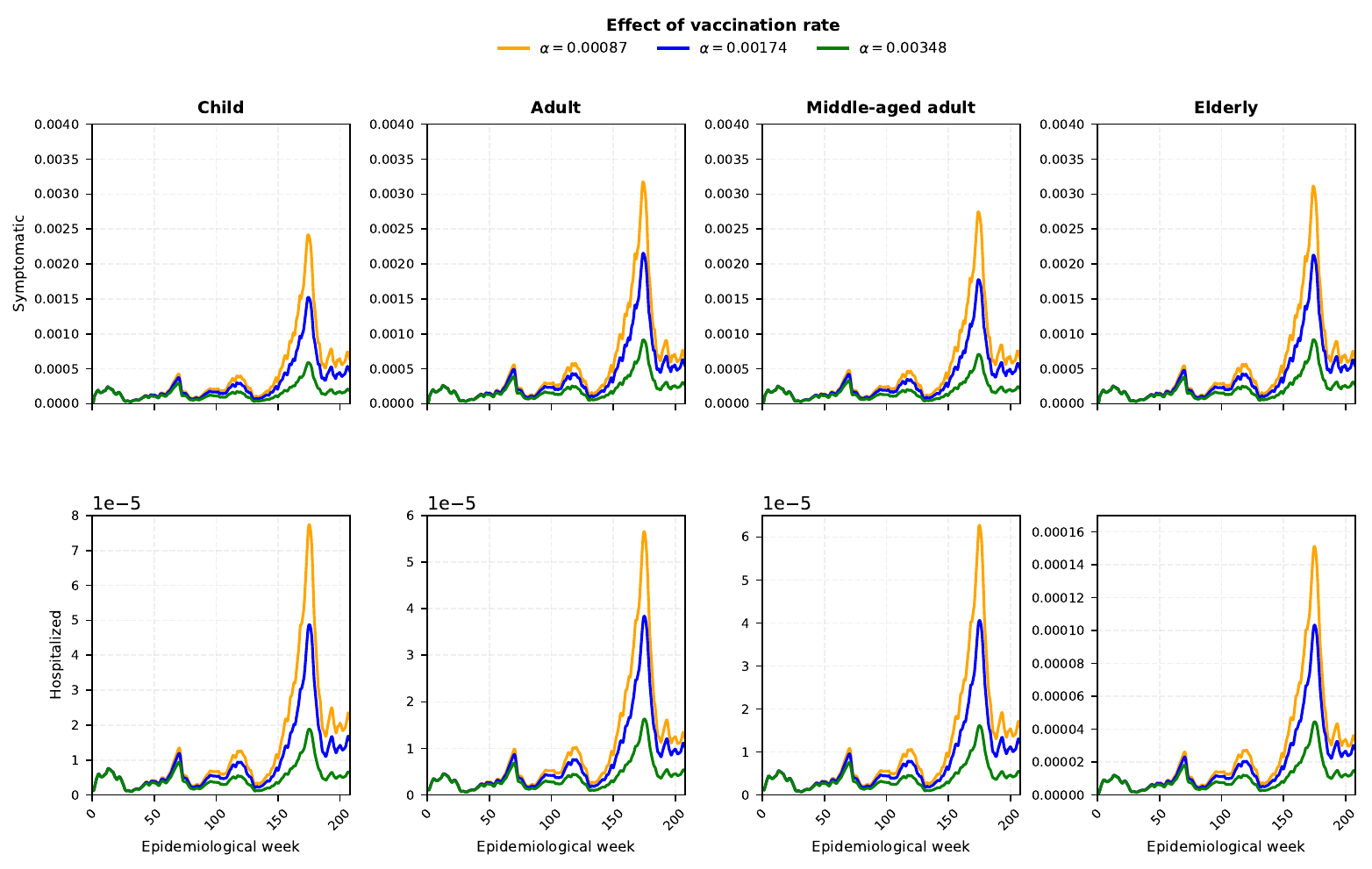}
    \caption[Effect of the vaccination rate]{\small
   \textbf{Effect of the vaccination rate.} Effects of different vaccination rates on the normalized incidence of symptomatic cases and hospitalized in the model.}
    \label{fig24}
\end{figure}

\begin{figure}[htbp]
    \centering
    \includegraphics[scale=0.5]{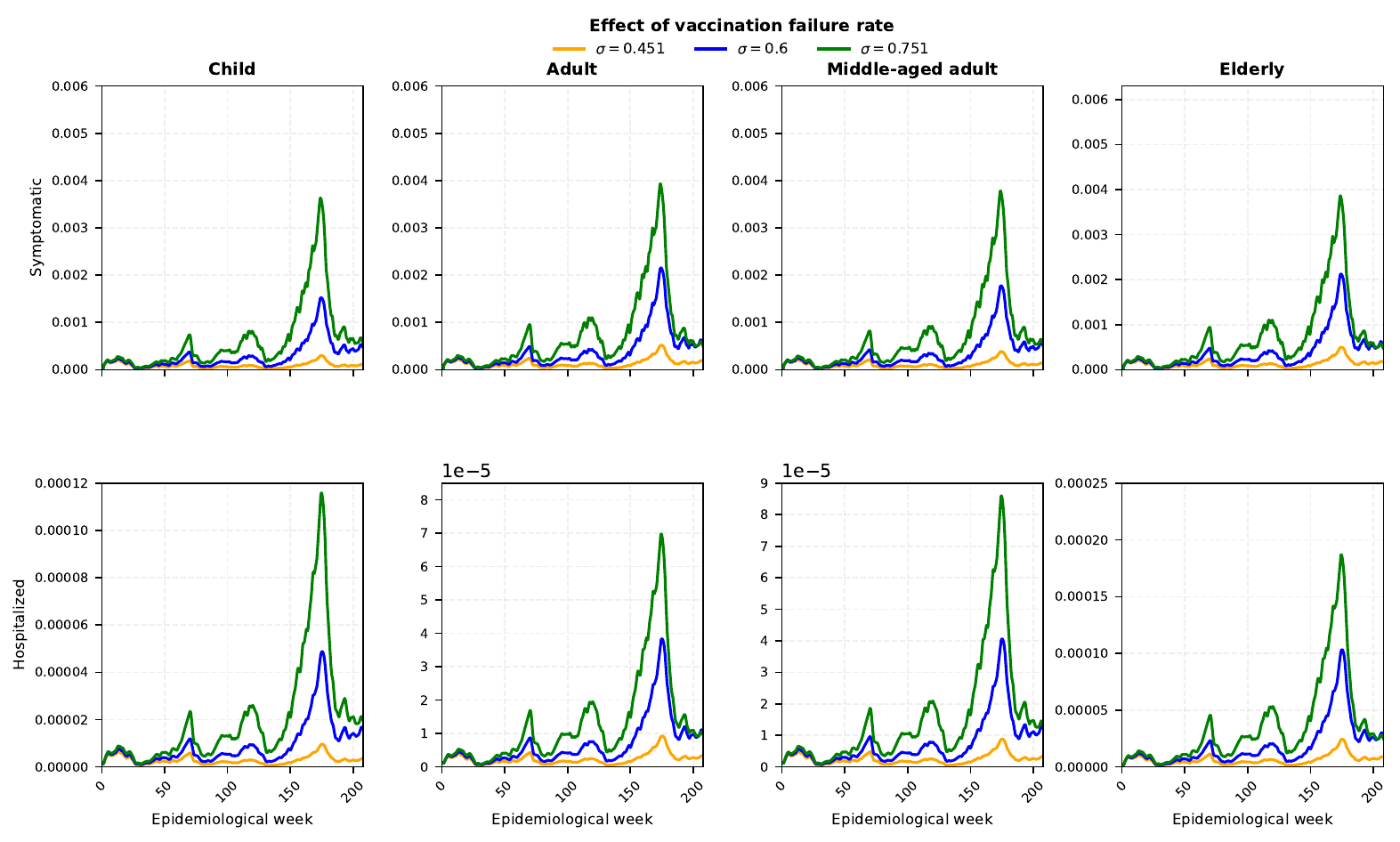}
    \caption[Effect of vaccination failure rate]{\small
    \textbf{Effect of vaccination failure rate}. Effect of varying the vaccination failure rate on the normalized incidence of symptomatic cases and hospitalized in the model.}
    \label{fig22}
\end{figure}

\begin{figure}[htbp]
    \centering
    \includegraphics[scale=0.5]{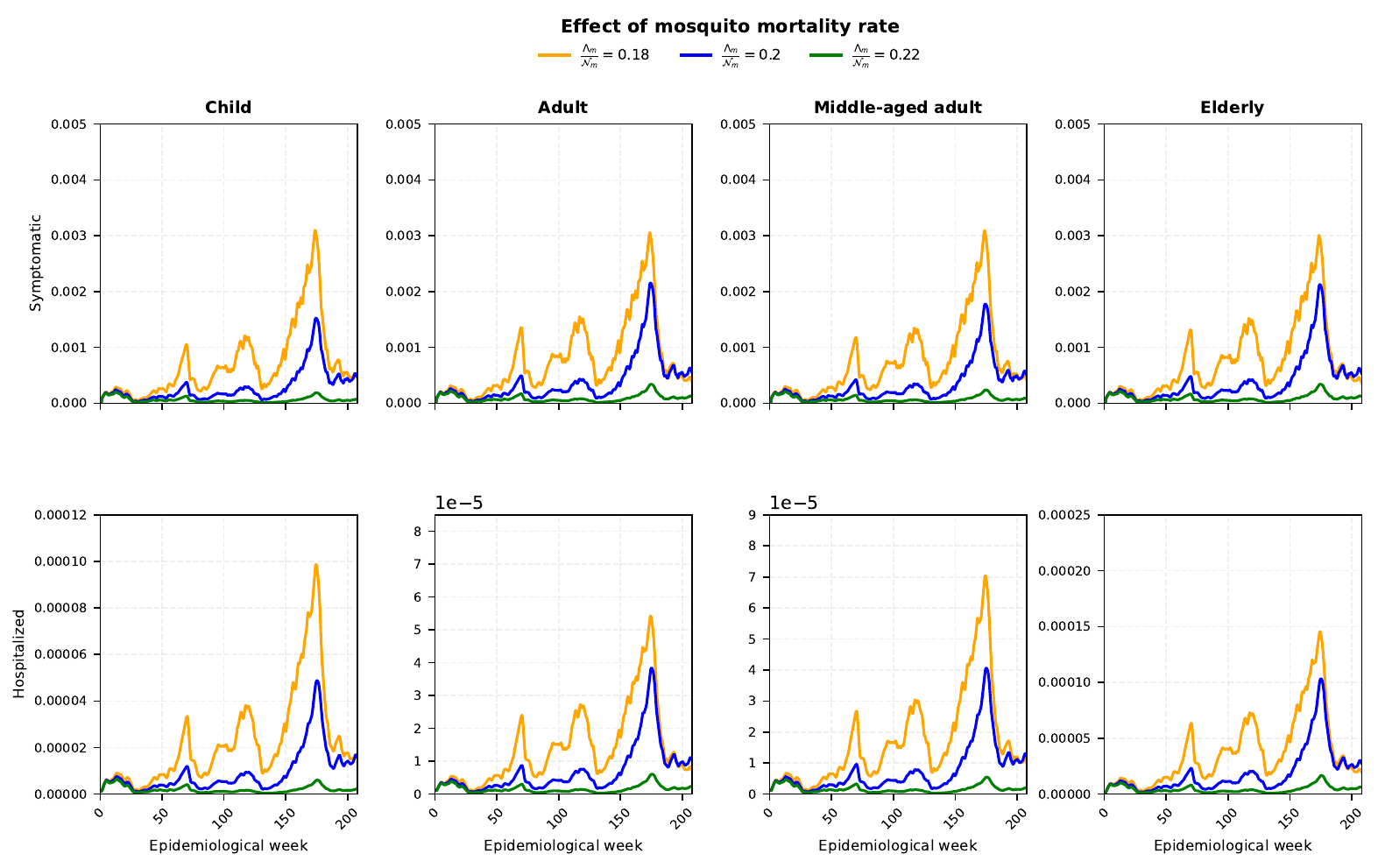}
    \caption[Effect of mosquito mortality rate]{\small
    \textbf{Effect of mosquito mortality rate $\boldsymbol{\frac{\varLambda_{m}}{\N_{m}}}$}. Effect of different mosquito mortality rates on the normalized incidence of symptomatic cases and hospitalized in the model.}
    \label{fig23}
\end{figure}

Let us now move to the effects of some parameters on the symptomatic individuals and on the hospitalized individuals.
For this study, we always consider the value of the parameters used in the previous subsection, and also two different values.
It means that one of the curves in the following figures corresponds to the one of Section \ref{subsec_trans_r}, 
the one with the temperature-humidity transmission rates, and the other two curves are obtained with the indicated value for the parameter
under investigation, and all other parameters unchanged. The study of the sensitivity of four parameters is reported in Figures
\ref{fig21} to \ref{fig23}.

In Figure \ref{fig21} the effect of the parameter $f$ is illustrated, namely the probability that infected individuals become symptomatic.
The influence of vaccination is shown in Figure \ref{fig24}. When the vaccination rate increases, we notice a significant decrease 
in the number of susceptible individuals who contract the disease and this reduces the number of infected humans. 
On the other hand, the effect of vaccination failure is illustrated in Figure \ref{fig22}. Following \cite{Xue2021} this parameter is quite big, 
but the aim is also to take into account the decay in time of the protection of the vaccination.
In Figure \ref{fig23}, we observe the effect of varying mosquito mortality rates on disease transmission. 
Increasing $\frac{\varLambda_{m}}{\N_{m}}$ shortens the infectious period of mosquitoes, which subsequently reduces the potential for disease transmission. From Figure \ref{fig24}, 

\subsection{Time varying effective reproduction numbers}

We finally compare the different approaches for computing the time-varying effective reproduction number,
as introduced in Section \ref{sec_tvern}. Recall that $\RD(t)$ was introduced in \eqref{section9.12} and reported in Figure \ref{fig6}. Alternatively, the effective reproduction numbers
can be computed based on the original expression \eqref{section8.0012} and by using the two possible approaches
for the value of the transmission rates: the data-based transmission rates or the temperature-based transmission rate.
All other parameters are similar for the computation. These time-varying effective reproduction numbers
are called respectively $\RMD(t)$ and $\RMC(t)$, see also the begining of Section \ref{sec_transmission_r}.

\begin{figure}[htbp]
    \centering
    \includegraphics[width=0.8\linewidth]{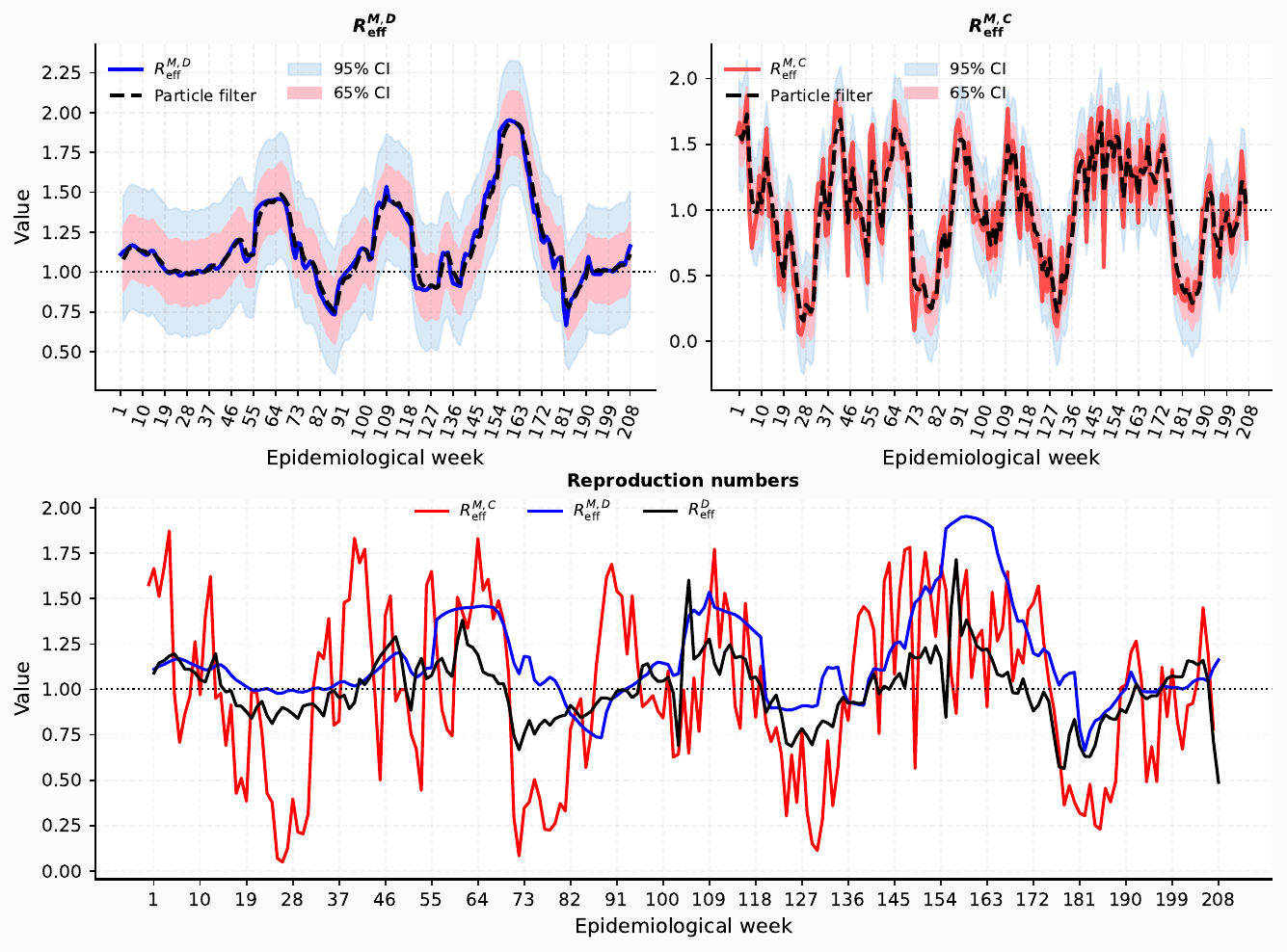}
    \caption[Time-varying effective reproduction numbers]{\small
    \textbf{Time-varying effective reproduction numbers}. This figure illustrates the comparison between model-based and data-based time-varying effective reproduction numbers. In the upper subplots, the sky blue region represents the 95\% confidence interval and the pink region shows the 65\% confidence interval.}
    \label{fig25}
\end{figure}

In Figure \ref{fig25}, the top subplots show the $\RMD(t)$ and $\RMC(t)$, along with the particle filter estimation 
and with its $65$\% and its $95$\% confidence intervals. The bottom subplot illustrates the comparison of the three values $\RD(t)$, $\RMD(t)$ and $\RMC(t)$. Clearly,  $\RD(t)$ and $\RMD(t)$ follow essentially the same pattern, since both use data. Let us emphasize that the former effective reproduction number  does not rely on any model, why the latter relies on our model. As a consequence, one observes that $\RMD(t)$ presents smoother variation compared to  $\RD(t)$.

On the other hand, $\RMC(t)$ follows trends which are similar to the two previous reproduction numbers. The main difference is that strong oscillations are taking place, which are certainly due to the varying weekly temperature. It also seems that reproduction number should be slightly rescaled: its values below $1$ are clearly much smaller than for the other two reproductive numbers, while its values above $1$ look slightly smaller (on average) than the other two reproduction numbers. For that purpose, the model provided in Section \ref{sec_thbased} for the computation of the transmission rate based on temperature and humidity could certainly be improved. 

\section{Predictions}\label{sec_pred}

Let us now check how our model and our approach can suitably forecast the dengue transmission dynamics.
These investigations are divided into two parts: a test part, and a purely prediction part.
For the test part, we use the period from January 1, 2025, to September 30, 2025. 
The prediction part starts on October 1, 2025 and covers 2026 as well.
For these two periods, we keep all medical parameters obtained and used for the period 2021--2024, 
as reported in Table \ref{table2}. On the other hand, we decided to reset the initial
conditions by January 1, 2025. The new initial conditions for that day are provided in 
Table \ref{table:initial_conditions_pred}.

\begin{table}[htbp]
\centering
\caption{Initial conditions for different age groups.}
\scalebox{0.6}{
\begin{tabular}{lcccccccc}
\hline
\hline
\textbf{Age Group} & \textbf{Susceptible} & \textbf{Vaccinated} & \textbf{Asymptomatic } & \textbf{Symptomatic} & \textbf{Hospitalized} & \textbf{Recovered} & \textbf{ Susceptible Mosquitoes} & \textbf{Infected Mosquitoes} \\ \hline \hline
\textbf{Child}              & 0.48   & 0.33  & 0.000018    & 0.000035  & 0.0000015  & 0.1899455  &  0.9985  &  0.0015  \\ 
\textbf{Adults}             & 0.48 & 0.33  & 0.000018    & 0.000035  & 0.0000015  & 0.1899455    & 0.9985  & 0.0015  \\ 
\textbf{Middle-Aged Adults} & 0.48 & 0.33  & 0.000018    & 0.000035  & 0.0000015  & 0.1899455       & 0.9985  & 0.0015  \\ 
\textbf{Elderly}            & 0.48 & 0.33  & 0.000018    & 0.000035  & 0.0000015  & 0.1899455   & 0.9985  & 0.0015  \\ \hline
\end{tabular}
}
\label{table:initial_conditions_pred}
\end{table}

We already observed that the transmission rates are some of the leading parameters for the propagation of the epidemic.
Recall that we have two transmission parameters: 
the mosquito-to-human transmission rate $\beta_{h}(t)$ and the human-to-mosquito transmission rate $\beta_{m}(t)$.
For that reason, we have decided to forecast these parameters with different approaches:
\begin{enumerate}
\item[$1)$] Constant $\beta_h$, as suggested by the literature and already reported in Table \ref{table2},
\item[$2)$] Weekly forecast of $\beta^{D}_{m}(t)$ based on the average values obtained during the four years 2021--2024, as shown in Figure \ref{fig8},
\item[$3)$] The temperature / humidity transmission rates as introduced in Section \ref{sec_thbased}.
\end{enumerate}

\begin{figure}[htbp]
    \centering
    \includegraphics[width=0.8\linewidth]{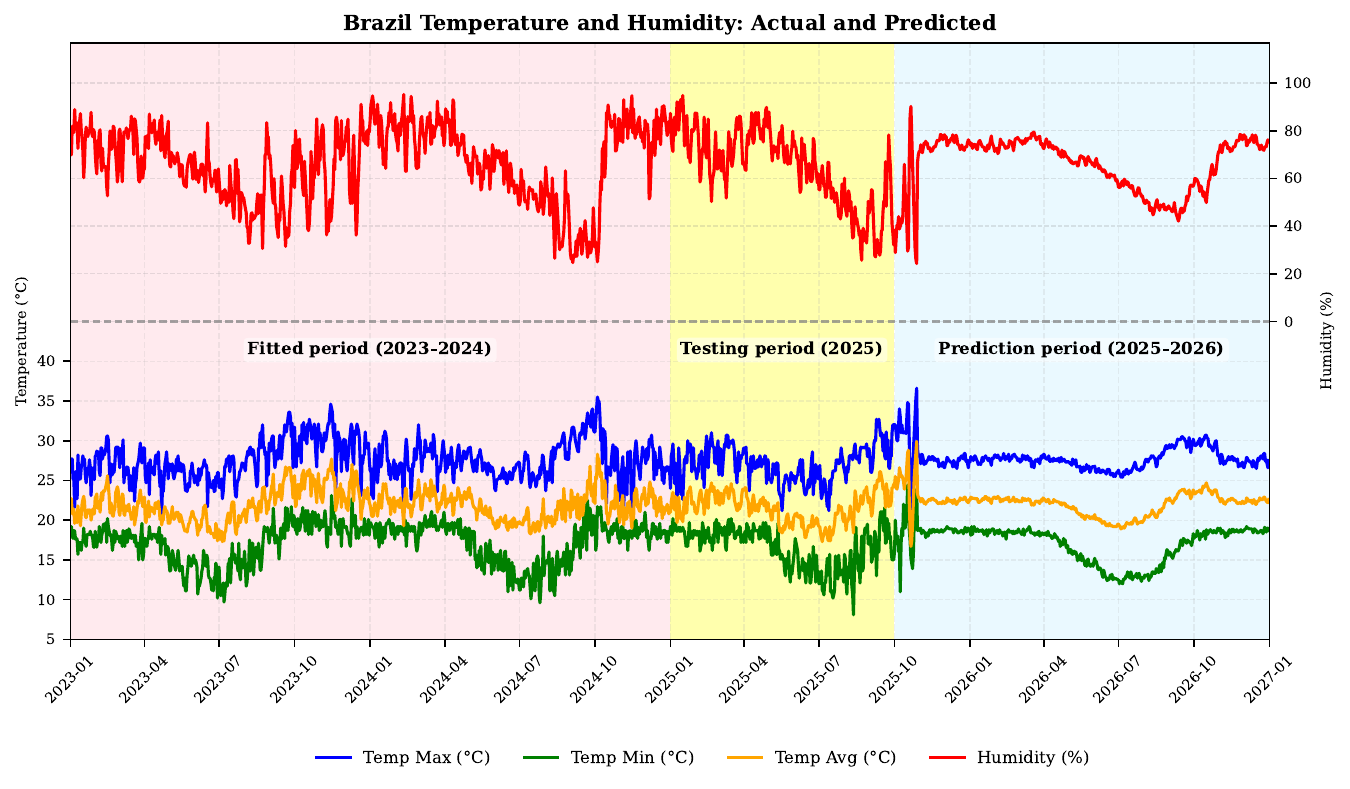}
    \caption[Temperature and humidity trends in Brazil from 2023 to 2026]{\small
    \textbf{Temperature and humidity trends in Brazil.} Comparison of temperature and  humidity patterns observed and predicted for Brazil during the period 2023–2026.}
    \label{fig_H_T_prediction}
\end{figure}

For the transmission rates based on temperature and humidity, we shall use the real data from the test period,
while we shall use some forecast data provided by \cite{Brazil_weather}. The data are indicated in Figure \ref{fig_H_T_prediction}.
Note that for comparison, we also recall the data for the period 2023--2024.
The transmission rate based on data is computed with the approach introduced in Section \ref{sec_trans_rate_data}.

\begin{figure}[htbp]
    \centering
    \includegraphics[width=0.8\linewidth]{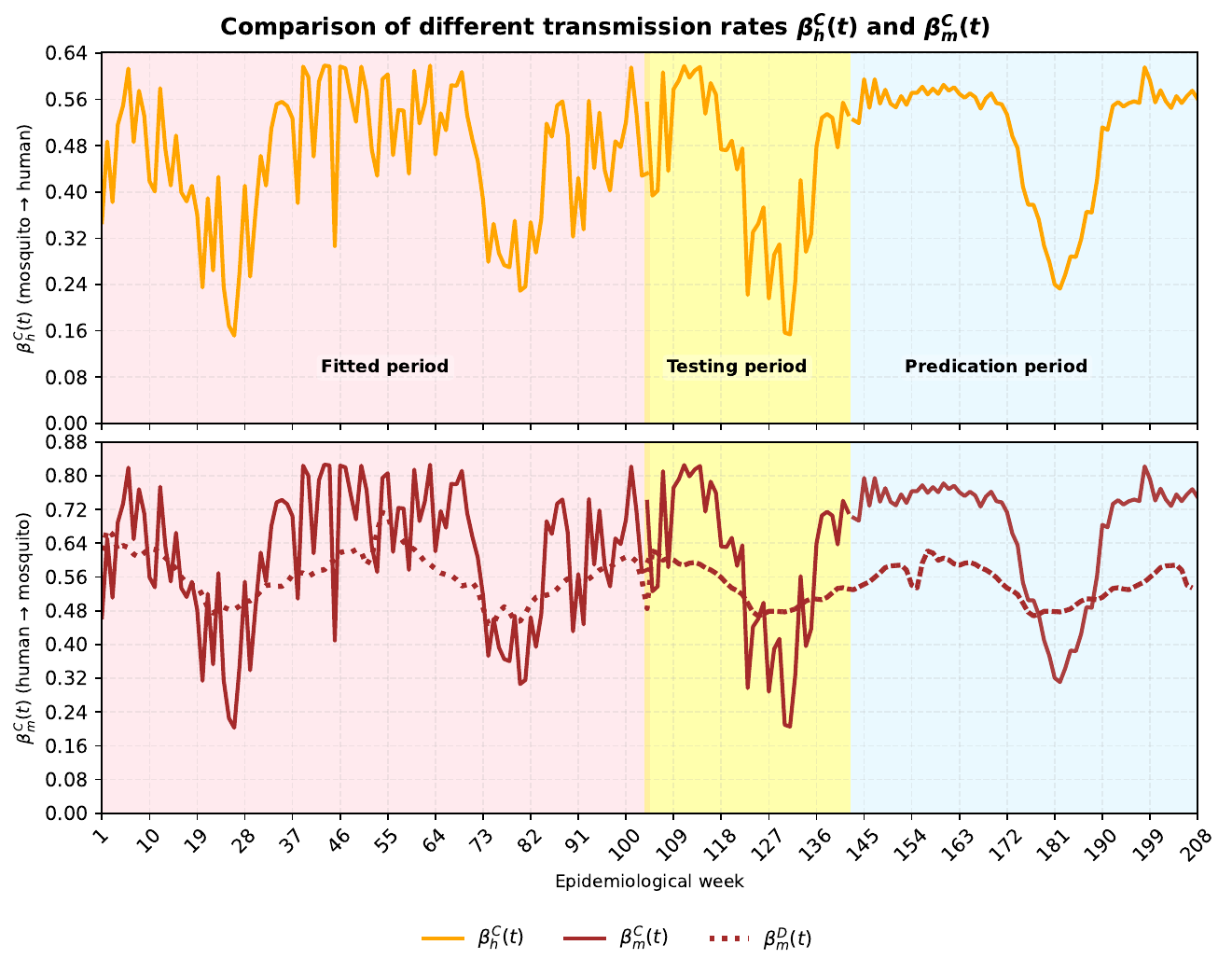}
    \caption[Estimated transmission rates in Brazil from 2023 to 2026]{\small
    \textbf{Estimated transmission rates in Brazil.} Comparison of the estimated different mosquito–human and human–mosquito transmission rates for Brazil across the years 2023–2026.}
    \label{trans_coef_23_26}
\end{figure}

The transmission coefficients are pictured in Figure \ref{trans_coef_23_26}. The test part and the 
purely prediction part are clearly indicated. For comparison, the values obtained for the period 2023--2024
are also indicated. 
For the test period, both temperature / humidity and compartmental data are available, which means
that the temperature / humidity based transmission rates $\beta^{C}_{m}(t)$ and $\beta^{C}_{h}(t)$ can be evaluated.
For the purely prediction part, the transmission rates are based on the forecast
temperature / humidity.
For $\beta^{D}_{m}(t)$ we also present the weekly average obtained from the period 2021--2024, as mentioned in $2)$ above.

\begin{figure}[htbp]
    \centering
    \includegraphics[scale=0.55]{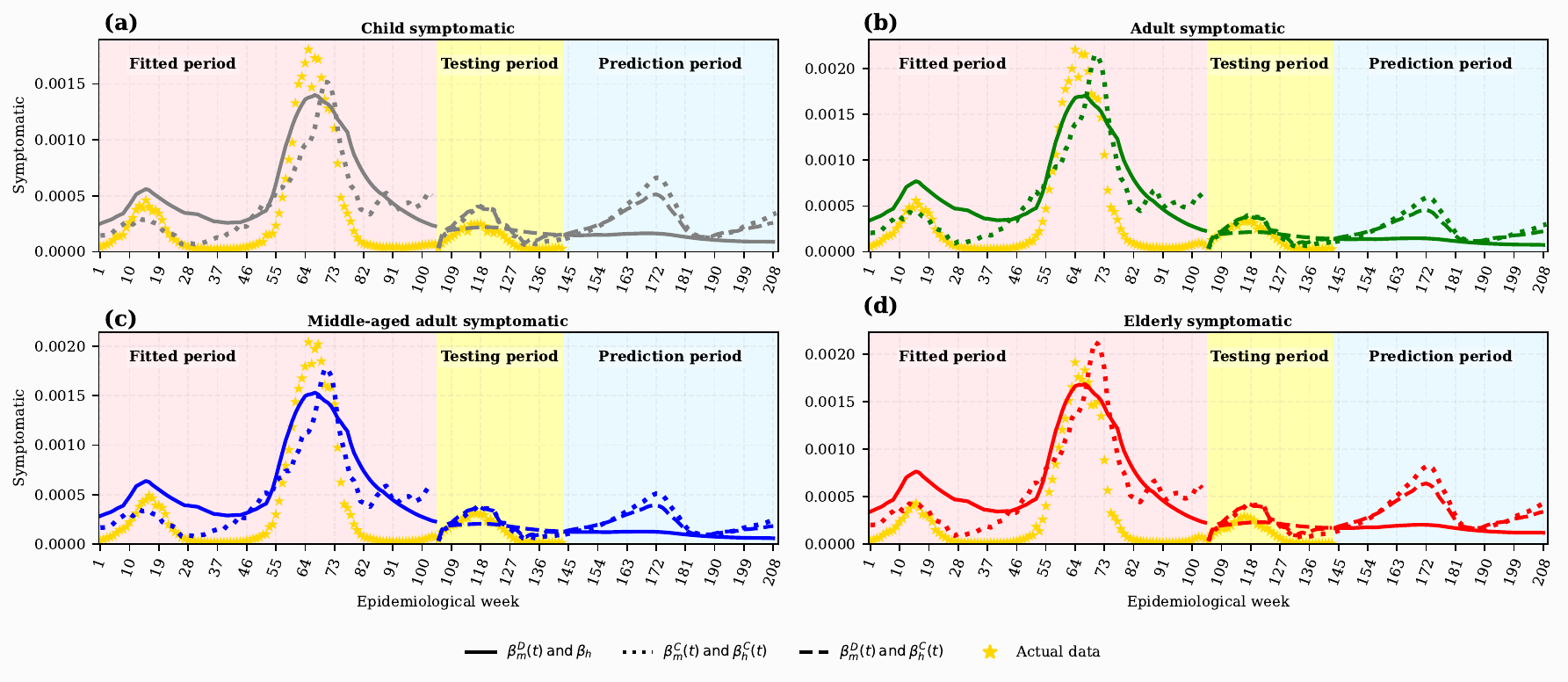}
      \caption[Actual vs. estimated vs. predicted values]{\small
\textbf{Comparison between the actual, estimated and predicted values.} 
This figure compares the actual, estimated and predicted normalized symptomatic dengue incidence in Brazil from 2023--2026.}
\label{fig_est_sympto}
\end{figure}

\begin{figure}[htbp]
    \centering
    \includegraphics[scale=0.55]{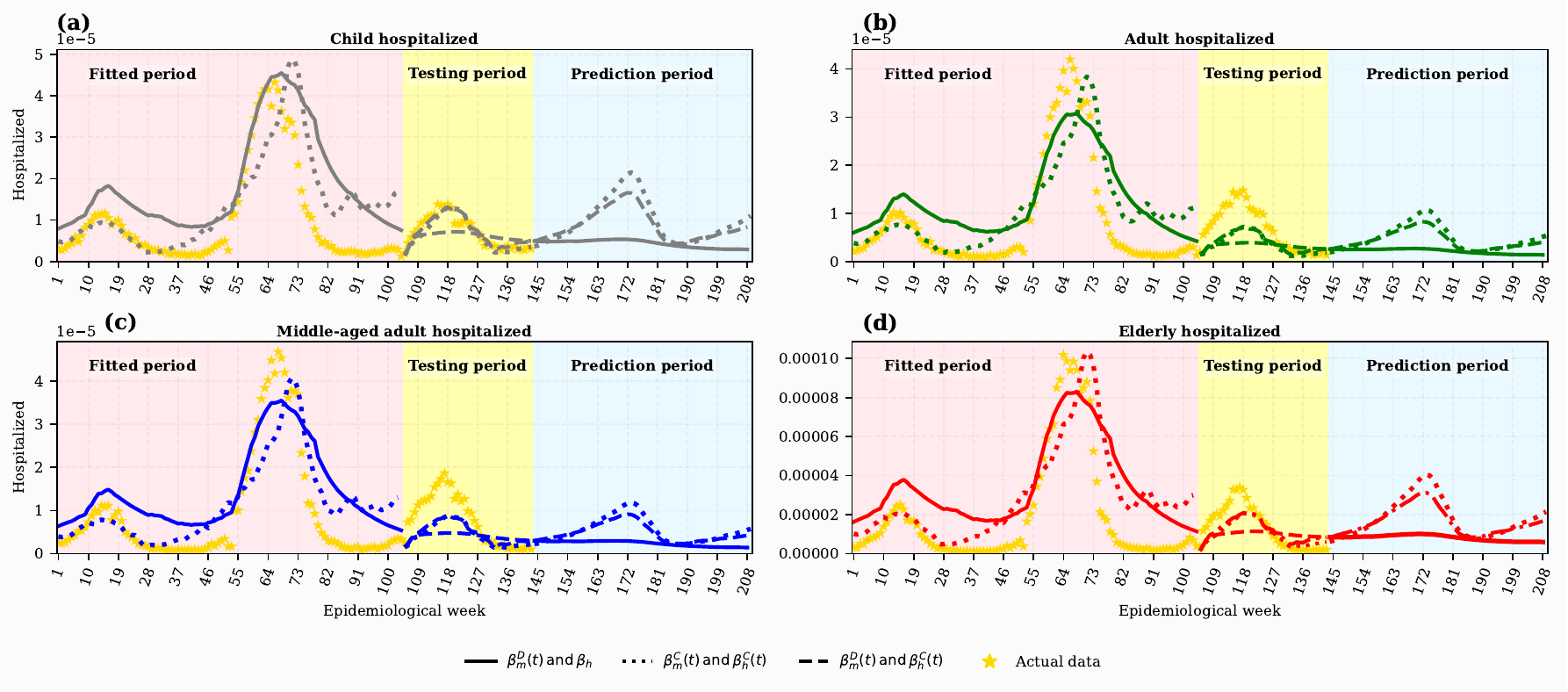}
   \caption[Actual vs. estimated vs. predicted values]{\small
\textbf{Comparison between the actual, estimated and predicted values.} 
This figure compares the actual, estimated and predicted normalized incidence of dengue hospitalized in Brazil from 2023--2026.}
\label{fig_est_hospi}
\end{figure}

In Figures \ref{fig_est_sympto} and \ref{fig_est_hospi} we provide the estimated symptomatic individuals
and hospitalized individuals for the different age groups and for test period and the purely prediction period.
For the test period from January 1, 2025, to September 30, 2025, the ratio of symptomatic individuals and of hospitalized
individuals can be computed with three approaches, based on the choice for the transmission rates:
\begin{enumerate}
\item[$i)$] $\beta^{C}_{h}$  and  $\beta^{D}_{m}(t)$ (weekly average on the period 2021--2024),
\item[$ii)$] $\beta_h=0.44$ (as suggested by the literature) and $\beta^{D}_{m}(t)$ (weekly average on the period 2021--2024),
\item[$iii)$] $\beta^{C}_{h}(t)$ and $\beta^{C}_{m}(t)$.
\end{enumerate}
The three corresponding curves are reported, as well as the true data.
For the purely prediction part, the transmission rates based on temperature / humidity use the corresponding
predicted data. 
As we can observe both on our prediction and on the true data, no real outbreak of dengue took place in 2025.
Note that the jumps in the curve on January 1, 2025, are due to the reset of the initial conditions, as already mentioned.

\begin{figure}[htbp]
    \centering
    \includegraphics[width=0.8\linewidth]{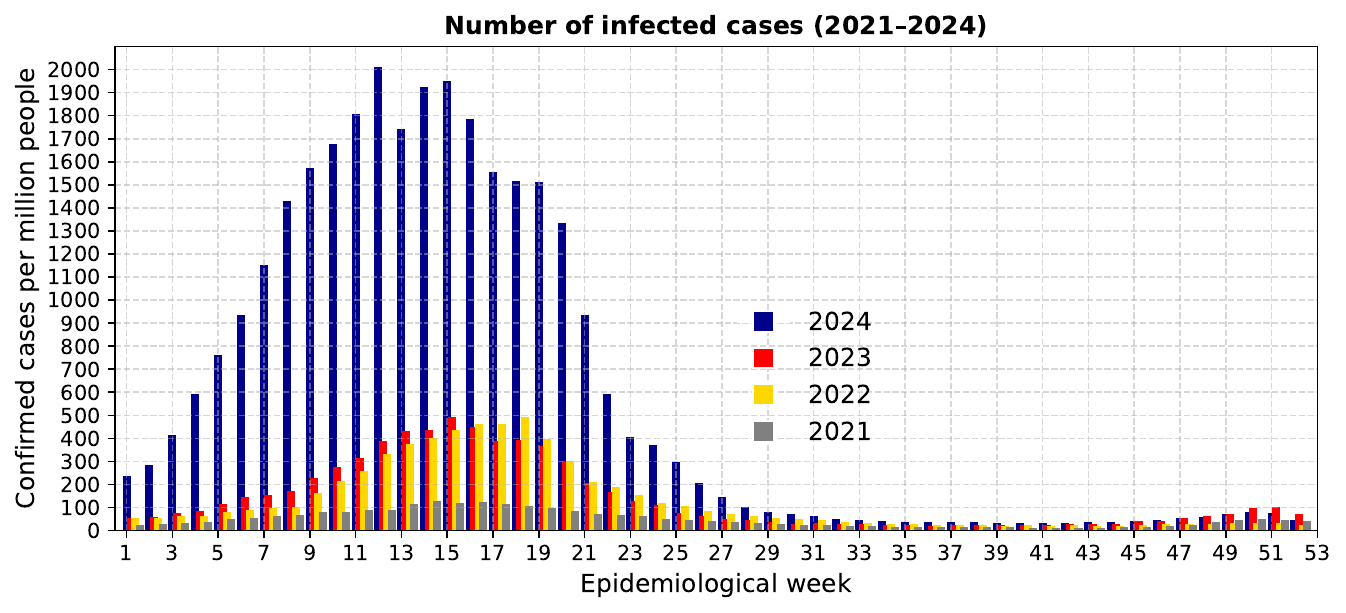}
    \caption[Infected cases]{\small
 \textbf{Infected cases.}
 Number of dengue infected cases in Brazil, 2021–2024.}
    \label{fig_superposed_outbreaks}
\end{figure}

The predictions based on the fixed $\beta_h$ and on $\beta^{D}_{m}(t)$ based on the weekly average on four years is not good. This can be understood based on the following observations: these parameters are really dependent on the seasons,
and a constant value can not reproduce any yearly variations. In addition, taking the average on four years has the drawback of
scaling down any natural variations. Indeed, the outbreaks do not take place exactly at the same period of the year, 
and therefore any averaging process will reduce the intensity of the oscillations. 
In Figure \ref{fig_superposed_outbreaks} we superpose the outbreaks for the period 2021--2024. It is clearly visible
that their intensity vary a lot, but also the periodicity of the outbreak is not exactly of one year. 

\begin{figure}[htbp]
    \centering
    \includegraphics[width=1.0\linewidth]{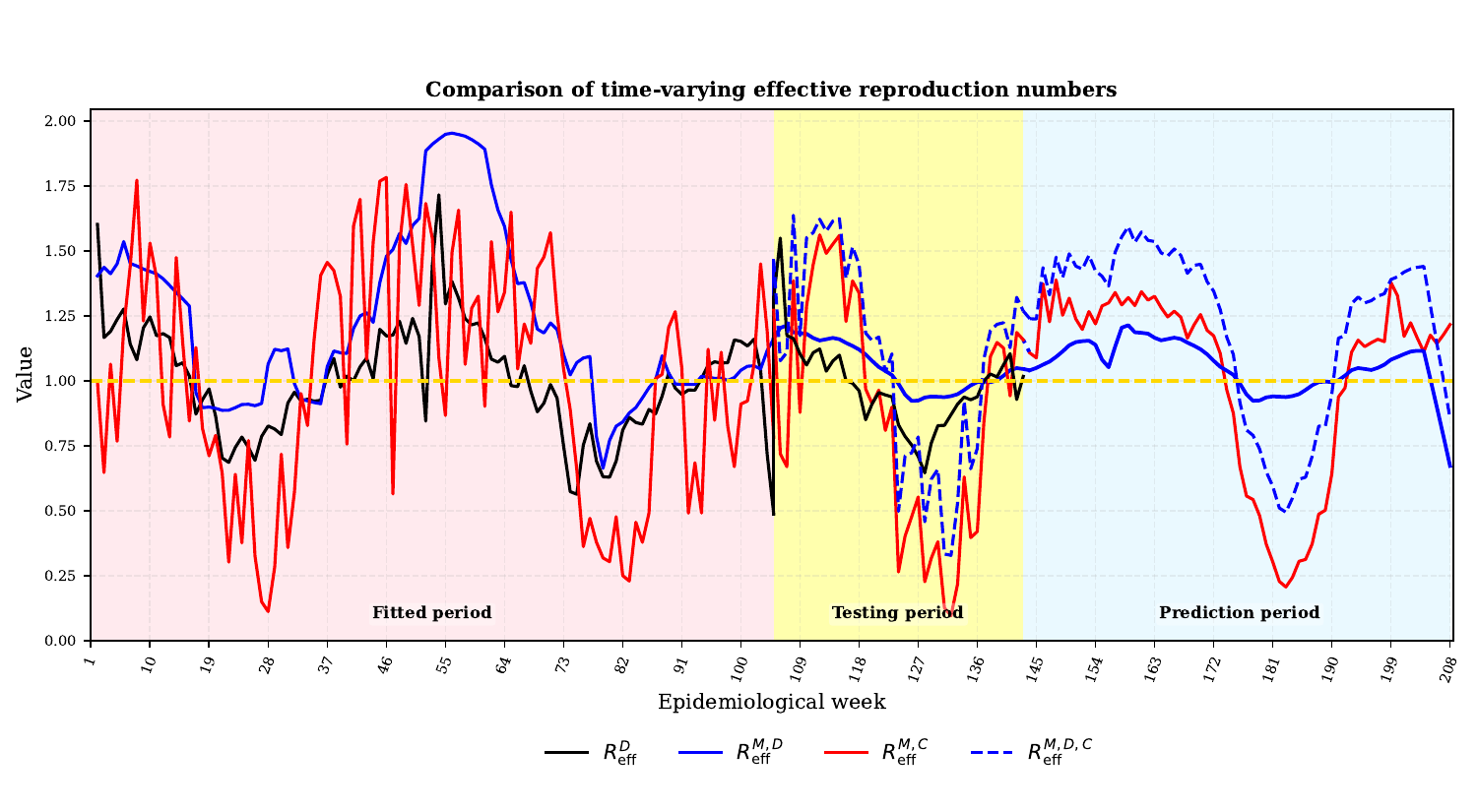}
    \caption[Time-varying Reproduction Number for Brazil from 2023 to 2026]{\small
    \textbf{Comparison of time-varying effective reproduction number for Brazil}. The comparison of different estimated time-varying effective reproduction numbers for Brazil for 2023 to 2026.}
     \label{fig_pred_Reff}
\end{figure}

Finally, we provide in Figure \ref{fig_pred_Reff} a comparison between the different effective reproduction numbers
obtained with the various transmission rates:
$\RD(t)$ (based on data only and computed with \eqref{section9.12}), $\RMC(t)$ (using $\beta^{C}_{m}(t)$ and $\beta^{C}_{h}(t)$),
$\RMD(t)$ (using $\beta^{D}_{m}(t)$ and $\beta_{h}$), and $\RMDC(t)$ (using $\beta^{D}_{m}(t)$ and $\beta^{C}_{h}(t)$).
For comparison, we also report the values obtained for 2023--2024.
We easily observe that the effective reproduction numbers based on temperature / humidity have
larger amplitudes and more variations, compared to the one obtained with the transmission rates either constant or based
on the weekly average on four years. A bit surprisingly, the effective reproduction number $\RMD(t)$ based on $\beta_{h}=0.44$ and
$\beta^{D}_{m}(t)$ obtained by the weekly average on four years is closer to the $\RD(t)$ obtained with the data only, 
at least for the first half of 2025. We finally observe that the dependence on the transmission rates is very involved: 
when $\beta^C_m(t)$ is smaller than $\beta^D_m(t)$, as visible in Figure \ref{trans_coef_23_26}, the reproduction number
$\RMC$ based on $\beta^C_m(t)$ takes the smallest value, while when $\beta^C_m(t)$ is larger than $\beta^D_m(t)$, 
$\RMC$ takes an intermediate value between $\RMD$ and $\RMDC$.
Despite the different of amplitudes, we observe that these various time-varying reproduction numbers cross the critical value $1$
on similar periods of time.

\section{Discussion and Conclusion}\label{sec_dis_con}

In this work, we develop a time-dependent and age-structured dengue transmission model 
that explicitly includes asymptomatic and symptomatic infections. Vaccination effects were also taken into account.
We allowed most of the model parameters to vary with time in order to capture true transmission rates
and seasonality dynamics.
One specific feature of this research is to derive renewal equations from the first principles 
that can be used to model the host-vector disease outbreaks.

In the first theoretical part of the investigations, we showed that renewal theory gives us rich analytical insights 
into the definition and computation of various time-dependent threshold quantities.
In particular, we derived the basic reproduction number $\Rfo$, the time-varying effective reproduction number $\Rf(t)$, 
the instantaneous growth rate, and established analytical relationships between these quantities.
Note that the basic effective reproduction number has been computed following three different approaches.
On the other hand, the basic reproduction number has been computed with two approaches, one involving the Euler-Lotka equation, and one based on the next generation matrix.

In the second part, we used data from Brazil to validate our model and to estimate several parameters.
Note that these data allowed us to consider only 4 age groups, and therefore the subsequent
investigations are always analyzed separately for these four groups.
Firstly, various medical parameters were obtained directly from the data for the period 2021--2024.
Transmission rates were also obtained with two approaches: one involving the data and our model,
one based on temperature and humidity only. These investigations on parameters are explained and reported 
in Sections \ref{sec_med_pre} and \ref{sec_transmission_r}. 
It is worth noting that the reliance on accurate parameter estimates is a limitation of most dynamical models 
of complex biological phenomena, since many parameters are difficult or costly to measure. 
  
In Section \ref{sec_21_24}, we provided simulations based on our model for the period 2021--2024,
with initial conditions for January 2021 reported in Table \ref{table:initial_conditions}, and with 
the medical parameters obtained before.
Figures \ref{fig17} and \ref{fig18} contain the results of these simulations,
with the two transmission rates already mentioned: In the first figure, the symptomatic population is provided, 
and in the second figure the hospitalized population is shown.
These model simulations are also compared with the corresponding data from  2021 to 2024 in Brazil.
A seasonality is clearly visible in these simulations and in the data, but the size of the yearly outbreak
is highly dependent on the specific year. 
With the simulations using temperature / humidity based transmission rates, we readily conclude 
that mosquito-borne pathogen transmission is highly sensitive to environmental fluctuations, 
especially to temperature and humidity. 
Our results highlight that climate variables can directly affect mosquito survival and 
also modify human behavior, both of which can influence epidemic outbreaks.

In Figures \ref{fig21}\,-\,\ref{fig23}, we discussed the influence of some parameters such as 
the probability of becoming symptomatic, the vaccination rate, vaccination failure, and the mosquito mortality rate.
All of them strongly affect the transmission dynamic, as shown on these figures. 
Still in Section \ref{sec_21_24}, we estimated and represented the time-varying effective reproduction number $\Rf(t)$ 
using three different approaches: (i) incidence data  $\RD(t)$; (ii) model-based with temperature and humidity dependent rates $\RMC(t)$,
(iii) model-based with data-derived rates $\RMD(t)$. All three showed consistent temporal patterns like initial rise, 
outbreak peaks and decline etc. 

In Section \ref{sec_pred} we finally provided some independent test of the model, and then some predictions.
By using the parameters estimated from the period 2021--2024 we tested the model on the first 39 weeks of 2025. 
Since the transmission rates are the dominant factors for the simulations, we simulated the evolution
based on three different sets of estimated transmission rates.
These simulations were reported in Figure \ref{fig_est_sympto} and \ref{fig_est_hospi}, and 
we observed that the model simulations fit the actual data well during this test period. 
After the testing period, we still provided predictions over a period of 15 months. 
For the temperature / humidity based transmission rates, we used the available predictions data for Brazil. 
Two simulations follow the same pattern, while one has another behavior which is explained in 
this section.

Clearly, the theoretical approach of the first part, together with the application to Brazil data in the second part
complement each other. Our time-dependent and age structured model provides interesting results, but it seems difficult
to have enough data for taking full advantage of the age structure. For example on the data from Brazil, it has been possible 
to create only four age group. Nevertheless, this flexibility can be useful, as emphasized when comparing
the population of Japan and Pakistan in Section \ref{rem_ideal_pop}.

Let us finally stress a few limitations of this study. Firstly, we assumed a fixed recruitment rate for mosquitoes, 
which may not be a realistic assumption. This rate should be temperature and humidity dependent, 
which could potentially lead to better fitting and predictions. Secondly, a gradual loss of the vaccination immunity 
could have been introduced. This would have prevented the accumulation of too many vaccinated individuals, 
as observed in some of our simulations. A better understanding of the variability of the weather conditions
in Brazil could have helped us to get a better understanding of the intensity of the dengue outbreaks~? 
For example, can one implement in our model large scale events which influence the epidemic~?
We have in mind phenomenon like El Ni$\tilde{\text{n}}$o which does have an impact, as reported in \cite{El_nino}.

\section{Appendix}\label{sec_app} 

\subsection{Solution of PDEs along the characteristic line}

This section is about the solution to PDEs along the characteristic line.

\begin{lemma}\label{App_PDE_sol}
Consider the first-order partial differential equation
\begin{equation}
\begin{cases}
\frac{\partial u(t,\zeta)}{\partial t} + \frac{\partial u(t,\zeta)}{\partial \zeta} = y(t,\zeta) - b(t,\zeta) u(t,\zeta), \\
u(t,0) = \Psi(t), \quad u(t^{0},\zeta) = \phi^{0}(\zeta), \\
\Psi(t^{0}) = \phi^{0}(0),
\end{cases}
\label{appendixa1}
\end{equation}
where $ b(t, \zeta)$, $y(t, \zeta)$, $\Psi(t)$, and $\phi^{0}(\zeta)$ are given continuous functions, and $t^{0}\leq t$. 
The general solution of this first-order PDE is given by
\begin{equation}
u(t, \zeta) = 
\begin{cases}
\Psi(t-\zeta)e^{-\int_{0}^{\zeta} b(\xi + t - \zeta, \xi) \d\xi} + \int_{0}^{\zeta} e^{-\int_{\eta}^{\zeta} b(\xi + t - \zeta, \xi) \d\xi} y(\eta + t - \zeta, \eta) \,\d\eta & \text{for }  \,t-t^{0} \geq \zeta, \\
\phi^{0}(t^{0}+\zeta - t) e^{-\int_{t^{0}}^{t} b(\xi, \xi + \zeta - t) \d\xi} + \int_{t^{0}}^{t} e^{-\int_{\eta}^{t} b(\xi, \xi + \zeta - t) \d\xi} y(\eta, \eta + \zeta - t) \,\d\eta & \text{for } \, t-t^{0}<\zeta,
\end{cases}
\label{appendixa18}
\end{equation}
or equivalently by 
\begin{equation}
u(t, \zeta) = 
\begin{cases}
\Psi(t-\zeta)e^{-\int_{0}^{\zeta} b(\xi + t - \zeta, \xi) \d\xi} + \int_{0}^{\zeta} e^{-\int_{\eta}^{\zeta} b(\xi + t - \zeta, \xi) \d\xi} y(\eta + t - \zeta, \eta) \,\d\eta & \text{for } t-t^{0} \geq \zeta, \\
\phi^{0}(t^{0}+\zeta - t) e^{-\int_{t^{0}+\zeta-t}^{\zeta} b(\xi+t-\zeta, \xi) \d\xi} + \int_{t^{0}+\zeta-t}^{\zeta} e^{-\int_{\eta}^{\zeta} b(\xi+t-\zeta, \xi) \d\xi} y(\eta+t-\zeta, \eta)\, \d\eta & \text{for } t-t^{0}<\zeta.
\end{cases}
\label{appendixa19}
\end{equation}
\end{lemma}

The method of characteristics is a technique used to solve hyperbolic partial differential equations. 
While it is most commonly applied to first order equations, it can also be used for other types of hyperbolic PDEs. 
The method involves finding special curves, known as characteristic curves, along which the original PDE 
is transformed into a system of ordinary differential equations. 
Once these ODEs are solved along the characteristic curves, the solutions of the ODEs can then be used to 
determine the solution of the original PDE.

\begin{proof}
We look for a solution to \eqref{appendixa1} by introducing the characteristic equations,
\begin{align}
\frac{\d t}{\d s} & =1,\label{appendixa2}\\
\frac{\d\zeta}{\d s} & =1,\label{appendixa3}\\
\frac{\d u}{\d s} & =y\left(t,\zeta\right)-b\left(t,\zeta\right)u\left(t,\zeta\right).\label{appendixa4}
\end{align}
Now, using \eqref{appendixa3} and \eqref{appendixa2}, we derive $\frac{\d t}{\d\zeta}=1$, which leads to $t-\zeta =x_{0}$.  
 We can also integrate the ODE \eqref{appendixa4} along the characteristics. However, by substituting the expressions into \eqref{appendixa4}, we obtain
\begin{equation}
\begin{aligned} 
& \frac{\d}{\d\zeta} u(\zeta + x_{0}, \zeta) = y\left(\zeta + x_{0}, \zeta\right) - b\left(\zeta + x_{0}, \zeta\right) u\left(\zeta + x_{0}, \zeta\right),\\
& u(\zeta + x_{0}, 0) = \Psi(\zeta + x_{0}).
\end{aligned}
\label{appendixa5}
\end{equation}
This can be rewritten as:
\begin{equation}
\begin{aligned} 
& \frac{\d}{\d\zeta} u(\zeta + x_{0}, \zeta) + b\left(\zeta + x_{0}, \zeta\right) u\left(\zeta + x_{0}, \zeta\right) = y\left(\zeta + x_{0}, \zeta\right).
\end{aligned}
\label{appendixa6}
\end{equation}
Now, by solving this ODE \eqref{appendixa6} using the integrating factor method with the integrating factor $ e^{\int_{0}^{\zeta}b\left(\xi+x_{0},\xi\right)\d\xi}$,
we obtain
\begin{equation}
\begin{aligned} & \frac{\d}{\d\zeta}\left(u(\zeta+x_{0},\zeta)e^{\int_{0}^{\zeta}b\left(\xi+x_{0},\xi\right)\d\xi}\right)=e^{\int_{0}^{\zeta}b\left(\xi+x_{0},\xi\right)\d\xi}y\left(\zeta+x_{0},\zeta\right).\end{aligned}
\label{appendixa7}
\end{equation}
Applying integration to both sides, we have, 
\begin{equation}
\begin{aligned} & u(\zeta+x_{0},\zeta)e^{\int_{0}^{\zeta}b\left(\xi+x_{0},\xi\right)\d\xi}-u(x_{0},0)=\int_{0}^{\zeta}e^{\int_{0}^{\eta}b\left(\xi+x_{0},\xi\right)\d\xi}y\left(\eta+x_{0},\eta\right)\,\d\eta.\end{aligned}
\label{appendixa8}
\end{equation}
Using the given condition $u(x_{0}, 0)=\Psi(x_{0})$, we get
\begin{equation}
\begin{aligned} & u(\zeta+x_{0},\zeta)=\Psi( x_{0})e^{-\int_{0}^{\zeta}b\left(\xi+x_{0},\xi\right)\d\xi}+e^{-\int_{0}^{\zeta}b\left(\xi+x_{0},\xi\right)\d\xi}\int_{0}^{\zeta}e^{\int_{0}^{\eta}b\left(\xi+x_{0},\xi\right)\d\xi}y\left(\eta+x_{0},\eta\right)\,\d\eta.\end{aligned}
\label{appendixa9}
\end{equation}
Combining the exponential terms, we infer that
\begin{equation}
\begin{aligned} & u(\zeta+x_{0},\zeta)=\Psi( x_{0})e^{-\int_{0}^{\zeta}b\left(\xi+x_{0},\xi\right)\d\xi}+\int_{0}^{\zeta}e^{-\int_{\eta}^{\zeta}b\left(\xi+x_{0},\xi\right)\d\xi}y\left(\eta+x_{0},\eta\right)\,\d\eta.\end{aligned}
\label{appendixa10}
\end{equation}
Now, putting $x_{0}=t-\zeta$, we obtain
\begin{equation}
\begin{aligned} & u(t,\zeta)=\Psi(t-\zeta)e^{-\int_{0}^{\zeta}b\left(\xi+t-\zeta,\xi\right)\d\xi}+\int_{0}^{\zeta}e^{-\int_{\eta}^{\zeta}b\left(\xi+t-\zeta,\xi\right)\d\xi}y\left(\eta+t-\zeta,\eta\right)\,\d\eta.\end{aligned}
\label{appendixa11}
\end{equation}
Similarly, from the characteristic \eqref{appendixa2} and \eqref{appendixa3}, we can write $\frac{\d\zeta}{\d t} = 1 \Rightarrow \zeta = t + x_{1}$. Therefore, we have
\begin{equation}
\begin{aligned} 
& \frac{\d}{\d t} u(t, t + x_{1}) = y\left(t, t + x_{1}\right) - b\left(t, t + x_{1}\right) u\left(t, t + x_{1}\right), \\
& u(t^{0}, t + x_{1}) = \phi^{0}(t + x_{1}).
\end{aligned}
\label{appendixa12}
\end{equation}
By using the method of integration factors, with the integration factor 
$ e^{\int_{t^{0}}^{t} b\left(\xi, \xi + x_{1}\right) \d\xi}$, we infer 
\begin{equation}
\begin{aligned} & \frac{\d}{\d t}\left(u(t,t+x_{1})e^{\int_{t^{0}}^{t}b\left(\xi,\xi+x_{1}\right)\d\xi}\right)=e^{\int_{t^{0}}^{t}b\left(\xi,\xi+x_{1}\right)\d\xi}y\left(t,t+x_{1}\right).\end{aligned}
\label{appendixa13}
\end{equation}
By applying integration to both sides from $t^{0}$ to $t$ and using the condition $u(t^{0},t^{0}+x_{1})=\phi^{0}(t^{0}+x_{1})$, we have
\begin{equation}
\begin{aligned} & u(t,t+x_{1})=\phi^{0}(t^{0}+x_{1})e^{-\int_{t^{0}}^{t}b\left(\xi,\xi+x_{1}\right)\d\xi}+e^{-\int_{t^{0}}^{t}b\left(\xi,\xi+x_{1}\right)\d\xi}\int_{t^{0}}^{t}e^{\int_{t^{0}}^{\eta}b\left(\xi,\xi+x_{1}\right)\d\xi}y\left(\eta,\eta+x_{1}\right)\,\d\eta.\end{aligned}
\label{appendixa14}
\end{equation}
By combining the exponential terms, we get
\begin{equation}
\begin{aligned} & u(t,t+x_{1})=\phi^{0}(t^{0}+x_{1})e^{-\int_{t^{0}}^{t}b\left(\xi,\xi+x_{1}\right)\d\xi}+\int_{t^{0}}^{t}e^{-\int_{\eta}^{t}b\left(\xi,\xi+x_{1}\right)\d\xi}y\left(\eta,\eta+x_{1}\right)\,\d\eta.\end{aligned}
\label{appendixa15}
\end{equation}
Now, substituting $x_{1}=\zeta - t$, we obtain
\begin{equation}
\begin{aligned} & u(t,\zeta)=\phi^{0}(t^{0}+\zeta - t)e^{-\int_{t^{0}}^{t}b\left(\xi,\xi+\zeta-t\right)\d\xi}+\int_{t^{0}}^{t}e^{-\int_{\eta}^{t}b\left(\xi,\xi+\zeta-t\right)\d\xi}y\left(\eta,\eta+\zeta-t\right)\,\d\eta.\end{aligned}
\label{appendixa16}
\end{equation}
Finally, we get
\begin{equation}
u(t, \zeta) = 
\begin{cases}
\Psi(t-\zeta)e^{-\int_{0}^{\zeta} b(\xi + t - \zeta, \xi) \d\xi} + \int_{0}^{\zeta} e^{-\int_{\eta}^{\zeta} b(\xi + t - \zeta, \xi) \d\xi} y(\eta + t - \zeta, \eta) \,\d\eta & \text{for } t-t^{0} \geq \zeta, \\
\phi^{0}(t^{0}+\zeta - t) e^{-\int_{t^{0}}^{t} b(\xi, \xi + \zeta - t) \d\xi} + \int_{t^{0}}^{t} e^{-\int_{\eta}^{t} b(\xi, \xi + \zeta - t) \d\xi} y(\eta, \eta + \zeta - t) \,\d\eta & \text{for } t-t^{0}<\zeta,
\end{cases}
\label{appendixa17}
\end{equation}
which can be written as 
\begin{equation}
u(t, \zeta) = 
\begin{cases}
\Psi(t-\zeta)e^{-\int_{0}^{\zeta} b(\xi + t - \zeta, \xi) \d\xi} + \int_{0}^{\zeta} e^{-\int_{\eta}^{\zeta} b(\xi + t - \zeta, \xi) \d\xi} y(\eta + t - \zeta, \eta) \,\d\eta & \text{for } t-t^{0} \geq \zeta, \\
\phi^{0}(t^{0}+\zeta - t) e^{-\int_{t^{0}+\zeta-t}^{\zeta} b(\xi+t-\zeta, \xi) \d\xi} + \int_{t^{0}+\zeta-t}^{\zeta} e^{-\int_{\eta}^{\zeta} b(\xi+t-\zeta, \xi) \d\xi} y(\eta+t-\zeta, \eta) \,\d\eta & \text{for } t-t^{0}<\zeta.
\end{cases}
\end{equation}
This completes the proof.
\end{proof}

\subsection{Particle Filter Method}\label{sec_particle}

The particle filter is a Sequential Monte Carlo (SMC) inferential technique based on repeated use of importance sampling. 
The construction of a particle filter in this disease model effectively combines the dynamics of the proposed compartmental model 
with available observational data to managing the errors in both sources.
The state vector $x_t$ represents the latent state of the system at time $t$ and $y_{1:t}$ 
represents the time series of observations up to time $t$. 
The latent variables $x_t$ are modeled as a Markov process with initial distribution $p(x_0)$ and transition probabilities 
$p(x_t| x_{t-1})$, see \cite[Sec.~2.2]{Papageorgiou2024} for more detail. 
Basically, the particle filter estimates the posterior distribution $p(x_{0:t}| y_{1:t})$ recursively over time 
$t = 1, 2, \ldots, N$, using $N_p$ particles $\left\{x_{t}^{(j)}\right\}_{j=1}^{N_p}$. 
The posterior is approximated as:
\begin{equation}
p(x_{0:t}| y_{1:t}) \approx \sum_{j=1}^{N_p} w_t^{(j)} \delta_{x_{t}^{(j)}}(x_{0:t}),
\end{equation}
where $w_t^{(j)}$ are the normalized weights and $\delta_{x_{0:t}^{(j)}}(\cdot)$ is the Dirac measure centered at $x_{0:t}^{(j)}$.
Given an approximation to the filtered distribution at time $t$ an approximation at time $t+1$ can be obtained by performing the following steps for each particle $j = 1, 2, \ldots, N_p$
\begin{itemize}
\item \textbf{Initialization}: Sample initial particles $x_0^{(j)} \sim p(x_0)$.
\item \textbf{Predication}: Obtain a new predication particle for each particle by propagating the model 
using the transition probabilities $\tilde{x}_{t}^{(j)} \sim p(x_{t} | x_{t-1}^{(j)})$.
\item \textbf{Weights}: Compute importance weights based on the likelihood of observation 
$\tilde{w}_{t}^{(j)} = p(y_{t} | \tilde{x}_{t}^{(j)})$ and normalize the weights
$w_{t}^{(j)} = \frac{\tilde{w}_{t}^{(j)}}{\sum_{k=1}^{N_p} \tilde{w}_{t}^{(k)}}$.
\item \textbf{Resampling}: Particles are sampled with replacement from $\{\tilde{x}_{t}^{(j)}\}_{j=1}^{N_p}$ 
according to their normalized weights $\{w_{t}^{(j)}\}$, producing the new set $\{x_{t}^{(j)}\}$. 
Particles with very small weights are likely to be disappear, while those with larger weights are likely to be duplicated multiple times.
\end{itemize}
This procedure is applied recursively beginning with uniform initial weights $w_0^{(j)} = 1/N_p$ and initial states $x_0^{(j)}$. For estimating parameters instead of the state‐evolution  model, we adopted a random‐walk model for particle filter estimation \cite{storvik2023}. We model $g_t$ as a random walk as
$$g_t = g_{t-1} + \epsilon_t, \qquad \epsilon_t \sim \mathcal{N}(0, 0.01).$$
 In this study, we used $N_p = 1000$ particles to approximate the posterior distribution.
 
\section*{\textbf{CRediT authorship contribution statement}}
\textbf{I. Haq}: Writing – original draft, Software, Methodology, Conceptualization; \textbf{S. Richard}: Writing – editing, Methodology, Conceptualization, Formal analysis 

\section*{\textbf{Data availability}}
   All the data used in this article are available at  \cite{Brazil_source}, \cite{Brazil_weather} and \cite{El_nino}.
\section*{\textbf{Code availability}}
The code that supports the findings of this study is available from the corresponding author upon reasonable request.
\section*{\textbf{Funding}}
This research received no specific grant from any funding agency in the public, commercial, or not-for-profit sectors.
\section*{\textbf{Competing interests}}
 We declare that we have no competing interests.

\end{document}